\documentclass[11pt]{article}
\usepackage[utf8]{inputenc}
\usepackage{amsmath}
\usepackage{fullpage}
\usepackage{graphicx}
\usepackage{subcaption}
\usepackage{booktabs}
\usepackage{lscape}
\usepackage{tablefootnote}
\usepackage{longtable}
\usepackage{multirow}
\usepackage{dsfont}
\usepackage{pdflscape}
\usepackage{bm}
\usepackage{vmargin}
\usepackage{float}
\usepackage[citecolor=blue,colorlinks=true,linkcolor=blue]{hyperref}
\usepackage[comma,super]{natbib}
\usepackage[parfill]{parskip}
\setmargrb{25mm}{25mm}{25mm}{25mm}
\usepackage[noblocks]{authblk}

\usepackage{url}

\DeclareMathOperator{\logit}{logit}

\title{\vspace*{-1cm}\sc Diversity of symptom phenotypes in SARS-CoV-2 community infections observed in multiple large datasets}
\author[1,2,3]{Martyn Fyles}
\author[4,5,6]{Karina-Doris Vihta}
\author[7,8]{Carole H Sudre}
\author[3]{Harry Long}
\author[1]{Rajenki Das}
\author[9,2]{Caroline Jay}
\author[10,11,12]{Tom Wingfield}
\author[3]{Fergus Cumming}
\author[3]{William Green}
\author[3]{Pantelis Hadjipantelis}
\author[3]{Joni Kirk}
\author[13,14]{Claire J Steves}
\author[5]{Sebastien Ourselin}
\author[15,16]{Graham F Medley}
\author[15,16,17,$\dagger$]{Elizabeth Fearon}
\author[1,2,18,$\dagger$,$\ast$]{Thomas House}
\affil[1]{Department of Mathematics, University of Manchester, Manchester, UK}
\affil[2]{The Alan Turing Institute for Data Science and Artificial
Intelligence, London NW1 2DB, UK.}
\affil[3]{United Kingdom Health Security Agency (UKHSA).}
\affil[4]{Nuffield Department of Medicine, University of Oxford, Oxford, UK.}
\affil[5]{Department of Engineering, University of Oxford, Oxford, UK.}
\affil[6]{National Institute for Health Research Health Protection Research
Unit in Healthcare Associated Infections and Antimicrobial Resistance at the
University of Oxford, Oxford, UK.}
\affil[7]{School of Biomedical Engineering \& Imaging Sciences, King's College
London, London, UK.}
\affil[8]{MRC Unit for Lifelong Health and Ageing, University College London,
London, UK.}
\affil[9]{Department of Computer Science, University of Manchester, Oxford
Road, Manchester, M13 9PL, UK.}
\affil[10]{Department of Clinical Sciences and International Public Health,
Liverpool School of Tropical Medicine, Liverpool L3 5QA, UK.}
\affil[11]{Tropical and Infectious Disease Unit, Liverpool University Hospitals
NHS Foundation Trust, Liverpool L7 8XP, UK.}
\affil[12]{WHO Collaborating Centre on Tuberculosis and Social Medicine,
Department of Global Public Health, Karolinska Institutet, 171 77 Stockholm,
Sweden.}
\affil[13]{Department of Twin Research and Genetic Epidemiology King's College London, London, UK.}
\affil[14]{Department of Ageing and Health Guy's and St Thomas' NHS Foundation Trust.}
\affil[15]{Centre for the Mathematical Modelling of Infectious Disease, London
School of Hygiene and Tropical Medicine, London WC1E 7HT, UK.}
\affil[16]{Department of Global Health and Development, London School of
Hygiene and Tropical Medicine, London WC1E 7HT, UK.}
\affil[17]{Institute for Global Health, University College London, WC1E 6BT, UK.}
\affil[18]{IBM Research, Hartree Centre, Daresbury WA4 4AD, UK.}
\affil[$\dagger$]{Contribution considered equal.}
\affil[$\ast$]{Corresponding Author: thomas.house@manchester.ac.uk}
\date{}

\begin{document}

\maketitle

\clearpage

\begin{abstract}
\noindent{}Variability in case severity and in the range of symptoms experienced has been apparent from the earliest months of the COVID-19 pandemic. From a clinical perspective, symptom variability might indicate various routes/mechanisms by which infection leads to disease, with different routes requiring potentially different treatment approaches. For public health and control of transmission, symptoms in community cases were the prompt upon which action such as PCR testing and isolation was taken. However, interpreting symptoms presents challenges, for instance, in balancing the sensitivity and specificity of individual symptoms with the need to maximise case finding, whilst managing demand for limited resources such as testing. For both clinical and transmission control reasons, we require an approach that allows for the possibility of distinct symptom phenotypes, rather than assuming variability along a single dimension. Here we address this problem by bringing together four large and diverse datasets deriving from routine testing, a population-representative household survey and participatory smartphone surveillance in the United Kingdom. Through the use of cutting-edge unsupervised classification techniques from statistics and machine learning, we characterise symptom phenotypes among symptomatic SARS-CoV-2 PCR-positive community cases. We first analyse each dataset in isolation and across age bands, before using methods that allow us to compare multiple datasets. While we observe separation due to the total number of symptoms experienced by cases, we also see a separation of symptoms into gastrointestinal, respiratory and other types, and different symptom co-occurrence patterns at the extremes of age. In this way, we are able to demonstrate the deep structure of symptoms of COVID-19 without usual biases due to study design. This is expected to have implications for the identification and management of community SARS-CoV-2 cases and could be further applied to symptom-based management of other diseases and syndromes. 
\end{abstract}

\section{Introduction} Since the identification of the SARS-CoV-2 virus, the COVID-19 pandemic has led to over 700 million confirmed cases and 7 million confirmed deaths. In response to this, one of the largest ever
public health responses has been mounted, with over 13 billion vaccine doses administered and a large variety of non-pharmaceutical interventions that have fundamentally changed behaviour and healthcare provision around the world since the start of 2020\citep{WHO_Sitrep,hale_global_2021,GoogleMobility}.

Understanding the clinical presentation and course of SARS-CoV-2 infection
has been central for transmission control, particularly in determining policy
for identification of cases for isolation and tracing of their
contacts\citep{Fyles:2021,Croziern1625}, and for prediction of clinical
outcomes.  COVID-19 cases can present with symptoms from a wide range of
categories: respiratory, systemic, cardiovascular and
gastrointestinal\citep{struyf_signs_2020}, with high variability in severity of disease between
individuals depending strongly on age, and on factors including
comorbidities\citep{williamson_opensafely_2020, clift_2020_living_risk}. In
addition, a significant proportion of infections are estimated to remain
asymptomatic\citep{buitrago-garcia_occurrence_2020}. Analyses of symptom clustering patterns amongst hospitalised patients have helped to improve clinical care\cite{millar_robust_2020}, while longitudinal clustering approaches contribute to early identification of cases more likely to experience severe outcomes\cite{sudre_symptom_2020}, though these cases will be those with the more severe disease course.

Targeted population transmission control policies that do not simply require the whole population to avoid contact, with all the damage that this entails, require identifying infectious cases. In the absence of regular population-wide screening approaches, case identification requires a symptoms-based approach, and this was a central pillar in the UK's COVID-19 response from May 2020 until April 2022. Over this time period, PCR testing was initiated when individuals from the broader population experienced at least one of; fever, new continuous cough or loss of taste or smell. Because PCR testing requires laboratories and associated staffing and logistical networks, there was a need to balance the sensitivity of symptom criteria for testing with specificity, given the varieties of infections and syndromes that could give rise to the relevant symptoms. Thus, the effectiveness of testing, contact tracing and isolation policies were dependent in part on the performance of the symptom criteria for testing. A number of single study analyses in the UK have sought to investigate and to improve upon these criteria, sequentially adding or dropping symptoms to better optimise the sensitivity and specificity trade-off\cite{Elliott2021.02.10,
Fragaszy2021.05.14}. While valuable, these approaches essentially assume a single phenotype, whereas it is possible that multiple phenotypes exist and, therefore, that the trade-off cannot be optimised in this manner.

Assessment of the diversity of genotypes and (endo-)phenotypes is a long-standing
tool in both infectious diseases and chronic non-communicable diseases, which
has been significantly accelerated by modern experimental and theoretical
techniques\citep{Hofmann:2011,Deliu:2017,Geifman:2018}. In particular, such
analysis often helps with the standard process of identifying multiple disease
aetiologies with the same presentation, or vice versa, a single disease with
highly variable outcomes. This latter distinction is particularly important for
COVID-19, where different courses of action, including public health
interventions, are taken depending on symptom status\citep{NHSTT:2021}. Beyond the acute phase of the disease, consideration of different potential `long COVID' phenotypes based on symptom status could help to identify more or less appropriate treatment approaches. 

Here, to investigate the presence of distinct COVID-19 symptom phenotypes, we investigate patterns of symptom \emph{occurrence}, \emph{co-occurrence} and \emph{clustering}
in PCR-positive symptomatic SARS-CoV-2 cases -- previously considered predominantly in
hospitalisation data heavily skewed towards more severe infections
\cite{Swannm3249,millar_robust_2020,sudre_symptom_2020} -- in four very large community-based datasets.

\section{Methods}

\subsection{Data}

\subsubsection{Population and setting}

We examine identified infections for the time period May 2020 to March 2021 in
the UK. Due to this data collection time period, and the effects of vaccination
on preventing disease, we expect the datasets to contain predominantly
unvaccinated individuals. These datasets are diverse in their sampling and data
collection methods and include (a) 1,637,965 symptomatic cases from `Pillar 2'
testing data from the National Health Service (NHS) Test and Trace system,
designed to capture cases in the general population; (b) 112,925 symptomatic
cases from the Second Generation Surveillance System (SGSS) in England's
national laboratory reporting system, which includes cases associated with
healthcare settings among patients and healthcare staff; (c) 52,084 symptomatic
self-reported cases from the COVID-19 Symptom Study (CSS), which uses a
smartphone app associated with \url{https://COVID.joinzoe.com/} to collect daily
symptom reports; and (d) 9,166 symptomatic cases from The Office for National
Statistics COVID-19 Infection Survey (CIS), a longitudinal study of a representative sample of UK households. 

\subsubsection{NHS Test and Trace routine testing data}

NHS Test and Trace data is further split into two parts: Pillar 2, cases
detected in the community, usually on the basis of symptoms to initiate
testing; and the Second Generation Surveillance System (SGSS), for people
tested in healthcare settings. In May 2020, the UK government made PCR testing
available for individuals who had one of the following symptoms: a new,
continuous cough; fever; loss of taste or loss of smell. These tests are
reported through Pillar 2, through which several different avenues to testing
are available. Individuals can book a test appointment through a government
website for either a drive-in or walk-through testing centre, where they
self-swab their nose and throat (under some supervision, with an adult carer
conducting the swabbing for children), with the swab then sent to a lab for PCR
testing. Alternatively, individuals can order home test kits where they
self-swab at home and post the kit back, with the swab again sent to a lab for
PCR testing. If the individual tests positive, their case is transferred to NHS
Test and Trace who contact cases to inform them of their result and ask them to
conduct a questionnaire including symptoms experienced. The questionnaire is
conducted either via a web form or over the phone with a trained contact
tracer. Since the end of 2020, Pillar 2 has also included positive cases
identified using rapid antigen tests among people not experiencing one of the
PCR test prompting symptoms. These tests also use a nasopharyngeal swab and are
conducted at the home, workplace or school and, if positive, are requested to be
followed up by a confirmatory PCR test (though policy has varied over time).
Reported positive cases from asymptomatic testing are also followed up by NHS
Test and Trace. 

The Second Generation Surveillance System (SGSS) dataset includes people who test because they work in or have been tested in a healthcare setting as a patient. This latter group includes both those in hospital because of severe COVID-19 symptoms, but also those in hospital for other reasons but receiving SARS-CoV-2 testing. Thus they are likely, to be more severe cases in the SGSS versus Pillar 2 data, but not exclusively. Again, individuals are swabbed and PCR tested, with their case transferred to NHS Test and Trace if testing positive for symptom reporting and contact tracing.

\subsubsection{COVID Symptom Study (CSS)}
The CSS is a participatory surveillance study collecting data via a smartphone app. It is led by Kings College London and Zoe Global Ltd and was initiated in March 2020 in the United Kingdom and the United States\citep{Drew1362}. Individuals are asked to report daily whether they are feeling `physically normal' that day and, if not, what symptoms they are experiencing. As well as demographic data that is collected upon sign-up, participants are also asked to self-report whether they have had any tests for SARS-CoV-2 infection and, if so, the date of the test and its result. Demographic data and data about underlying conditions are collected at first registration. Participants can also proxy-report for children or for others they care for (e.g elderly adults they care for). 

As well as enabling individuals to self-report COVID-19 testing that they have undertaken via the UK’s routine testing programmes or surveillance studies, the CSS invites individuals to complete a PCR test via routine testing if they 1) have made at least one report of no symptoms in the previous week and 2) report a new symptom not on the list to prompt symptomatic testing (e.g sore throat). This means that we might expect the CSS reporting to be less dominated by the symptoms required to initiate symptom-based testing than the Pillar 2 routine testing dataset. 

\subsubsection{ONS COVID-19 Infection Survey (CIS) }
The CIS is a UK population-representative survey of households randomly selected continuously since April 2020 from address lists and previous surveys\citep{ONSKoen}. Households are followed longitudinally with weekly visits for the first month and monthly visits for 12 months from enrolment. A fieldworker attends enrolled households each visit for testing for household members aged 2 years and above and to conduct an interview including, among other topics, demographic data (reported at the first visit) and symptoms experienced over the previous 7 days. At each visit, participants conduct a nose and throat swab under the supervision of a fieldworker. These swabs are sent for PCR testing, and the result is communicated to participants. At the same time as swabbing, all participants are also interviewed by the fieldworker to complete a symptom questionnaire.
 
 \subsection{Data extraction and preparation}

From each dataset, we extract all PCR-positive individuals and associate them
with symptoms experienced within a time window of the test appropriate for the
dataset. More detail about each dataset, data collection and extraction are given in the Supplementary Materials. For the $i$-th individual and $a$-th symptom, we let $X_{ia} = 1$ if
the symptom is present during the time window around the positive test and
$X_{ia} = 0$ otherwise. For a dataset with $n$ individuals measuring $p$
symptoms, we can then construct an $n\times p$ matrix $[X_{ia}]$, where the rows of
this matrix form a set of $n$ length-$p$ feature vectors for individuals,
$\{\mathbf{y}_i \}$, and the columns form a set of $p$ length-$n$ feature
vectors for symptoms, $\{\mathbf{x}_a \}$, each of which can then be used as
input for unsupervised learning algorithms. In addition to descriptive analysis
of the data, we used three complementary approaches to looking at clustering
and co-occurrence of symptoms.

\subsubsection{Sample populations}

The dates over which cases are collected from each study are shown in
Table~\ref{tab:dates} and in total cover the period from April 2020 to March
2021, with the largest overlap between November 2020 and January 2021. We make
no exclusions based on age or other characteristics.

From each dataset, we include only cases that report at least one symptom within
the symptom reporting window around a positive test (detailed below and listed
in Table~\ref{tab:symptom_questions}). For NHS Test and Trace data, positive
cases who are never reached and interviewed post-testing are not included in
this dataset. The definition of `symptomatic' necessarily varies across the
datasets because there are differences in the full list of symptoms asked
about. Symptoms that were not core dataset variables and were instead recorded
by manual entry were not included. For each dataset, we chose to include all
dataset symptoms from each study (except for `write-in' symptoms), rather than
excluding symptoms that were not common across all. This was with the intention
of maximising the amount of symptom information available for analysis. We also
extracted demographic information.

It is expected that symptom data from the same PCR-positive cases is captured across the NHS Test and Trace, CIS and CSS datasets. Explicit deduplication of individuals across datasets was not performed but is expected to have no impact on the findings.

The proportion of symptomatic cases varies significantly between datasets, reflecting their different sampling. NHS Test and Trace Pillar 2 and CSS both have the highest number of symptomatic cases, which is not surprising given that both datasets mainly focus on symptom-initiated testing. The NHS Test and Trace SGSS dataset has the next highest proportion of symptomatic cases. We expect to see some asymptomatic screening in SGSS populations, which may explain the decrease in symptomatic cases when comparing Pillar 2 to SGSS. In the CIS study, we see a much smaller proportion of symptomatic cases, likely due to the sampling strategy being independent of symptoms, therefore resulting in asymptomatic and pre-symptomatic individuals testing positive and being included in the study.

\begin{figure}[H]
    \centering
    \includegraphics[width=\linewidth]{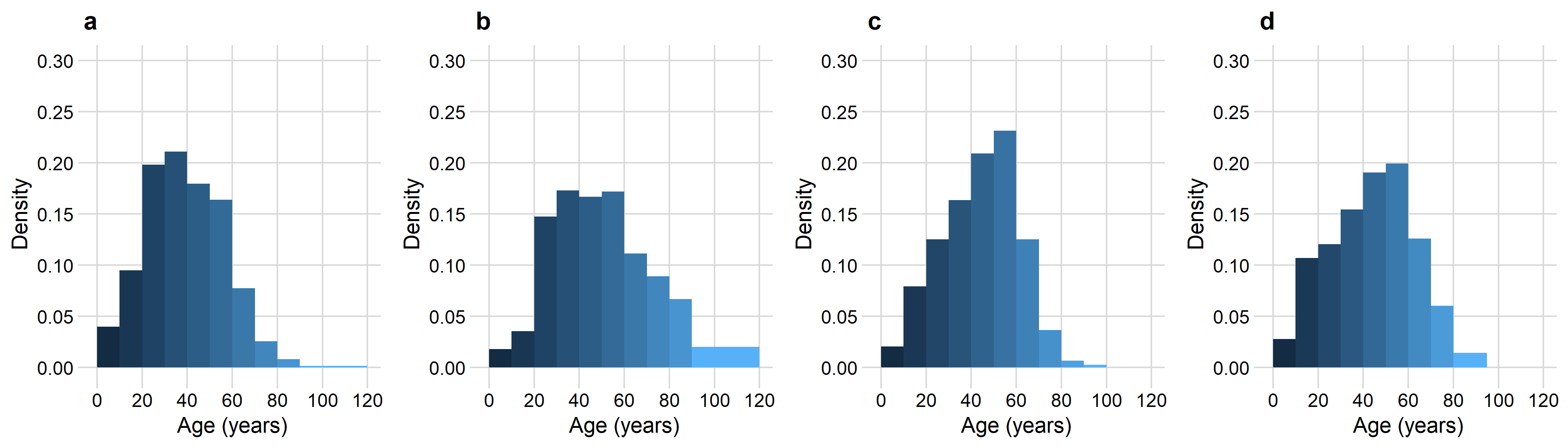}
    \caption{Histograms showing the age density for each dataset. \textbf{a}.
Pillar 2, \textbf{b}. SGSS, \textbf{c}. COVID Symptom Study, \textbf{d}.
COVID-19 Infection Survey. } \label{fig:AgeDistributions}
\end{figure}

The Pillar 2 routine testing contributed by far the largest number of cases to
the study, with CIS the fewest. While all datasets contained a slight female
majority, with CSS the largest (61.8\%), there was some variability in the age
distribution of cases (Figure~\ref{fig:AgeDistributions}); Pillar 2 routine testing was the youngest,
while SGSS included the oldest groups. This is likely to be because SGSS more
heavily represents a hospitalised population. CSS and CIS are UK-wide, while
NHS Test and Trace data contains cases testing in England.

The characteristics of the infected subpopulation relative to the general UK population have likely changed over time for a multitude of reasons; different levels of restrictions and lockdown across different localities, vaccination coverage and uptake, varying prevalence, weather, levels of outdoors mixing, incentives to ignore social distancing, workplace/school closures and changing availability of testing. Moreover, each study and route of data collection results in different samples of the infected population. The CIS is a population-based household sample and thus should be broadly representative (participation biases aside), those discovered through routine testing (NHS Test \& Trace) may overrepresent a population adherent to testing guidance, those prone to more severe infections and the sub-populations with the highest prevalence and testing-seeking behaviour. For the CSS's app-based reporting, then the sub-populations with high levels of smartphone ownership and compliance are likely to be over-represented.

\subsubsection{Symptom data}

Data is collected at the level of the symptoms experienced by an individual,
and for the majority of datasets we have a binary outcome of whether an
individual experienced a symptom or not. Exact symptom questions and lists are
given in Table~\ref{tab:symptom_questions}. In CSS, an individual is able to
choose from several levels of fatigue: ``none", ``mild" or ``severe". Our
planned analyses are designed to work with binary data, and as a result, we map
multiple levels into a binary outcome variable. When performing this mapping,
we choose to merge levels together, with the aim of making the symptoms as
comparable as possible to what is reported in other datasets. Datasets with a
binary fatigue variable report 40-60\% of cases, which is consistent with most
cases only reporting severe fatigue; if we included mild fatigue then we find
that close to 80\% of cases report fatigue which is inconsistent with what is
reported in the other datasets.

\subsubsection{Symptom reporting windows}

The symptom reporting window and its timing relative to the positive test
varies across the datasets. For CIS, participants are asked about symptoms in
the previous 7 days prior to testing. For cases contacted and interviewed by
NHS Test and Trace (Pillar 2 and SGSS), individuals are asked to report
symptoms that they are currently experiencing. For CSS, individuals are
prompted to report symptoms daily but for this dataset, we include all symptoms
reported in the 14 days before and 14 days after the date a positive test is
reported (note this does not mean that all participants report symptoms with
that level of frequency).

From the time of infection, individuals usually have a few days before they become symptomatic, while test sensitivity also varies over the course of infection, peaking around the time or just before symptom onset. Previous studies have also found patterns in the types of symptoms that present earlier versus later in the course of an infection\citep{sudre_symptom_2020}. Across each of the included datasets, the time in an individual's infection at which they are tested on average and over which they are asked to report symptoms varies. For CIS time of testing over the course of infection should be random over the period at which someone will test PCR-positive; for data primarily from symptomatic testing, it should be a few days post-symptom onset (reflecting a delay between onset and testing, test result and follow-up interview with Test and Trace). For CSS, the time of testing for many will reflect symptomatic testing in the community and some proportion of individuals with particular symptom reporting patterns are asked to obtain a test through NHS Test and Trace symptomatic testing routes.

\subsubsection{Symptom classification}
To aid interpretation we classify symptoms according to their clinical characteristics. These classifications were made a priori in consultation with an infectious diseases clinician (TW) with experience in caring for people with COVID-19 and without input from observed clustering patterns. We included systemic symptoms, lower respiratory, upper respiratory, gastrointestinal, altered state symptoms and `other' symptoms that did not fit into any of these categories.

\subsubsection{Research ethics}
The secondary analyses described in this paper received ethical approval from the London School of Hygiene and Tropical Medicine (22752). The COVID Symptom Study was approved by the Partners Human Research Committee (Protocol 2020P000909) and King's College London ethics committee (REMAS ID 18,210, LRS-19/20–18,210) and the CIS received ethical approval from the South Central Berkshire B Research Ethics Committee (20/SC/0195). All methods were performed in accordance with the relevant guidelines and regulations. Informed consent was obtained from all subjects and/or their legal guardian(s).

\subsection{Analysis}

We describe the frequency with which each symptom was reported in each dataset, categorising them using our symptom classification. We then perform three unsupervised learning techniques, each with a different but complementary aim. Our goal is to understand patterns of symptom co-occurrence and if there is any evidence of symptom clustering, as multiple distinct clusters would be evidence for the existence of distinct COVID-19 symptom phenotypes.

\subsubsection{Jaccard Distance}

We use a variety of methods to understand the behaviour of symptoms and the
analyses are sometimes performed on the Jaccard distance matrix of symptoms. 
Letting $\mathbf{x}_i$ be the feature vector constructed from the presence or absence of symptom $i$ in cases, which has $k$-th element $x_{ik}$,
the Jaccard distance between two such vectors is defined as
\begin{equation}
 D_{\mathrm{Jac}}(\mathbf{x}_i,\mathbf{x}_j) =
 1 - \frac{\sum_{k=1}^n x_{ik}x_{jk}}{\sum_{k=1}^n (x_{ik} + x_{jk} - x_{ik}x_{jk})}
\, . \label{jacdef}
\end{equation}
The simple interpretation of Jaccard distance is then; the proportion of cases who did not experience both symptoms $i$ and $j$, given that they experienced at least one of symptoms $i$ or $j$. In the case of missing data, the Jaccard distance is computed using only the subset of individuals for which there is no missing data for either symptoms $i$ and $j$.

\subsubsection{Hierarchical clustering}

This method starts with a set of symptoms, and the feature vector for each is
constructed from their presence or absence in individuals with a positive test
and report of at least one symptom (i.e. those positive cases not excluded as
asymptomatic). The Jaccard distance is used as an appropriate metric for such
binary data. Clusters of symptoms are agglomeratively joined on the dendrogram
produced on the basis of the maximum distance between cluster members (called
‘complete linkage’). Symptoms with a low shortest distance between each other
on the final dendrogram tend to co-occur, and those with a long distance are
not often both present. Clusters can also be identified by ‘cutting’ the
dendrogram at a given distance.

\subsubsection{Logistic PCA (LPCA)}

LPCA is an extension of principal component analysis (PCA) to binary data, and
reduces the dimension of the symptom space in a manner that preserves the maximum level of variance between individuals (rather than
symptoms)\citep{landgraf2020dimensionality}. The projection values of symptoms
onto lower-dimensional basis are called loadings, and these demonstrate the
directions in which individual phenotypes most commonly vary. In practice, the
first component is likely to have relatively even contributions from each of
the symptoms, and will represent an overall severity of illness at the
individual level, with subsequent components demonstrating more subtle ways in
which symptoms can vary.

Given an $n\times p$ binary data matrix $\bm{X} = [X_{ia}]\in \{0, 1\}^{(n\times p)}$, our aim is to find a low dimensional representation of the natural parameter matrix $\bm{\Theta} = [\theta_{ia}]$, where $\mathds{P}(X_{ia} = 1) = \logit^{-1}(\theta_{ia})$. This is achieved by finding $\bm{\hat\Theta}_k$, a rank-$k$ approximation of $\bm\Theta$ such that the Bernoulli deviance, $D_{\mathrm{Ber}}(\bm{X};\bm{\hat\Theta}_k)$ is minimised. This is conceptually related to logistic regression models, as these also attempt to minimise the Bernoulli deviance. In practice, the minimisation is solved over $\bm{U} \in \mathds{R}^{k\times p}$, such that $\bm{UU}^{\top} = \bm{I}$, with $\bm{\hat\Theta}_k = \bm{UU}^{\top}\bm{X}$. The column vectors of $\bm{U}$ are the loadings onto the principal components.

As with all dimensionality reduction techniques, we need to choose the number of dimensions in our low-dimensional approximation. We follow the recommendation of \citet{landgraf2020dimensionality}, and examine the change in the Bernoulli deviance as we increase $k$. Consider a rank-$0$ approximation, where $\bm{\hat\Theta}_0 = \bm{1}_n\bm{\hat\mu}^{\top}$ for $\hat\mu \in \mathds{R}^n$. That is to say that the natural parameter matrix contains a constant value in every column. This is treated as the null model, to which all other models are compared.

For a model with $k$ components, the proportion of Bernoulli deviance explained relative to the null model is given by 
\begin{equation}
    P(k) = 1 - \frac{D_{\mathrm{Ber}}(\bm{X};\bm{\hat\Theta}_k)}{D_{\mathrm{Ber}}(\bm{X};\bm{1}_n\bm{\hat\mu}^{\top})}
\, .
\end{equation}
If $k=p$, then $D_{\mathrm{Ber}}(\bm{X};\bm{\hat\Theta}_k) = 0$, as the model is saturated and $\logit^{-1}\hat\theta_{ia} = X_{ia}$, thus resulting in $P(d) = 1$. This means that $P(k)$ can be interpreted similarly to standard PCA, in the sense that $P(k)\times 100\%$ of the variance is explained by the first $k$ components. The marginal Bernoulli deviance, $M(k)$, is the change in the Bernoulli deviance explained by adding the $k^{th}$ component, for $k\geq1$, defined as
\begin{equation}
    M_{\mathrm{Ber}}(k) := P(k) - P(k-1) =
\frac{D_{\mathrm{Ber}}(\bm{X};\bm{\hat\Theta}_k) - D_{\mathrm{Ber}}(\bm{X};\bm{\hat\Theta}_{k-1})}%
{D_{\mathrm{Ber}}(\bm{X};\bm{1}_n\bm{\hat\mu}^{\top})} \, .
\end{equation}
When selecting the number of components in our low-dimensional representation of the data, we primarily focus on the marginal Bernoulli deviance and aim to find the largest $k$ such that for $k' > k$ the marginal Bernoulli deviance decreases rapidly. We also examine the proportion of Bernoulli deviance explained - if this gets close to 1, then that suggests we have selected too many components and are over-fitting.

In practice, two hyperparameters need to be chosen for logistic PCA: the number of components $k\in \mathds{N}$, and $m\in\mathds{R}_+$ which controls the magnitude of the loadings. The optimal choice of $m$ varies depending upon $k$ and is selected by leave-one-out cross-validation for a range of proposed $m$ values.

\begin{figure}[H]
    \centering
    \includegraphics[width = 0.7\textwidth]{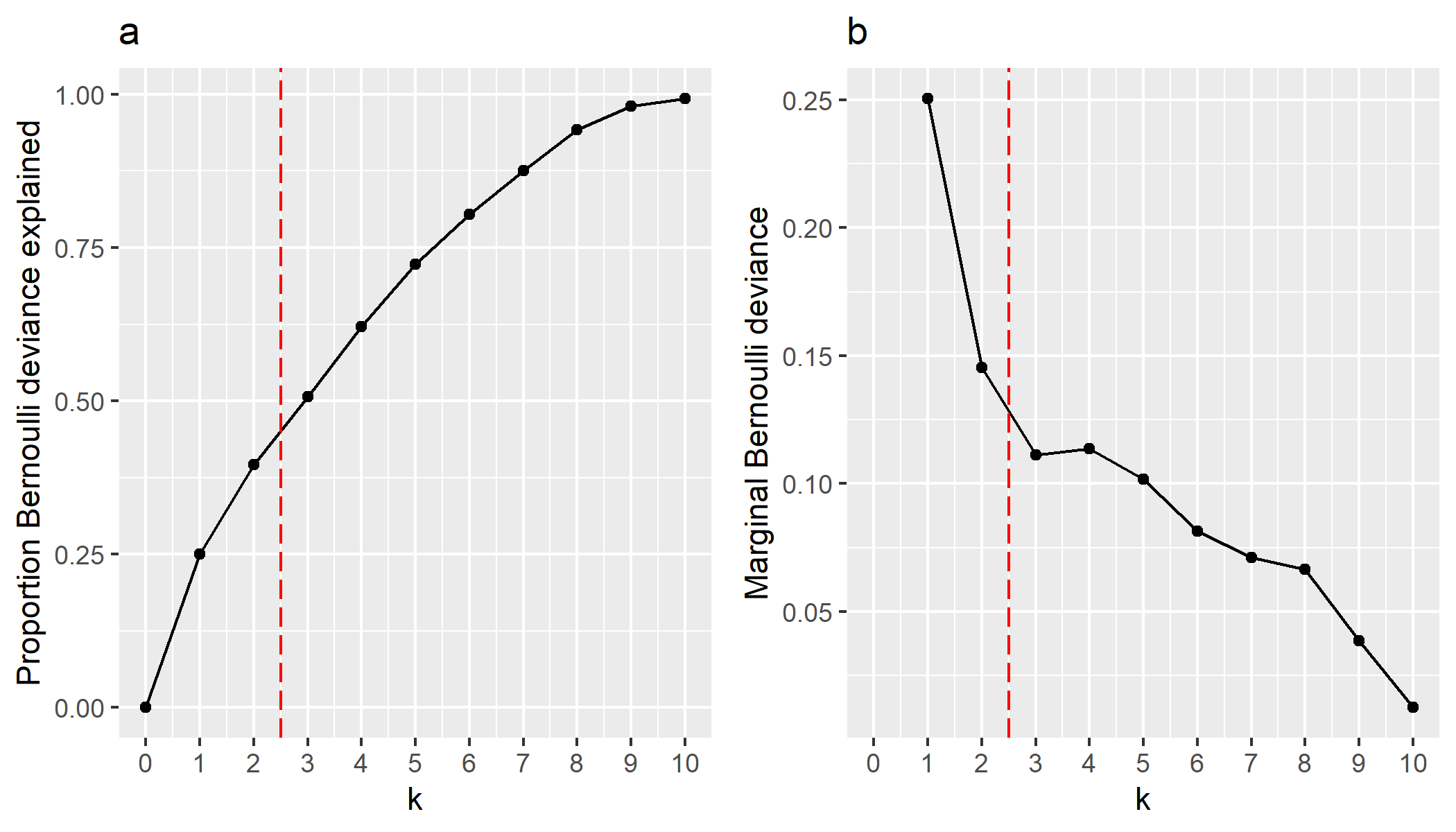}
    \caption{\textbf{a}. the proportion of the Bernoulli deviance explained using an LPCA model with $k$ components. \textbf{b}. the proportion of the Bernoulli deviance explained by adding the $k^{th}$ component to the model. In this example, we would select $k=2$ as the true number of components, as indicted by the vertical dashed red line.}
    \label{fig:LPCA model selection}
\end{figure}

An example of the model selection is plotted in Figure \ref{fig:LPCA model selection}. In the plotted example, we would choose $\hat{k} = 2$, indicated by the vertical dashed line. This is due to the first two components, having a significantly higher marginal Bernoulli deviance than all models with $k>2$ components. The marginal Bernoulli deviances for models where $k\in\{3,\dots,8\}$ have small differences between successive values of $k$, making it hard to favour one model over the other. For $k>8$, the marginal Bernoulli deviance does decrease rapidly; however, at this point we have explained close to 100\% of the Bernoulli deviance and are overfitting the model at this point. Hence, for this example we choose $\hat k = 2$.

If we select $k=2$ components, then approximately 33\% of the Bernoulli deviance of the saturated model is explained. In classic principal component analysis, ideally, the model would find a number of components that explains as close to 100\% of the variance while not overfitting to noise. In logistic principal component analysis, however, if a model explains close to 100\% of the Bernoulli deviance relative to the null Model, then this is indicative of dramatic over-fitting. For example, the saturated model where $k=d$ will exactly reproduce the input data and explains 100\% of the Bernoulli deviance. The true natural parameter matrix will not explain 100\% of the Bernoulli deviance, as it tells us about the probability of a symptom occurring; this will lead to a non-zero Bernoulli deviance. As such, our goal is not to explain 100\% of the Bernoulli deviance, relative to the null model.

Model selection plots for the number of components $\hat k$ can be found in the code repository for this paper (\url{https://github.com/martyn1fyles/COVIDSymptomsAnalysisPublic}). During model selection, we found that in the majority of cases we would definitively choose $\hat k = 2$. In a small number of cases, it was ambiguous whether $\hat k=1$ or $\hat k=2$. In these ambiguous cases, we have opted to take $\hat k = 2$. Our reasons for doing so are: a priori, we believe at least two dimensions are necessary to capture the range of COVID-19 presentations; across all other datasets, we found that $\hat k = 2$, which we believe increases the prior likelihood that $\hat k = 2$ in the ambiguous cases. However, due to this uncertainty, we have made available in the code repository LPCA plots where we have taken $\hat k = 1$. Unlike in traditional PCA, LPCA components are dependent upon the total number of components selected, and as such PC1 for example differs depending upon the total number of components selected. This is why we must rerun the analysis when we take $\hat k = 1$ and present these results separately to results produced where we set $\hat k =2$.

\subsubsection{UMAP (Uniform Manifold Approximation and Projection)}

UMAP is a technique for dimension reduction of complex data based on pairwise distances between symptoms. In contrast to the other methods, it is designed to achieve good separation between unknown classes in the low dimensional space, and as such complements the other machine learning methods used above.

The specifics of the UMAP algorithm are mathematically complex, however, we will provide a brief overview of the algorithms' strategy. For further details on UMAP, we refer readers to the original UMAP paper\cite{UMAP} and for practical demonstrations and visualisations we refer readers to \cite{mcinnes_umap_2021, understanding_umap}. The first step of UMAP is to construct an object that describes the shape of the data in high-dimensional space. This object is, effectively, a weighted network where the edge weights represent the probability that two points are connected in the high dimensional space. Points in the high dimensional data are connected, based upon some locally defined notion of distance; if there are regions of the high dimensional space that are dense with points, then the algorithm requires points in that region to be very close when connecting them. If there are regions of the high dimensional space that are sparse, then larger distances are acceptable when connecting points. This locally defined notion of distance is important to ensure that all points in the high-dimensional space form a single connected component. Given this high-dimensional representation of the data, UMAP then attempts to find a network in the low-dimensional space that approximates the representation of the data in the high-dimensional space - this is referred to as an \emph{embedding} and is what we later visualise. Particular attention is given to attempts to preserve the distances between points - but only the distances between points that are connected. By only preserving the most important distances - those between points that are close in the high dimensional space - UMAP is able to produce good low dimensional representations of the high dimensional dataset, particularly when there are complex geometries involved.

The specifics of the UMAP algorithm result in some nuances when interpreting UMAP embeddings. Firstly, the embedding attempts to provide the shape of the data in the high-dimensional space. If the data is clustered in high dimensional space, then the data will also be clustered in the low dimensional space. However, due to the locally varying notion of distance, the compactness of the clusters will not be captured. In addition, if two points in the high-dimensional space are far away from each other, then they will not be connected, and as a result, UMAP will not attempt to preserve the distance between these points. As such, the exact distance between faraway points should not be over-interpreted. The rotation of UMAP plots is not important, only the overall shape of the data. Finally, UMAP is a stochastic algorithm, and there can be minor differences between runs.

To compute our embeddings, we computed the Jaccard distance matrix of symptoms and provided this as an input to UMAP, which we configured to embed the symptoms into a 2-dimensional Euclidean space.

For our datasets, it is also of interest to partition these into different segments of data, such as different age groups, and to compare how the clustering of symptoms changes across each of these segments. Performing this comparison is not straightforward using the base UMAP algorithm, however, as the resulting embeddings can preserve distances of the high dimensional data structure of each segment while looking visually distinct due to rotation or different layouts. To remedy this, AlignedUMAP is an extension of the base UMAP algorithm that attempts to solve this problem by enabling better comparisons between UMAP embeddings of different segments of data. Effectively, AlignedUMAP attempts to find an optimal embedding for each segment of a dataset, and then, as a secondary objective, to minimise the distance between embeddings of adjacent segments in the low dimensional space. This results in embeddings for different segments of data that can directly be compared to each other; we are able to see how distances in the high dimensional space change between segments.

We use AlignedUMAP to produce aligned embeddings for each dataset where we 1) partition the data into three large age bands of key groups (0-17 years, 18-54 years and 55 years), and 2) partition the data into 10-year intervals. 

Additionally, we also use AlignedUMAP to align the embeddings produced by different datasets. Each dataset reports a different selection of symptoms, and consequently, the alignment between different datasets is based upon only a core group of symptoms that are shared across all datasets. By aligning the embeddings produced for different datasets, we facilitate easier comparison between datasets, and we can explore if datasets share a common underlying structure.

There are a wide number of UMAP hyperparameters that can be adjusted, and finding the optimal combination is not a solved problem to our knowledge. As part of a sensitivity analysis, we have opted to produce two UMAP outputs for each dataset, one configured to produce a tight clustering of symptoms, and another configured to produce a loose clustering of symptoms. This is achieved by changing the number of neighbouring points that UMAP considers when constructing the high-dimensional representation of the data. As a result, smaller values of the \emph{n\_neighbours} parameter will configure UMAP to focus on local structures, and it may not capture the global structure - this produces what we refer to as a tight clustering and produces well-separated clusters. Setting the \emph{n\_neighbours} parameter to higher values will configure UMAP to focus less on the local structures of the symptoms but produce a more general clustering of the data -- this produces what we refer to as a loose clustering. Both loose and tight UMAP embedding will capture different parts of the symptom topology and produce complementary analyses. When we produce the loose UMAP embeddings that focus more upon the global structure of the data, we take \emph{n\_neighbours} = 4, and for the tight UMAP embeddings that focus more upon the local structure of the data, we take \emph{n\_neighbours = 2}. When using AlignedUMAP for the fine age strata, we align each segment of data with the two prior segments, and the two slice post the current segments.  The numerical sizes of different age segments used are shown in Table~\ref{tab:asaa}. We discuss how other UMAP hyperparameters were selected in our Supplementary Materials S1.2, however, we briefly summarise that these other hyperparameters either had less impact on the resulting embeddings, provided that they were not set to extreme values, or could be reasonably selected after due consideration.

\subsubsection{Analyses summary}

We first run hierarchical clustering, LCPA and AlignedUMAP for all the included
cases from each dataset. The UMAP Alignment is performed on common symptoms across datasets for these results, as an attempt to synthesize common structures across datasets.

AlignedUMAP is then run when each dataset is stratified into 10-year intervals and then plotted in 3D space. Here, the alignment is performed across different age strata. 

Moving on, we stratify each dataset into the different age groups (0-17 years, 18-54 years and 55 years and older) and rerun hierarchical clustering, LPCA and AlignedUMAP with the alignment performed within the dataset across the three age strata. For only three age strata, it is not necessary to plot the results in 3D space as was required for the results from the finer age strata. Given the large number of plots, the results from our age-stratified findings are presented in the supplementary materials.

\subsection{Pre-hoc considerations for comparison across datasets}

Because of the different sampling of positive cases and the resulting sample
composition, data collection methods, and symptom questions across the
datasets, we expect potential differences in findings arising from several
causes.

\subsubsection{Sampling}

The majority of routinely detected community cases in the UK were detected via symptom-prompted tests, particularly prior to the widescale availability of rapid antigen testing for asymptomatic individuals in the Spring of 2021. Thus we expect Pillar 2 to over-represent individuals with at least one of cough, fever and loss of taste and/or smell. This bias is also likely to exist within the CSS as a majority of self-reported tests would also have been performed because they met the symptom criteria for routine community testing, though the study also invited a proportion of regular app-user participants to test based on reporting other symptoms. These biases are not present within the ONS study sample.

\subsubsection{Data collection method}

Across all datasets, symptoms are assessed via self-report, including fever. The experience of symptoms and their description is likely to vary across individuals and across demographic characteristics, such as by gender, ethnicity, region, and age. People are likely to report symptoms differently whether they are doing so via an in-person interview, a weekly or bi-weekly survey or via a daily symptom tracking app, and the design of the app or questionnaire interface, as well as the preceding questions, will likely affect reporting. The majority of studies examining the efficacy of symptom self-report have focused on psychiatric disorders. These have generally found agreement between patient self-report and clinician assessment, although this varies from 60\% to 90\%\citep{31035214,29172673,33416510}. In major depressive disorder, self-reported symptoms are more severe than clinician-assessed symptoms\citep{31035214}. When self-tracking for health and fitness purposes, BMI is systematically under-reported\citep{31293064}. Knowledge of test status could also affect symptom reporting, though this will be less of an issue in the CIS dataset, where individuals will not yet have received their test results. Some studies involve reporting on behalf of others, particularly children or adults receiving care, and communicating the subjective experience of symptoms might be challenging in these cases. When reporting symptoms related to cancer treatment, a dyad (parent and child) approach to reporting symptoms was found to be more effective and preferable to child self-reporting or parent proxy reporting alone\citep{32567173}.

\subsubsection{Phase of infection}

The symptom reporting window around positive test time varies across the different datasets. There is evidence from previous studies\citep{sudre_symptom_2020} that some symptoms tend to appear earlier in infection while some appear later. We also know that people who test negative, who are not included in this dataset, report a wide range of symptoms that are not related to SARS-CoV-2 infection\citep{Elliott2021.02.10}; widening the symptom reporting window around a test date might include symptoms that are non-specific to the SARS-CoV-2 infection. Our approach collapses across time and these variations in the reporting window could affect our findings regarding symptom frequency and clustering. While there is no way of varying this for the routinely collected NHS Test and Trace data, we do conduct sensitivity analyses to examine a wider symptom reporting window around the day of testing for the ONS dataset, making it more comparable to CSS. We arbitrarily define positive episodes as a new positive occurring more than 90 days after an index positive or after 4 consecutive negative tests and consider symptoms reported in [-7,+35] days around the index positive. We do not find that this wider symptom window affects our clustering and co-occurrence findings.    

\subsubsection{Epidemic phase}

The characteristics of cases differ over the course of the epidemic, for example by age, region, socioeconomic characteristics or variant of SARS-CoV-2 infection, which in turn could plausibly affect the symptoms experienced and the likelihood that they are reported. Some positive cases could be from single or double-vaccinated individuals, particularly from later time periods in Winter/Spring 2021.
Similar to our AlignedUMAP embeddings for age-stratified data, we could also produce AlignedUMAP embeddings for time-stratified data, allowing us to investigate how symptom co-occurrence patterns change over time. This would be  of particular interest as vaccination effects build, or as a new variant with a different disease profile becomes dominant. The requirement of such an analysis is that each time strata has a sufficient number of points such that the estimated Jaccard distance matrix is not subject to significant uncertainty. An initial exploration of this analysis was performed for Pillar 2 and SGSS datasets, by stratifying into week-long strata; however, no significant changes to the symptom co-occurrence patterns were observed during this time period.

\section{Results}

\subsection{Hierarchical clustering}
We first performed hierarchical clustering using complete linkage
\cite{Hastie:2009} and the Jaccard distance between symptom vectors
as defined in Equation~\eqref{jacdef},
with results shown in Figure~\ref{fig:Dendro}. This figure shows the matrix of such distances as
a heatmap, with a dendrogram to its right. We read these dendrograms from right
to left, with splitting points representing points at which the algorithm
suggests a separation of symptoms into groups on the basis of their occurrence
in infected individuals.

Of the plots, the CIS data in panel \textbf{d} shows the clearest signal of separation of symptoms under this analysis method: gastrointestinal symptoms form a separate symptom grouping, joining the rest of the hierarchy only at the 
highest level; the distinctive loss of taste and smell joins the tree at the
next; and the remaining symptoms join individually at remaining levels. In
Pillar 2 and SGSS data (Figure~\ref{fig:Dendro} panels \textbf{a} and \textbf{b}) a similar pattern is observed, except for
additional complexity associated with uncommon symptoms in $\leq 5\%$ of
positives and for Pillar 2 loss of smell or taste joining at a similar point on
the tree to upper respiratory tract symptoms. For the CSS data in panel \textbf{c},
we see that shortness of breath and hoarse voice, symptoms not collected in
other studies, appear before gastrointestinal symptoms join the tree.\\

\subsection{Logistic Principal Component Analysis}

Secondly, we performed Logistic Principal Component Analysis (LPCA), an
extension of Principal Component Analysis to binary data
\cite{landgraf2020dimensionality}. This method is used to project the set of
individual feature vectors $\{\mathbf{y}_i \}$ for each dataset onto (in our
case two) components that sequentially are as close to the original set of
vectors as possible. The results of this analysis are shown in
Figure~\ref{fig:LPCA}, and show quite strikingly consistent patterns across
datasets, despite the various biases and data collection techniques employed.

The first strong signal in the data is that the first principal component
involves all symptoms in the same direction, meaning that the closest
one-dimensional description of community symptoms is the number of symptoms
experienced. The second principal component, with some exceptions that vary by
dataset, suggests that a source of variation is a negative correlation between
upper respiratory tract symptoms and systemic (Pillar 2, SGSS and CIS) and
gastrointestinal symptoms (SGSS, CIS and CSS). The overall interpretation of
these results show that a parsimonious description of COVID-19 symptoms at the
individual level can be provided by quantifying the total number of symptoms
experienced, followed by the relative contribution of upper respiratory
symptoms versus systemic or gastrointestinal symptoms to the total number of
symptoms experienced. The contribution of upper respiratory versus systemic and
gastrointestinal symptoms is also seen and in fact, strengthened when examining
the age-stratified data (children 0-17 years, adults 18-54 years and elder
adults aged 55 years and older, see Supplementary Materials).\\

\subsection{Uniform Manifold Approximation and Projection}
Having different symptoms identified by taking a symptom-level view of
clustering as in the hierarchical analysis, and an individual-level view of
co-occurrence as in LPCA, is explained by questions these methods address. LPCA
attempts to find a description of the overall variation of the symptoms of
individuals within the dataset, while hierarchical clustering groups by
suitably defined co-occurrence to find natural clusters of symptoms within the
dataset. Our third main analysis method aims to provide an overall picture by
considering low-dimensional embeddings of the data based on the structure of
interactions encoded in the datasets. In particular, Uniform Manifold
Approximation and Projection (UMAP) and associated algorithms\citep{UMAP,
mcinnes_umap_2021} produces a low-dimensional embedding using the local structure
of the data (i.e.\ groups of commonly co-occurring symptoms) and provided the
intrinsic dimension of the system is not too large, and can capture some of the
global structure of the data (i.e.\ the relationships between such groups of
data points). The result is that symptoms which commonly co-occur are placed
close to each other in the outputted low-dimensional embeddings.
Hyperparameters are important for UMAP, so we performed the analysis for two
different hyperparameter choices: one that focuses more on the global structure
(shown in Figure~S2); and one that focuses less on the global structure and
attempts to preserve more of the local structure of the data (shown in
Figure~S3). 

To more explicitly compare findings across datasets, we extend the UMAP analyses
above by using the AlignedUMAP algorithm\citep{mcinnes_umap_2021}. AlignedUMAP takes several
different datasets as inputs and finds the optimal embedding for each inputted
dataset, subject to the loose constraint that data points that are shared
between datasets are placed in similar positions in the low-dimensional
embeddings. These are produced through a trade-off between finding the optimal
embedding for individual datasets, and aligning the embedding of shared
symptoms across datasets. By aligning embeddings, we gain several useful
insights, most importantly that an embedding can be directly compared with the
others it was aligned against, allowing better assessment of similarities and
differences.

We produce embeddings of each dataset that are aligned based on the core
symptoms shared by all the datasets in our analysis: cough, diarrhoea, fatigue,
fever, headache, muscle ache, and sore throat. These, shown in Figure~\ref{fig:umap}, allow us to explore whether datasets shared a common underlying structure of symptom co-occurrence.

Inspection of the embeddings with alignment based on the core symptoms shared
by different datasets provides some evidence of a broad structure shared across
all datasets. The embeddings produced can be broadly described by a central
cluster of systemic symptoms, and cough. Lower respiratory tract symptoms are
typically placed nearby, in particular with shortness of breath often being
placed close to fatigue. The upper respiratory tract symptoms (sore throat,
rhinitis, sneezing) are typically placed further away from gastrointestinal
symptoms, with the exception of lost/altered smell or taste symptoms. On most
plots, the gastrointestinal symptoms exist as a tail or are slightly separated
from the main central group of systemic symptoms. The main exception to this is
the CSS data, although we would caution against over-interpretation of this plot
since, in contrast to the other studies, CSS has a more diverse set of symptoms,
meaning that the task we have set of aligning the symptoms common to CSS and other
datasets, at the same time as preserving relationships between those symptoms and
ones unique to CSS, is inherently challenging. Overall, therefore, we argue that
the UMAP results complement the LPCA
analysis, which suggested that individuals separate between those who
experience upper respiratory tract symptoms or those who experience a mixture
of systemic and gastrointestinal symptoms.

As we did with hierarchical clustering and LPCA, we stratified each dataset
based on age bands that represent children and adolescents, adults and
elders, and produced aligned embeddings for ease of comparison (see
Supplementary Materials). However, AlignedUMAP allows us to directly compare
more embeddings than is possible for dendrograms or symptom loadings, as
there exist explicit relationships between the embeddings. We perform an
additional analysis where we again age-stratify each dataset into 10-year
strata and produce aligned embeddings. These embeddings can then be visualised
in 3-dimensional space to describe how patterns of symptom co-occurrence
change as age increases, see Figure~\ref{fig:au}, where linear interpolation has been used
to connect the different embeddings from each ten-year age strata. In supplementary Figures \ref{fig:marginal_umap_p2, fig:marginal_umap_sgss, fig:marginal_umap_css, fig:marginal_umap_cis}, we also provide 2D marginal plots of the embeddings. Across all
datasets, we observe changes to the local structure, indicated by the splitting
of the rope/ribbon-like structures for the youngest age strata (under 10 years
old), and for the older age strata (around 70 years old). The changes indicate
that, despite the attempt to align symptoms in adjacent embeddings, the
symptom-co-occurrence patterns of the data have changed too substantially for
that to be achieved.

This is clearest in the CIS dataset, where some gastrointestinal symptoms
(diarrhoea, nausea/vomiting, abdominal pain) are separated out from the main
body of symptoms for the youngest and older age strata. In Pillar 2 and SGSS,
we find the formation of new clusters of symptoms, in the older age strata with
a first cluster containing vomiting and nausea, and a second cluster containing
headache, sore throat, muscle ache and joint pain. For the CSS dataset,
separation into two main symptom clusters is observed, with one cluster
containing: abdominal pain, muscle ache, headache, sore throat, chest pain, and
cough, and with the second cluster containing loss of appetite, altered/loss of
smell, diarrhoea, hoarse voice, slightly separated shortness of breath,
fever, delirium and fatigue.

For the under-10s, the produced embeddings typically consist of small clusters
of symptoms. The CIS dataset is the exception by again separating out
gastrointestinal symptoms from the main body of symptoms. Inspection of the
Jaccard distance matrices for the youngest age strata suggests that a possible
explanation may be that fewer total symptoms are reported for young children.
The observed clusters in the embeddings appear to consist mainly of pairs, or
triplets of symptoms that do commonly co-occur, e.g.\ rhinitis and sneezing.
However, the level of co-occurrence between these distinct small clusters is
very small, leading to separation in the low dimensional embeddings.\\

\section{Discussion}
In summary, we have shown that considerable complexity and variation exists in
COVID-19 symptoms in community infections. We find that the primary source of
variation is in the number of symptoms experienced by a case, but conditional
on this there are various ways to be ill that provide a more fine-grained
description of phenotypes. In particular, we find evidence for the separation
between upper respiratory and systemic symptoms, both including commonly
reported symptoms, and between upper respiratory and gastrointestinal symptoms,
though the latter is less common overall. While the deep structure of the symptom
clustering was similar across the middle range of age groups, we found some
evidence that patterns of symptom reporting changed among the youngest and
oldest, though further work may be required to understand whether this is due
to symptom reporting differences, or differences in the symptoms experienced.

While there are some differences in our findings across the four datasets, this
is unsurprising given their different case sampling designs, data
collection methods, symptom reporting windows and specific symptom data
collected. Routinely tested cases, for instance, will be selected based on the
symptoms that qualify cases for testing (Pillar 2), leading to lower expected
variation in the presence of these symptoms compared to cases identified via
random sampling. Indeed, the broad consistency of findings across these
datasets, which derive from routine, representative household and participatory
surveillance methods respectively increases our confidence that our findings
are robust. 

Our findings have implications for case identification and associated public health guidance in the community, particularly school settings, and high-risk settings such as care
homes or hospitals. The existence of phenotypes would suggest that the
one-size-fits-all symptom-based criteria for symptomatic testing like that used in the UK from 2020-2022 may be sub-optimal, especially in these
sub-populations where multiple phenotypes are most likely. Differences by age
could imply that symptomatic testing criteria should be tailored for different
settings, though this would need to be balanced with what is feasible and
understandable for the public. Further, it may be the case that the different characterisation of cases could inform clinical outcomes, for example the finding that cases can be described by the contribution of upper respiratory symptoms versus systemic or gastrointestinal symptoms to the total number of symptoms experienced. We find that the symptom clustering patterns amongst the oldest age groups diverged from the middle age groups, which is of potential clinical relevance given the strong age-related risk of severe disease.   

Routinely collected datasets in this study include symptom information only from positive SARS-CoV-2 cases, meaning that we cannot evaluate the specificity of symptom testing criteria combinations informed by the symptom co-occurrence structures we have identified here, and this limits our direct evaluation of the symptomatic testing policies that were employed during 2020-2022. However, studies that examined the optimal combination of
symptoms to initiate testing of symptomatic community cases
\cite{Elliott2021.02.10, Fragaszy2021.05.14} may have been implicitly assuming the
existence of a single phenotype - to ensure that a symptom testing criteria is
optimal, the possible existence of multiple phenotypes and the wide spectrum of disease must be considered.  Emphasis should be placed on the extent of symptom variation across COVID cases in communication with the public. This messaging is critical for the initiation of transmission control interventions, including isolation, and in helping the public to manage risks, including transmission to more vulnerable contacts. Given that our datasets only consist of positive SARS-CoV-2 cases, we are unable to explore the clustering of symptoms in the ``background" landscape of symptoms that are caused by other infections, allergies or environmental conditions. Such an analysis could further elicit whether there are clusters of symptoms that are well-distinguished from other background symptoms and could further inform optimal symptom testing criteria, however, we leave this as future work given that the majority of datasets analysed here do not contain data on individuals testing negative for SARS-CoV-2.

As well as optimising response with respect to symptoms upon acute infection, another key question involves the role of comorbidities, including chronic conditions\citep{clift_2020_living_risk}. Unfortunately, information on these is not collected in as consistent and systematic a manner across large datasets as information on short-term symptoms. As such, adjustment for comorbidities is not possible without major additional data collection and/or linkage, which would be likely to be a fruitful direction for future research, for example, using the secure linkage methodology of \citet{williamson_opensafely_2020}.

With vaccination, re-infections and ongoing SARS-CoV-2 evolution, as
well as the resurgence of other previously suppressed respiratory infections,
understanding the variability of COVID-19 symptoms presentation is critical in
planning community intervention for control of transmission, identification of
cases potentially requiring greater care, and possibly understanding long term
presentation of the disease\citep{antonelli_risk_2021}. Beyond even the
current pandemic, the application of unsupervised learning analyses, such as
this one, in conjunction with clinical, epidemiological and behavioural
understanding is likely to yield important insights for other infectious
diseases.

\section*{Acknowledgments}

\noindent{}\textbf{Funding:} \textit{CSS funding:} ZOE provided in-kind support
for all aspects of building, running, and supporting the ZOE app and service to
all users worldwide. Support for this study was provided by the National
Institute for Health Research (NIHR)-funded Biomedical Research Centre based at
Guy’s and St Thomas’ (GSTT) NHS Foundation Trust. This work was supported by
the UK Research and Innovation London Medical Imaging \& Artificial
Intelligence Centre for Value-Based Healthcare (104691). Investigators also
received support from the Wellcome Trust (WT203148/Z/16/Z, WT213038/Z/18/Z, and
W212904/Z/18/Z), Medical Research Council (MRC; MR/V005030/1 and MR/M004422/1),
British Heart Foundation, Alzheimer’s Society, EU, NIHR, COVID-19 Driver Relief
Fund, Innovate UK, the NIHR-funded BioResource, and the Clinical Research
Facility and Biomedical Research Centre based at GSTT NHS Foundation Trust, in
partnership with Kings College London. This work was also supported by the
National Core Studies, an initiative funded by UK Research and Innovation,
NIHR, and the Health and Safety Executive. The COVID-19 Longitudinal Health and
Wellbeing National Core Study was funded by the MRC (MC\_PC\_20030).
\textit{CIS funding:} The ONS CIS is funded by the Department of Health and
Social Care with in-kind support from the Welsh Government, the Department of
Health on behalf of the Northern Ireland Government and the Scottish
Government. \textit{Individual funding:} MF is supported by The Alan Turing
Institute under the EPSRC grant EP/N510129/1. K-DV is supported by the National
Institute for Health Research Health Protection Research Unit (NIHR HPRU) in
Healthcare Associated Infections and Antimicrobial Resistance at the University
of Oxford in partnership with Public Health England (PHE) (NIHR200915). RD and
TH are supported by the Engineering and Physical Sciences Research council
(Award numbers 2373157 and EP/V027468/1). EF is supported by the Medical
Research Council award MR/S020462/1; MF, EF, TW and TH are supported by the
Medical Research Council award MR/V028618/1; TH is supported by the JUNIPER
consortium (MR/V038613/1), the Royal Society (INF/R2/180067) and Alan Turing
Institute for Data Science and Artificial Intelligence. SO was supported by the
French government, through the 3IA C\^ote d’Azur Investments in the Future
project managed by the National Research Agency (ANR-19-P3IA-0002). CHS was
supported by the Alzheimer’s Society Junior Fellowship (AS-JF-17–011). TW is
supported by grants from the Wellcome Trust, the Medical Research Council, and
the Foreign Commonwealth and Development Office Joint Global Health Trials
(MR/V004832/1 and 209075/Z/17/Z), the Medical Research Foundation (MRF-131-0006-RG-KHOS-C0942),
and the Swedish Research Council.

The views expressed are those of the authors and not necessarily those of the
National Institute for Health Research, UK Health Security Agency or the
Department of Health and Social Care.  The funders/sponsors did not have any
role in the design and conduct of the study; collection, management, analysis,
and interpretation of the data; preparation, review, or approval of the
manuscript; or decision to submit the manuscript for publication. 

\noindent{}\textbf{Author contributions:} All authors contributed to collection
and processing of data, choice and implementation of analysis methods, and
writing of the paper.

\noindent{}\textbf{Competing interests:} None declared.

\noindent{}\textbf{Data and materials availability:} Datasets are too sensitive
for public release, and can be accessed by researchers through secure research
environments. The first of these is the Secure Anonymised Information Linkage
(SAIL) Databank, with information for researchers wishing to access this
resource at \url{https://saildatabank.com}. The second of these is the Office
for National Statistics' Secure Research Service (SRS), with information for
researchers wishing to access this resource at
\url{https://www.ons.gov.uk/aboutus/whatwedo/statistics/requestingstatistics/secureresearchservice}.

Code and datasets that have been approved for publication are
available at: \url{https://github.com/martyn1fyles/COVIDSymptomsAnalysisPublic}.

\textit{SAIL acknowledgement:} This study makes use of anonymised
data held in the Secure Anonymised Information Linkage (SAIL) Databank.
\textit{ONS acknowledgement:} This work contains statistical data from ONS
which is Crown Copyright. The use of the ONS statistical data in this work does
not imply the endorsement of the ONS in relation to the interpretation or
analysis of the statistical data. This work uses research datasets which may
not exactly reproduce National Statistics aggregates.  We would like to
acknowledge all the data providers who make anonymised data available for
research.

\clearpage

\section*{Tables}

\begin{table}[H]
\caption{\label{tab:dates}Descriptive statistics of the population in each dataset}
\centering
\begin{tabular}{lcccc}\toprule
 Variable                   & \multicolumn{4}{c}{Dataset}                       \\
                            & Pillar 2  & SGSS  & CSS   & CIS   \\
\midrule
 \emph{Data collection variables} &         &           &           &           \\
 Start date                 & 29/11/2020    & 29/11/2020& 11/05/2020& 28/04/2020\\
 End Data                   & 28/03/2021    & 28/03/2021& 11/01/2021& 13/03/2021\\
 Location of participants   & England       & England   & UK        & UK        \\ \midrule
 \emph{Age variables}       &               &           &           &           \\ 
 Mean (years)               & 38            & 48        & -         & 43        \\
 Median (years)             & 37            & 47        & 40-49     & 45        \\
 IQR (years)                & 26-51         & 32-62     & -         & 29-57     \\
 \midrule
 \emph{Sex breakdown}       &               &           &           &           \\
 Male                       & 736,906 (45.0\%)& 42,355 (40.3\%)     & 23,540 (38.2\%)   & 4,142 (45.2\%)    \\
 Female                     & 875,545 (55.0\%)& 62,808 (59.7\%)     & 38,051 (61.8\%)   & 5,024 (54.8\%)    \\
 Intersex                   & -             & -         & 3         &  -        \\
 N/A                        & -             & -         & 29        &  -        \\
 \midrule
 \emph{Sample sizes}        &               &           &           &           \\
 Total dataset size         & 1,898,273     & 179,550   & 61,623    & 27,903    \\
 Symptomatic cases          & 1,637,965 (86.3\%)    & 112,925 (62.9\%)   & 52,084 (84.5\%) & 9,166 (32.8\%)\\
 Asymptomatic cases         & 260,308 (13.7\%)      & 66,625 (37.1\%)    & 9,539 (15.5\%)  & 18,737 (67.2\%)\\
\end{tabular}
\end{table}

\clearpage

\begin{center}
\begin{longtable}{p{0.15\textwidth}p{0.25\textwidth}p{0.25\textwidth}p{0.25\textwidth}}
\caption{Symptom survey questions by dataset.} \label{tab:symptom_questions} \\
\hline
Variable        & \multicolumn{3}{c}{Dataset}         \\
 \hline 
                            & {Test and Trace (Pillar 2 \& SGSS)} & CIS & CSS \\ \hline

\endfirsthead

\multicolumn{4}{l}%
{{\it \tablename\ \thetable{} -- continued.}} \\
\hline
Variable        & \multicolumn{3}{c}{Dataset}         \\
 \hline 
                            & {Test and Trace (Pillar 2 \& SGSS)} & CIS & CSS \\ \hline
\endhead

\hline \multicolumn{4}{r}{{\it Continued on next page.}} \\ 
\endfoot

\endlastfoot

Symptomatic                     & Are you experiencing any of the following symptoms? Please select at least one. (cases may select ``I have no symptoms") & Have you had any of the following symptoms in the last 7 days? & Are you feeling physically normal today? (I feel physically normal; I do not feel physically normal)\\ \hline
Abdominal pain & - & Abdominal pain & Do you have an unusual abdominal pain? \\ \hline
Altered consciousness & Altered consciousness & - & - \\ \hline
Altered/loss of smell & - & - & Do you have a loss of smell/taste?\\ \hline
Chest pain & - & - & Are you feeling an unusual chest pain or tightness in your chest?\\ \hline
Cough & A new, continuous cough& Cough & Do you have a persistent cough? \\ \hline
Delirium & - & - & Do you have any of the following symptoms: confusion, disorientation, or drowsiness? \\ \hline
Diarrhoea & Diarrhoea & Diarrhoea & Are you experiencing diarrhoea? \\ \hline
Fatigue & Extreme tiredness & Weakness/tiredness & Are you experiencing unusual fatigue? (mild; severe)* \\ \hline
Fever & High temperature or fever (higher than 38$^{\circ}$C) & Fever & Do you have a fever? \\ \hline
Headache & Headache & Headache & Do you have a headache?\\ \hline
Hoarse voice & - & - & Do you have an unusually hoarse voice? \\ \hline
Joint pain & Joint pain & - & -\\ \hline
Loss of appetite & Loss of appetite & - & Have you been skipping meals? \\ \hline
Loss of smell & - & Loss of smell & - \\ \hline
Loss of smell or taste & Loss or change to your sense of smell or taste (you cannot smell or taste anything, or things smell or taste different to normal) & - & - \\ \hline
Loss of taste & - & Loss of taste & - \\ \hline
Muscle ache & Muscle ache & Muscle ache & Do you have unusual strong muscle pains? \\ \hline
Nausea & Feeling sick (nausea) & - & -\\ \hline
Nausea / vomiting & - & Nausea/vomiting & - \\ \hline
Nose bleed & Nose bleed & - & - \\ \hline
Rash & Rash & - & - \\ \hline
Rhinitis & Runny nose & - & - \\ \hline
Seizures & Seizures & - & - \\ \hline 
Shortness of breath & - & Shortness of breath & Are you experiencing unusual shortness of breath? (no; yes mild symptoms/ slight shortness of breath during ordinary activity;  yes significant symptoms -breathing is comfortable only at rest; yes, severe symptoms/ breathing is difficult even at rest)**\\ \hline
Sneezing & Sneezing & - & - \\ \hline
Sore throat & Sore throat & Sore throat & Do you have a sore throat? \\ \hline
Vomiting & Vomiting & - & -
\end{longtable}
\end{center}

\clearpage

\begin{table}
\caption{\label{tab:asaa}Sample sizes for each strata of the AlignedUMAP embeddings in the main paper}
\centering
\begin{tabular}{ccccc}\toprule
 Age strata                 & \multicolumn{4}{c}{Dataset}                       \\
                            & Pillar 2  &SGSS  & CSS   & CIS   \\
\midrule
0-9                         & 70,051        & 2,759     & 1,256     & 255       \\
10-19                       & 154,848       & 3,966     & 4,891     & 979       \\
20-29                       & 323,244       & 16,250    & 7,716     & 1,106     \\
30-39                       & 343,935       & 19,398    & 10,075    & 1,416     \\
40-49                       & 292,823       & 18,673    & 12,896    & 1,746     \\
50-59                       & 267,361       & 19,221    & 14,263    & 1,827     \\
60-69                       & 125,840       & 12,453    & 7,709     & 1,153     \\
70-79                       & 41,814        & 9,963     & 2,261     & 552       \\
80-89                       & 12,889        & 7,488     & 396       & 120       \\
90-99                       & 2,222         & 2,158     & -       & -         \\
\end{tabular}
\end{table}

\clearpage

\section*{Figures}

\begin{landscape}
\thispagestyle{empty}
\begin{figure}
\begin{center}
\vspace*{-0.5cm}
\hspace*{-1.1cm}
\includegraphics[width=1.2\linewidth]{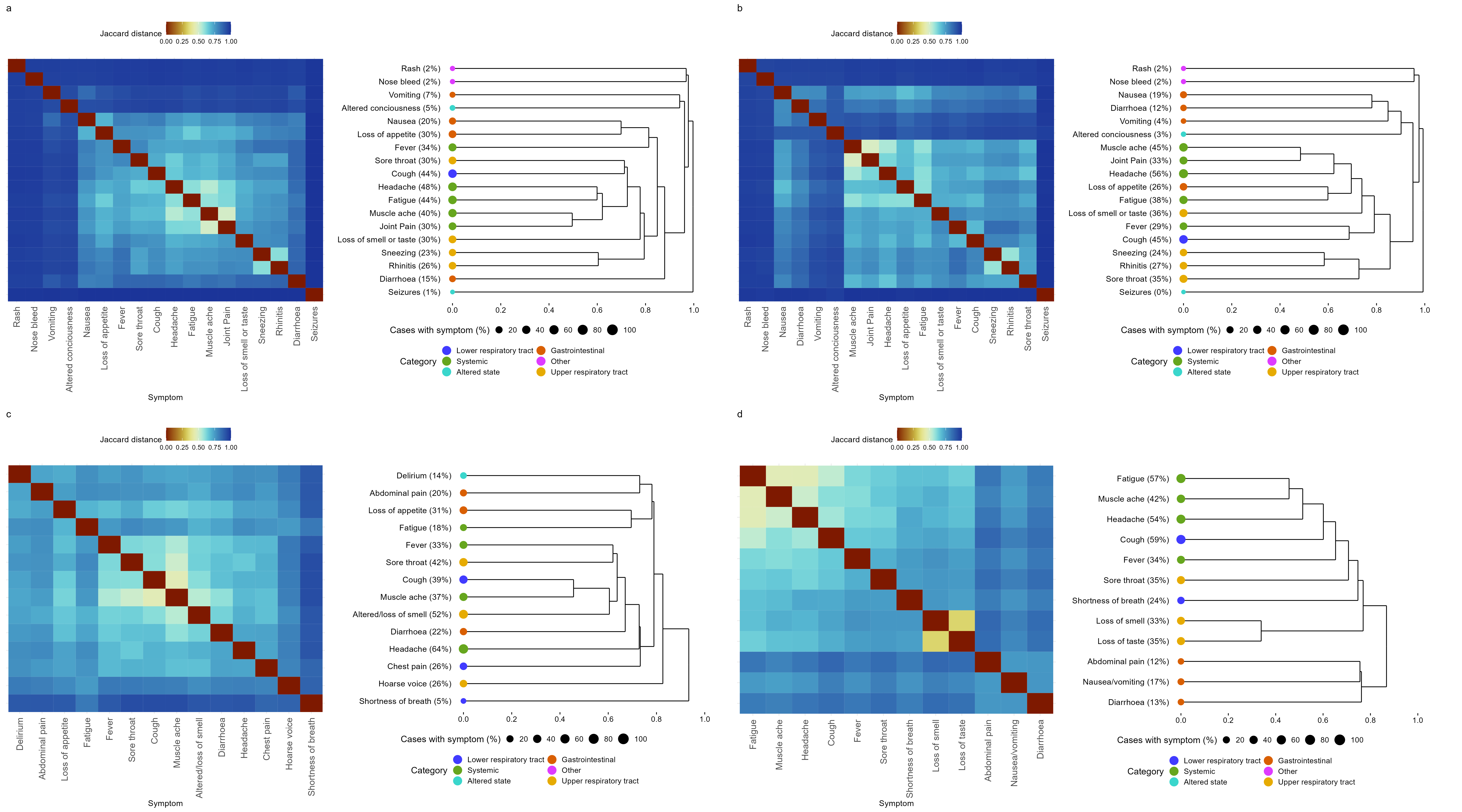}
\end{center}
\caption{Jaccard distance matrices between symptoms adjacent
to associated dendrograms obtained through hierarchical clustering under
complete linkage. The symptom category is denoted using coloured points at the
roots of the dendrogram. The central columns give the name of the symptom with
the percentage of symptomatic cases who exhibit the symptom in the dataset.
\textbf{a}. Pillar 2, \textbf{b}. SGSS, \textbf{c}. COVID Symptom Study,
\textbf{d}. COVID-19 Infection Survey. \label{fig:Dendro}}
\end{figure}
\end{landscape}

\begin{figure}
\begin{center}
\thispagestyle{empty}
\vspace*{-1.5cm}
\includegraphics[width=\linewidth]{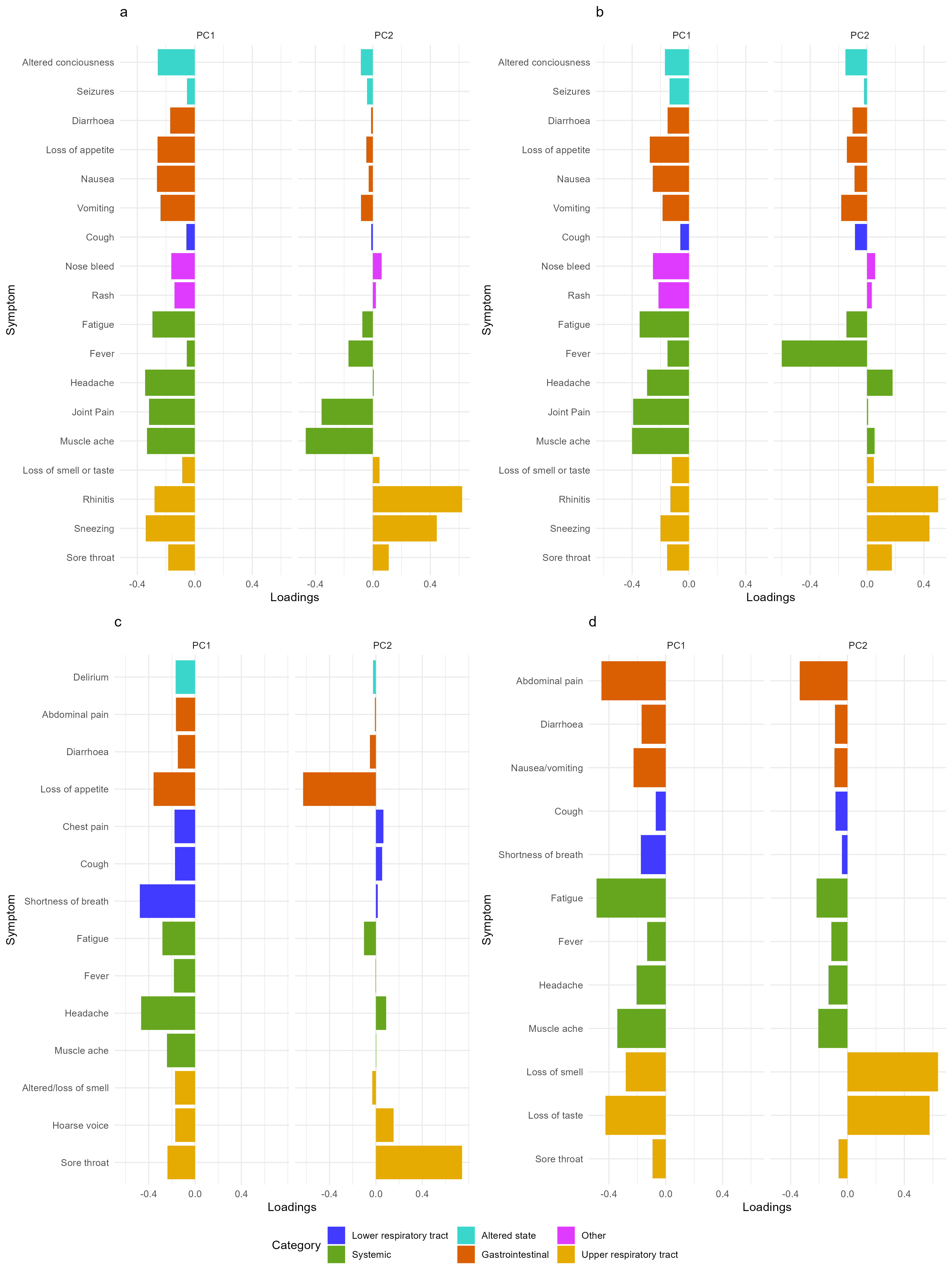}
\end{center}
\caption{Logistic Principal Components Analysis (LPCA)
results. For each dataset, elements of the principal components are visualised
as vertical bar plots. Each vector is insensitive to overall multiplication by
$-1$. Symptom categories are labelled by colours.
\textbf{a}. Pillar 2, \textbf{b}. SGSS, \textbf{c}. COVID Symptom Study,
\textbf{d}. COVID-19 Infection Survey. \label{fig:LPCA}}
\end{figure}

\begin{figure}
\vspace*{-2cm}
\thispagestyle{empty}
\begin{center}
\includegraphics[width=1.2\linewidth]{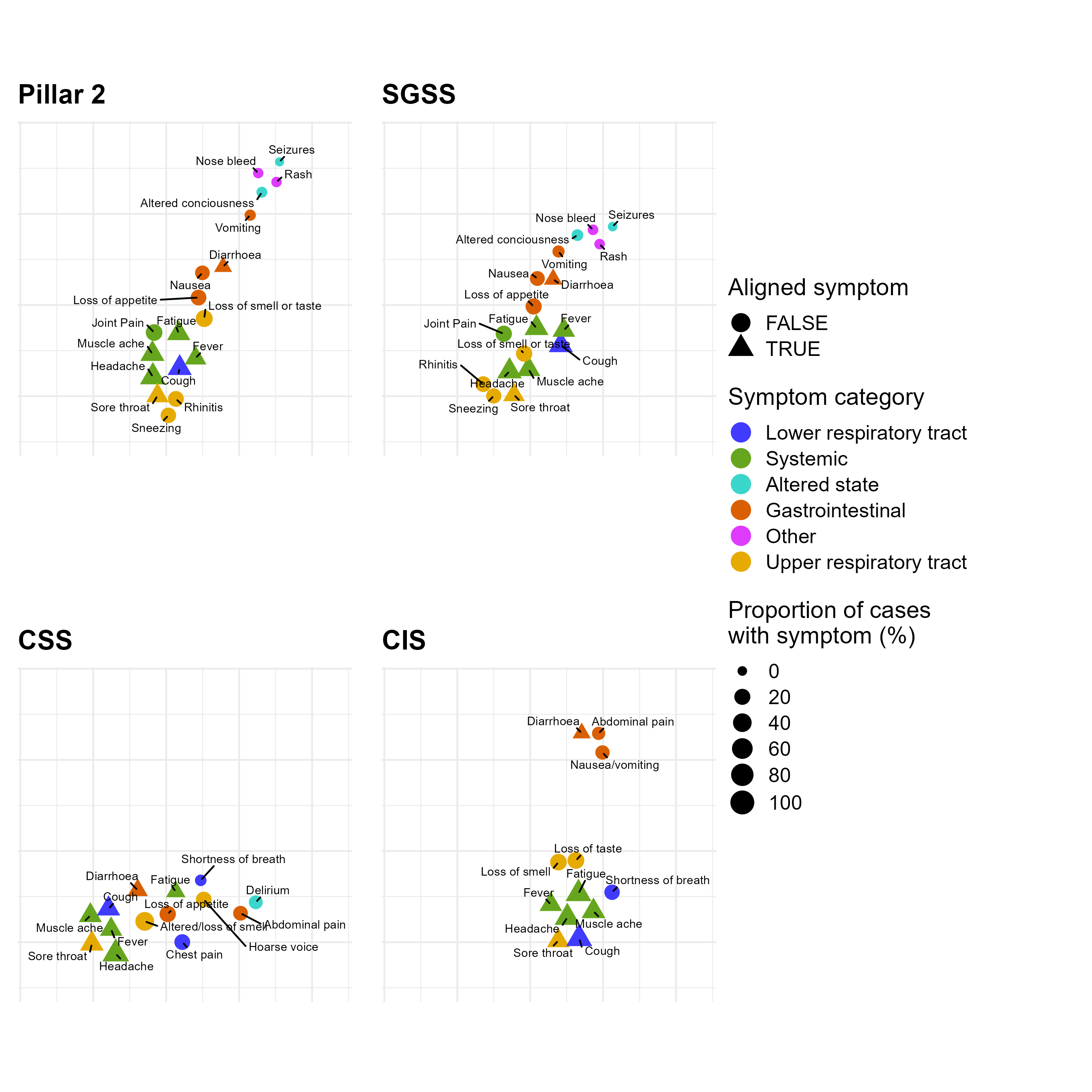}
\caption{AlignedUMAP embeddings of SARS-CoV-2 symptoms. For
each dataset, an optimal embedding of the symptoms into 2D Euclidean space is
found, subject to the following loose constraint: if a symptom is common to all
datasets, then it should be placed in roughly the same position across all
datasets. This alignment allows for easier comparison, and investigation of
shared symptom structures across all datasets. Point size is proportional to
the proportion of cases that develop a given symptom. Symptoms that are common
to all datasets, and are aligned between distinct datasets are plotted as
triangles. For this embedding the parameters were chosen to capture more of the
global structure of symptoms and produces less well-defined clusters.
\textbf{a}. Pillar 2, \textbf{b}. SGSS, \textbf{c}. COVID Symptom Study,
\textbf{d}. COVID-19 Infection Survey. \label{fig:umap}}
\end{center}
\end{figure}

\clearpage
\thispagestyle{empty}

\begin{figure}
  \centering
\vspace*{-2.5cm}
  \includegraphics[width = 0.80 \textwidth]{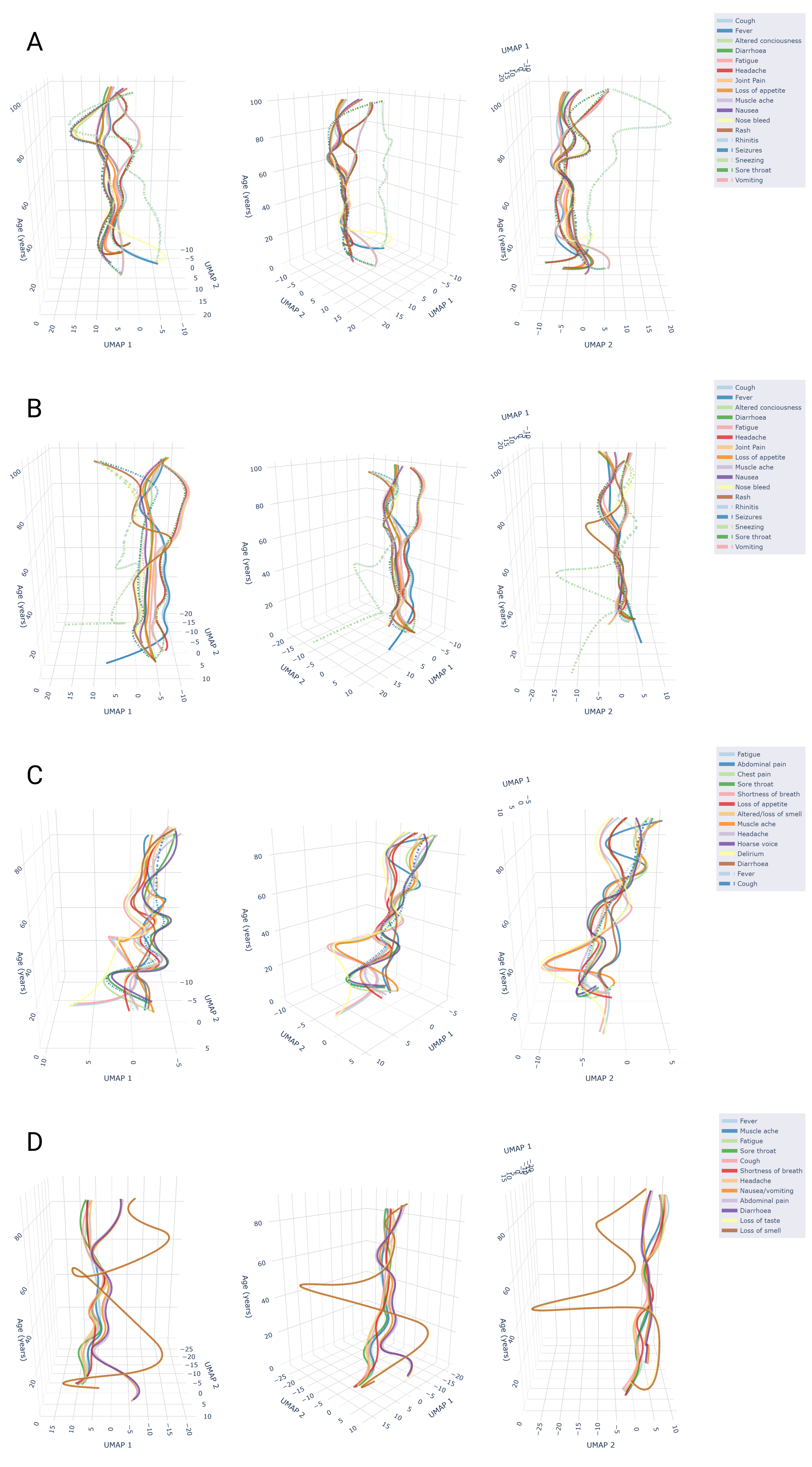}
\caption{AlignedUMAP embeddings of SARS-CoV-2 symptoms across
several datasets. Each dataset has been age-stratified into strata of length 10
years. For each strata, an optimal two-dimensional embedding into Euclidean
space of the symptoms is found, subject to the loose constraint that each
symptom is placed in a similar location in adjacent embeddings. Linear
interpolation is used to connect the embedding of each strata, allowing for a
3-dimensional visualisation of how the co-occurrence patterns of symptoms
change with age. For each 3D embedding, we take three images at 45 degree rotations. 
\textbf{a}. Pillar 2, \textbf{b}. SGSS, \textbf{c}. COVID Symptom Study,
\textbf{d}. COVID-19 Infection Survey. \label{fig:au}}
\end{figure}

\clearpage

\renewcommand{\thetable}{S\arabic{table}}
\renewcommand{\thefigure}{S\arabic{figure}}
\renewcommand{\thesection}{S\arabic{section}}
\setcounter{section}{0}
\setcounter{table}{0}
\setcounter{figure}{0}

\begin{center}
\textbf{\LARGE Supplementary Materials}
\end{center}

\section{Supplementary Text}
\subsection{Symptom frequencies}

All datasets include only cases reporting at least one symptom for these
analyses. The most commonly reported symptom across all datasets was headache,
with approximately half of the cases in the Pillar 2, SGSS and CIS datasets
reporting them, and almost two-thirds of those from CSS, (see Figure~3 from the
main text and Figure~\ref{fig:SymptomFreqs} below). The frequency of systematic
symptom reports is high across the datasets. Fever, a systemic symptom intended
to prompt isolation and testing in the UK, was experienced by less than one-third of all symptomatic cases. Cough, another isolation and testing initiating
symptom (when new and continuous, which was not captured in these datasets), was
also common (39\% to 59\%). NHS Test and Trace did not include any other lower
respiratory tract symptoms, but in CIS, shortness of breath was experienced by
24\% and by 5\% in CSS, while 26\% of those symptomatic cases participating in
CSS reported chest pain (not collected in other datasets). Each dataset
includes information about altered/loss of smell and/or taste but collected
this differently, though all variations were commonly reported. Altered/loss of
smell was most frequently reported in the CSS (52\%), while loss of taste and
smell separately (CIS) and in combination (NHS Test \& Trace) was reported by
over 30\%. These symptoms also trigger isolation and testing. Sore throat was a
common upper respiratory symptom in all datasets (30\% to 42\%). Sneezing and
rhinitis, only collected by Test and Trace, were reported by around one-quarter
of symptomatic cases. Gastrointestinal symptoms tended to be less frequent than
systemic and respiratory but were not unusual (mainly reported by 10-20\%
though less frequently for vomiting alone), with the exception of loss of
appetite, which was reported by between one-quarter and one-third of cases in Test
and Trace and CSS, datasets in which it was collected. Symptoms that we
described as `altered state' were rarer and not collected in CIS. Rash and
nosebleeds were reported by 2\% of symptomatic cases in Test and Trace, but not
collected in CIS or CSS.

\subsection{UMAP Hyperparameter Selection\label{UMAP hyperparameter selection}}

The base UMAP algorithm has four main hyperparameters; \texttt{n\_neighbours}, \texttt{n\_components} and \texttt{min\_dist}. In the main text, we discuss how we have chosen to vary \texttt{n\_neighbours} between two values to produce ``loose" (\texttt{n\_neighbours} = 4) and ``tight" (\texttt{n\_neighbours} = 2) clusterings, which focus more on either local or global structure. These two values of \texttt{n\_neighbours} were selected as they produced distinct embeddings that demonstrated different aspects of the high-dimensional data structure. More extreme values of \texttt{n\_neighbours} produced embeddings that did not appear to perform well at describing the structure of the data.

The number of dimensions of the embedding is set by n\_components, which we have fixed at two dimensions. During the development of this paper, we explored using 3D embeddings however we found that they were not sufficiently different to the 2D embeddings to warrant their inclusion, particularly given the increased challenge of visualising a 3D embedding in a paper. The next parameter is min\_dist, which provides a minimum distance between points in the produced embeddings. The role of this parameter is largely to improve the readability of UMAP embeddings by preventing points from being overplotted in embeddings which can make them difficult to read. We find that setting this parameter to large values can impact the quality of the embedding by artificially enforcing large distances between points. Therefore to set this parameter, we started off with the smallest possible value and increased until the produced embeddings were sufficiently readable. Finally, we configured our UMAP algorithm to embed into a Euclidean space, and while it is possible to vary this, we did not believe that embedding to a non-Euclidean space would help elicit further information.

AlignedUMAP inherits all the parameters of the base UMAP algorithm, and we apply the same arguments for how to set these parameters. In addition, AlignedUMAP has two new hyperparameters; \texttt{n\_slices} and \texttt{alignment\_strength}. Recall that AlignedUMAP first produces embeddings for different segments of a dataset, such as different age groups, and then attempts to minimise the distance between the embeddings between different segments of the data to produce embeddings that can directly be compared for different segments of the dataset. Assume without loss of generality that a dataset can be partitioned into ordered segments $S_1, S_2, S_3, \dots$. The parameter \texttt{n\_slices} controls how many of the neighbouring segments with be used in the Alignment process. For example, if \texttt{n\_slices} = 1, then $S_i$ will be aligned with $S_{i-1}, S_{i+1}$, it's immediate neighbours. If n\_slices = 2, then $S_i$ will be aligned with $S_{i-2}, S_{i-1}, S_{i+1},S_{i+2}$, it's two neighbours on either side. The effect of increasing \texttt{n\_slices} is to increase the smoothness of the AlignedUMAP embedding, akin to increasing the width of a histogram bin. For large values of \texttt{n\_slices}, the AlignedUMAP embeddings are over-smoothed, and for small values of \texttt{n\_slices} the embedding is under-smoothed. Additionally, it would not make sense to align the 0-10 age group with the 50-60 age group, where we would expect there to be different symptom occurrence patterns. Therefore, we prefer a smaller value of \texttt{n\_slices} to prevent this. We find that \texttt{n\_slices} = 2 is an ideal value that demonstrates the varying changing symptom occurrence patterns. At \texttt{n\_components} = 1, the embeddings are very noisy and at \texttt{n\_components} $\geq$ 3, the embeddings are over-smoothed into a single cluster of symptoms across all embeddings. Performing AlignedUMAP necessitates a small trade-off between finding the optimal embedding for a given segment, and aligning it with its neighbours - this trade-off is controlled by the \texttt{alignment\_strength} parameter. For this parameter, we generally prefer small values to ensure that we are still finding the optimal embedding for each segment of data, with the additional benefit of the embedding being aligned with the embeddings of adjacent segments. Therefore, to set this parameter, we slowly increased the value of this parameter until the embeddings were aligned, via visual inspection.

We believe that these are suitable methods for setting the UMAP parameters. For readers that are interested in exploring the effects of these parameters further, we provide the necessary data and code in our repository to reproduce this analysis.

\subsection{UMAP results without alignment between datasets}

Looking at Figure~\ref{fig:all_datasets_UMAP_loose}, we see a global structure
similar to what we observe in the main paper using the AlignedUMAP algorithm.
The embeddings of most datasets can be described by a central cluster of
systemic and lower respiratory tract symptoms. Upper respiratory tract
symptoms, such as rhinitis, sneezing, hoarse voice, and sore throat, are typically
placed close to the systemic symptoms cluster, with the exception of loss of
smell and taste symptoms. Gastrointestinal symptoms are often placed further
away from the upper respiratory tract symptoms and often form a tail leading
to some of the rarer symptoms. We note that these embeddings synthesise the
results we observed from the LPCA loadings, where the second loading suggested
that cases could be separated based on whether they predominantly
experienced upper respiratory tract symptoms or systemic and gastrointestinal
symptoms. The relatively low rates of occurrence of gastrointestinal symptoms
explains their appearance high in the hierarchical tree, while the higher
frequency of systemic and respiratory symptoms explains their relative
importance in LPCA loadings, within the general structure revealed by UMAP.

We repeat the UMAP analysis without alignment, this time with the algorithm
tuned to focus more on the local structure of the data and less on the global
structure of the data. As shown in Figures~\ref{fig:all_datasets_UMAP_tight},
this produces a better separation of the symptoms into clusters in the low
dimensional embeddings, however, some of the relationships between these clusters
may be lost. In the resulting embeddings, we observed several pairs of symptoms
that commonly co-occur but appear to be distinct from the main cluster of other
symptoms, notably sneezing and rhinitis in the Pillar 2 and SGSS datasets,
headache and sore throat in the CSS dataset, and loss of smell and taste in the
CIS dataset. The remaining symptoms are often packed into two tight clusters. For
Pillar 2 and SGSS, a clear separation between systemic and upper respiratory
tract symptoms, and the less frequently occurring gastrointestinal, altered
state and other symptoms is observed. Similarly, gastrointestinal are placed
into their own cluster in CIS, and in CSS with the exception of loss of
appetite. Focusing more on the local structure can make the resulting
embeddings more variable between datasets, as the choice of symptoms included
in the dataset appears to make more of a difference. We note that the
embeddings focusing more on the local structure can be more variable between
repeats, however, they do highlight small local structures in the data.  The
aligned UMAP results in the main paper focus more on local structure, however,
the requirement to align several related slices of the datasets appears to make
these results more consistent between runs.

Looking at Figures~\ref{fig:all_datasets_UMAP_loose} and
\ref{fig:all_datasets_UMAP_tight}, we see a global structure to the
relationship between symptoms that synthesises other results. This is clearest
in the CIS data, where we can draw a line from gastrointestinal through
systemic, to respiratory tract symptoms, but with sore throat closer to cough
than it is to loss of taste and smell. Such a line could be interpreted as
describing a spectrum of COVID-19 symptoms. In the other datasets, this pattern
is complicated by other types of symptoms, which typically occur closest to
gastrointestinal. The relatively low frequency of these symptoms explains their
appearance high in the hierarchical tree, while the higher frequency of
systemic and respiratory infections explains their relative importance in LPCA
components within the general structure revealed by UMAP.

\subsection{Age stratified findings}

We repeated our main analyses - hierarchical clustering, Logistic PCA and AlignedUMAP - on each dataset, stratified by broad age groups: children (0-17 years), adult (18-54 years) and elder adults (55+ years), Supplementary Figures \ref{fig:pillar 2 age stratified heirarchical clustering}-\ref{fig:ONS Age stratified UMAP (tight)}. 

Broadly, we did not find strong differences in the clustering and co-occurrence patterns of symptoms across age groups and studies. The unstratified findings reflect more strongly the middle age category (18-54 years), which accounts for the majority of the sample in each dataset. It is possible that symptom data collection, particularly among young children, which relies upon caregiver reports, could contribute to explaining some differences observed. 

The clear separation of gastrointestinal symptoms and loss of taste and smell is observed across the age strata in the CIS, Supplementary Figure \ref{fig:ONS age stratified heirarchical clustering}, with minor differences in the order at which some other individual symptoms join the tree (e.g. shortness of breath among children and sore throat amongst elder adults). In Pillar 2 and SGSS datasets, Supplementary Figures \ref{fig:pillar 2 age stratified heirarchical clustering} and \ref{fig:SGSS age stratified heirarchical clustering} respectively, across age groups the rarer symptoms separate earlier from other symptoms, with some later separation between systemic and upper respiratory symptoms observable. Patterns did not differ greatly across the age strata. Across age strata, symptoms among cases in the CSS, Supplementary Figure \ref{fig:Zoe age stratified heirarchical clustering}, show shortness of breath and delirium (rare symptoms) separating early, followed by some gastrointestinal symptoms (diarrhoea and abdominal pain) and, most clearly among adults 18-54, splitting between systemic and gastrointestinal symptoms and primarily lower and upper respiratory symptoms.

For all age-stratified LPCA analyses plotted in Supplementary Figures \ref{fig:pillar 2 age stratified LPCA}-\ref{fig:Zoe age stratified LPCA}, the first principle component essentially describes variation in severity, followed by characterisation according to either upper respiratory (loss of taste and smell) or upper respiratory symptoms. For CIS, plotted in Supplementary Figures \ref{fig:ONS age stratified LPCA}, cough had a high loading on the second component among children but not adults or elder adults, pointing the opposite direction to upper respiratory symptoms. The presence of gastrointestinal symptoms was more important in describing cases among elder adults, compared to children, with adults aged 18-54 years in between.

Similar patterns of separation between upper respiratory, systemic and gastrointestinal symptoms are seen across age groups when examining the UMAP embeddings when hyperparameters were selected that produce well-separated clusters, Supplementary Figures \ref{fig:Pillar 2 Age stratified UMAP (tight)}-\ref{fig:ONS Age stratified UMAP (tight)}. Despite the age strata being coarser here than in the results of the main paper, Fig ~4, we do observe similar structural changes to the data: in the children's age strata, we often observe the formation of several small clusters of symptoms; in the adults' age strata, the embeddings tend to resemble a larger cluster; and in the elders' age strata, the embeddings again start to fragment into two smaller clusters of symptoms. The structural changes are less striking than in the results in Fig. 4, where finer age slices are used. However, this is expected, given that the coarser age strata used in Supplementary Figures \ref{fig:Pillar 2 Age stratified UMAP (tight)}-\ref{fig:ONS Age stratified UMAP (tight)} make it harder for UMAP to detect structural changes to patterns of symptom co-occurrence that occur over small changes in age.

The results from tuning the UMAP algorithm to focus more on global structure are plotted in Supplementary Figures \ref{fig:Pillar 2 Age stratified UMAP}-\ref{fig:ONS Age stratified UMAP}. Unlike in embeddings that focus more on the local structure of the dataset, we do not observe a strong separation of symptoms into several small clusters in the youngest or separation into two main clusters in the elderly population. This is to be expected, as focusing more on the global structure results in an embedding that attempts to describe more of the spectrum of the disease, and less on small groups of commonly co-occurring symptoms, providing a complementary analysis. Our interpretation is that, in the youngest and oldest age groups,  patterns of co-occurrence of reported symptoms do change, particularly for pairs of symptoms, however, we do not observe significant changes to the overall spectrum of the disease, which can still be broadly described by number of symptoms experienced, and then the relative contribution of upper respiratory tract symptoms, or gastrointestinal symptoms. Across Pillar 2, SGSS and CIS, we consistently observe a central cluster of systemic and lower respiratory tract symptoms. Upper respiratory tract symptoms are clustered close to the systemic symptoms, but further away from the gastrointestinal symptoms. The CSS dataset is the most different, where shortness of breath, fatigue and delirium are clustered close to gastrointestinal symptoms, but further away from the main cluster of systemic, upper respiratory tract and lower respiratory tract symptoms.

\clearpage

\section{Supplementary Figures}

\begin{figure}[H]
    \centering
    \includegraphics[width=\linewidth]{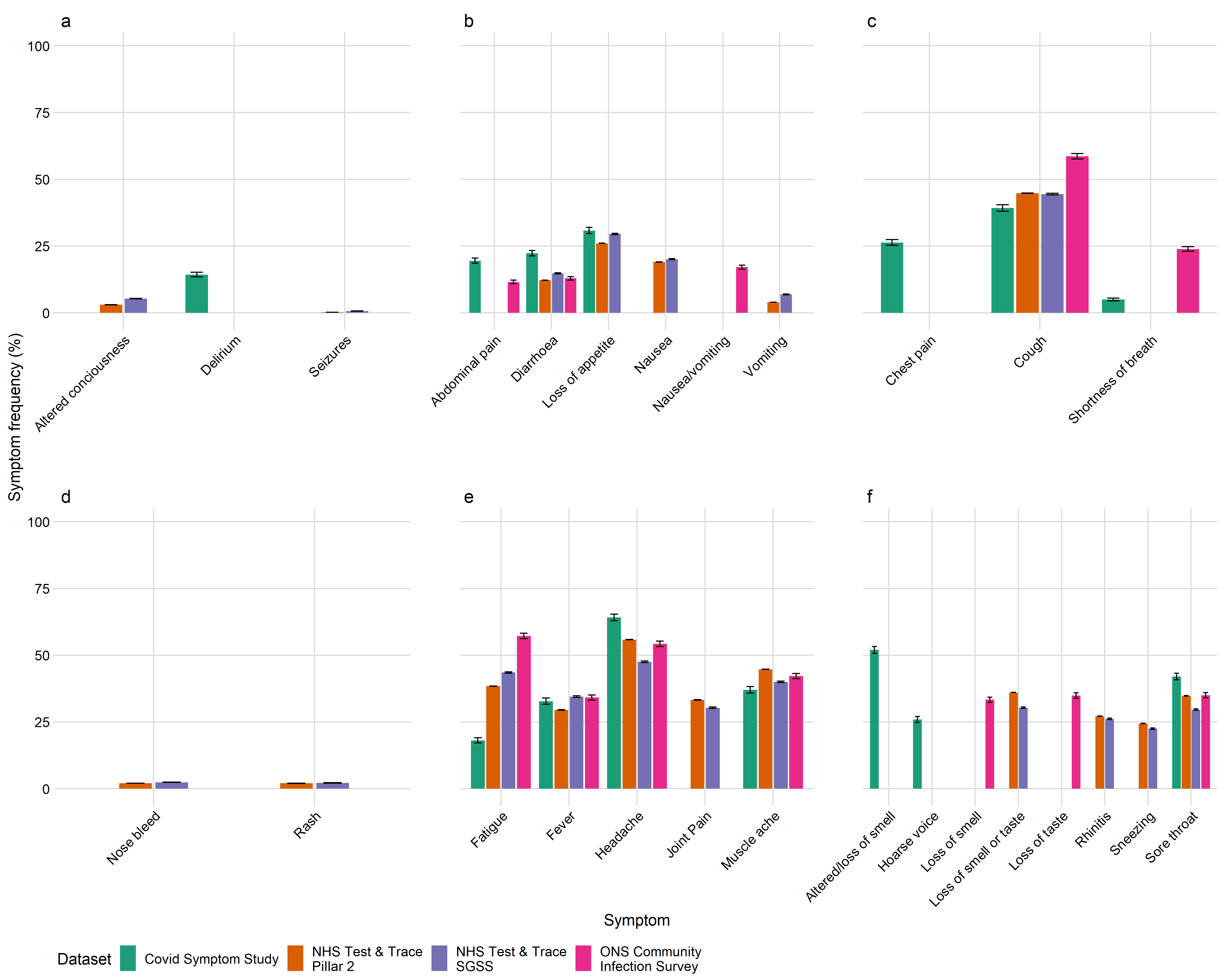}
    \caption{A plot containing the proportion of cases that develop a symptom across datasets. Each dataset records a different set of symptoms, and in some datasets multiple symptoms are considered to be one variable. Each subplot contains a different category of symptoms. \textbf{a}, altered state symptoms. \textbf{b}, gastrointestinal symptoms. \textbf{c}, lower respiratory tract symptoms. \textbf{d}, other symptoms. \textbf{e}, systemic symptoms. \textbf{f}, upper respiratory tract symptoms.}
    \label{fig:SymptomFreqs}
\end{figure}

\begin{figure}[H]
    \centering
    \includegraphics[width=\linewidth]{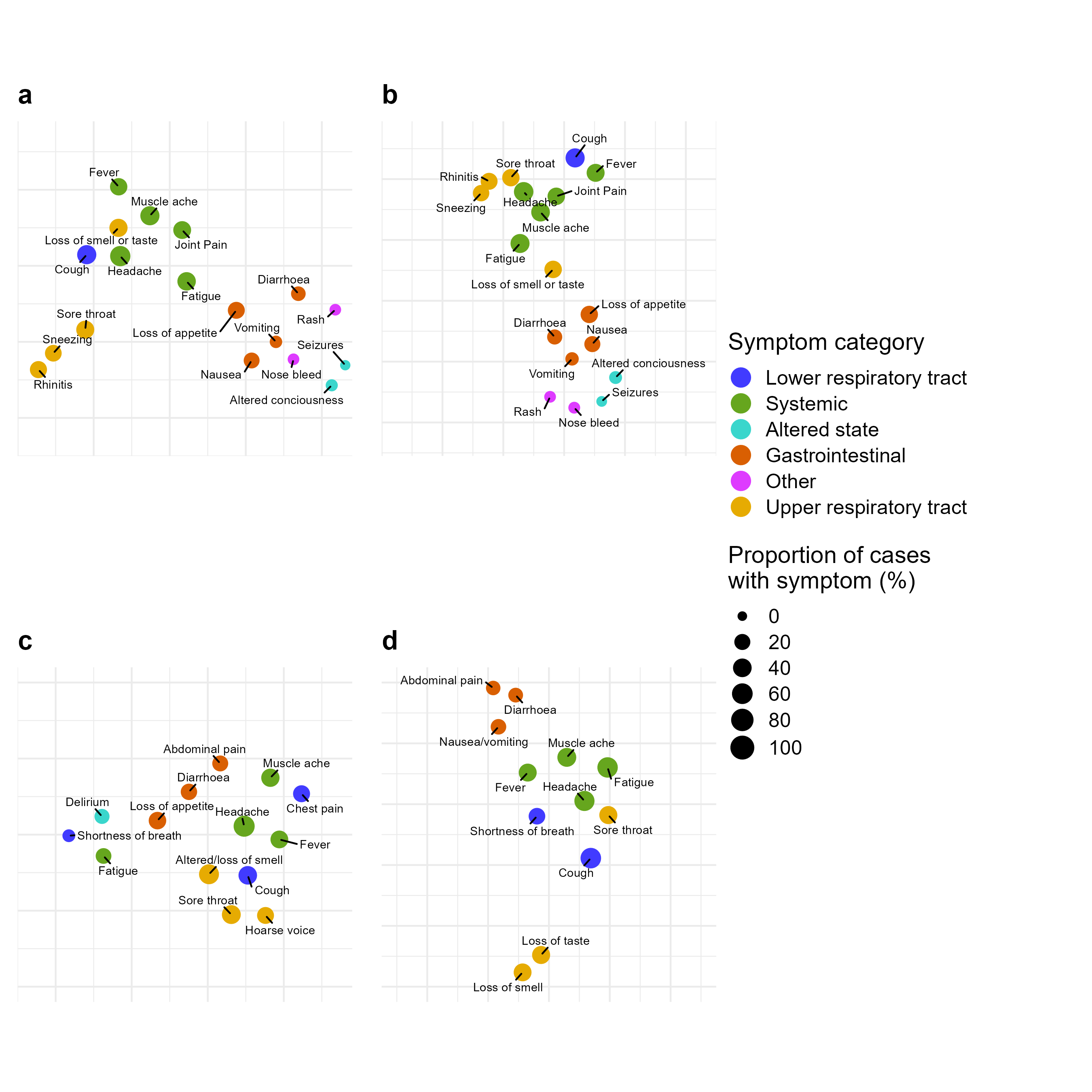}
    \caption{UMAP embeddings of SARS-CoV-2 symptoms. The algorithm
attempts to place combinations of symptoms that commonly co-occur close to each
other. Point size is proportional to the proportion of cases that develop a
given symptom. For this embedding, the parameters were chosen to capture more of
the global structure of symptoms and produces less well-defined clusters, and it was performed without any alignment between datasets. \textbf{a}. Pillar 2., \textbf{b}. SGSS, \textbf{c}. COVID Symptom Study, \textbf{d}. COVID-19 Infection Survey.}
    \label{fig:all_datasets_UMAP_loose}
\end{figure}

\begin{figure}[H]
    \centering
    \includegraphics[width=\linewidth]{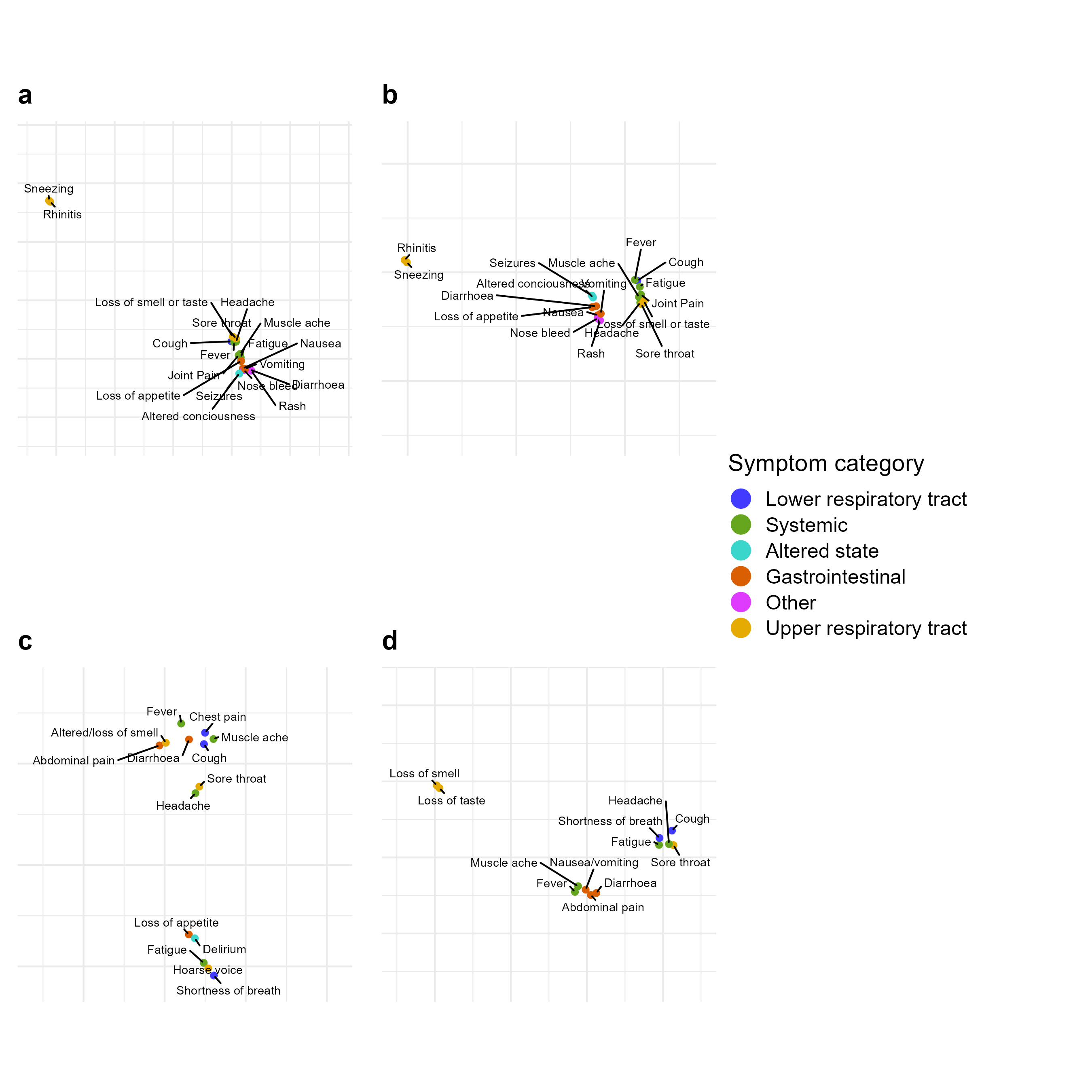}
    \caption{UMAP embeddings of SARS-CoV-2 symptoms. The algorithm
attempts to place combinations of symptoms that commonly co-occur close to each
other. For this embedding, the parameters were chosen to capture more of
the local structure of symptoms and produces less well-defined clusters, and it was performed without any alignment between datasets. \textbf{a}. Pillar 2, \textbf{b}. SGSS, \textbf{c}. COVID Symptom Study, \textbf{d}. COVID-19 Infection Survey.}
    \label{fig:all_datasets_UMAP_tight}
\end{figure}

\begin{figure}[H]
    \centering
    \includegraphics[width=0.7\linewidth]{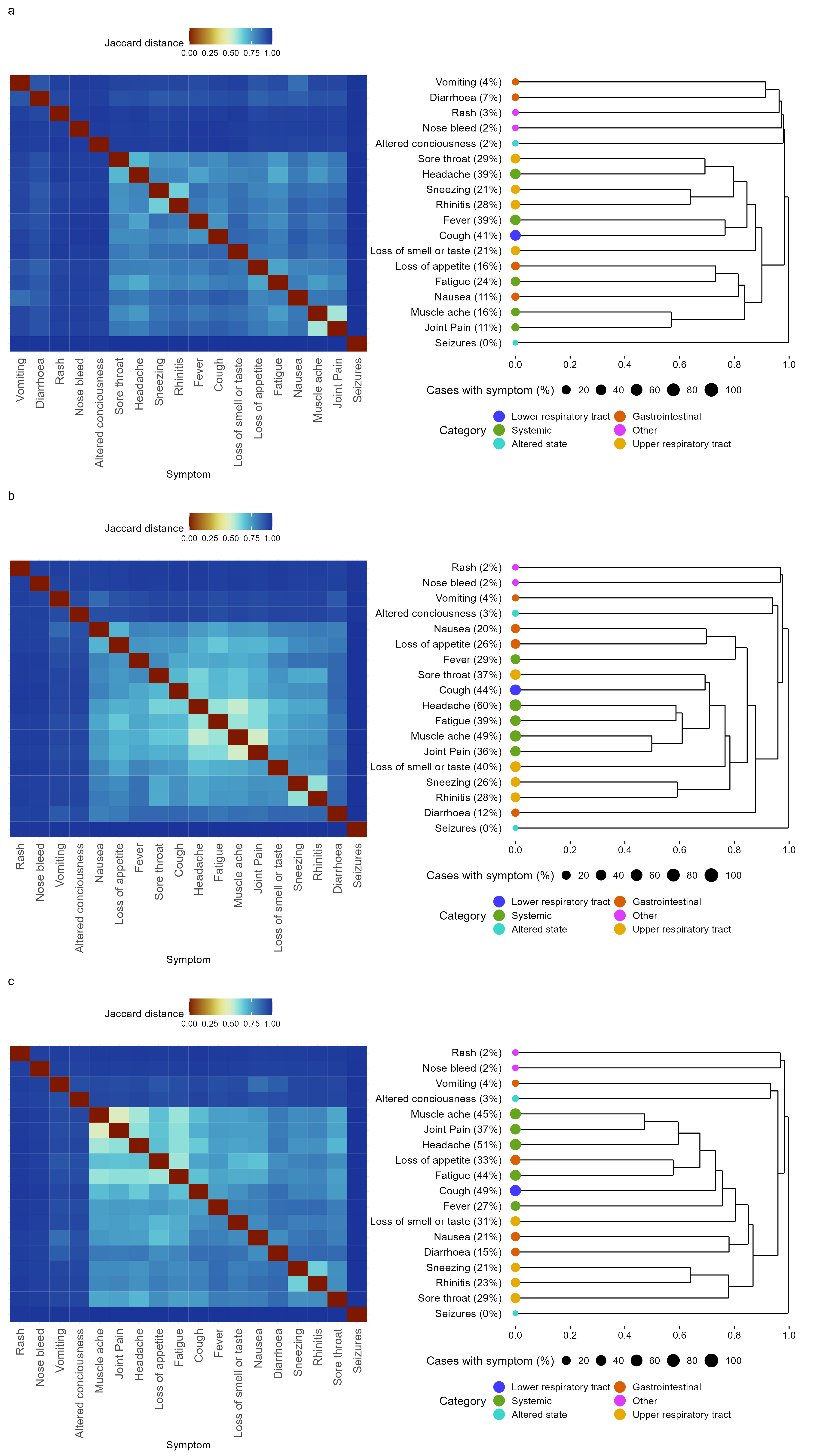}
    \caption{Hierarchical clustering of the Pillar 2 dataset with age stratification. Jaccard distance matrices between symptoms adjacent
to associated dendrograms were obtained through hierarchical clustering under
complete linkage. The symptom category is denoted using coloured points at the
roots of the dendrogram. The central columns give the name of the symptom with
the percentage of symptomatic cases who exhibit symptoms in the dataset. \textbf{a}. Children, \textbf{b}. Adults, \textbf{c}. Elders.}
    \label{fig:pillar 2 age stratified heirarchical clustering}
\end{figure}

\begin{figure}[H]
    \centering
    \includegraphics[width=0.7\linewidth]{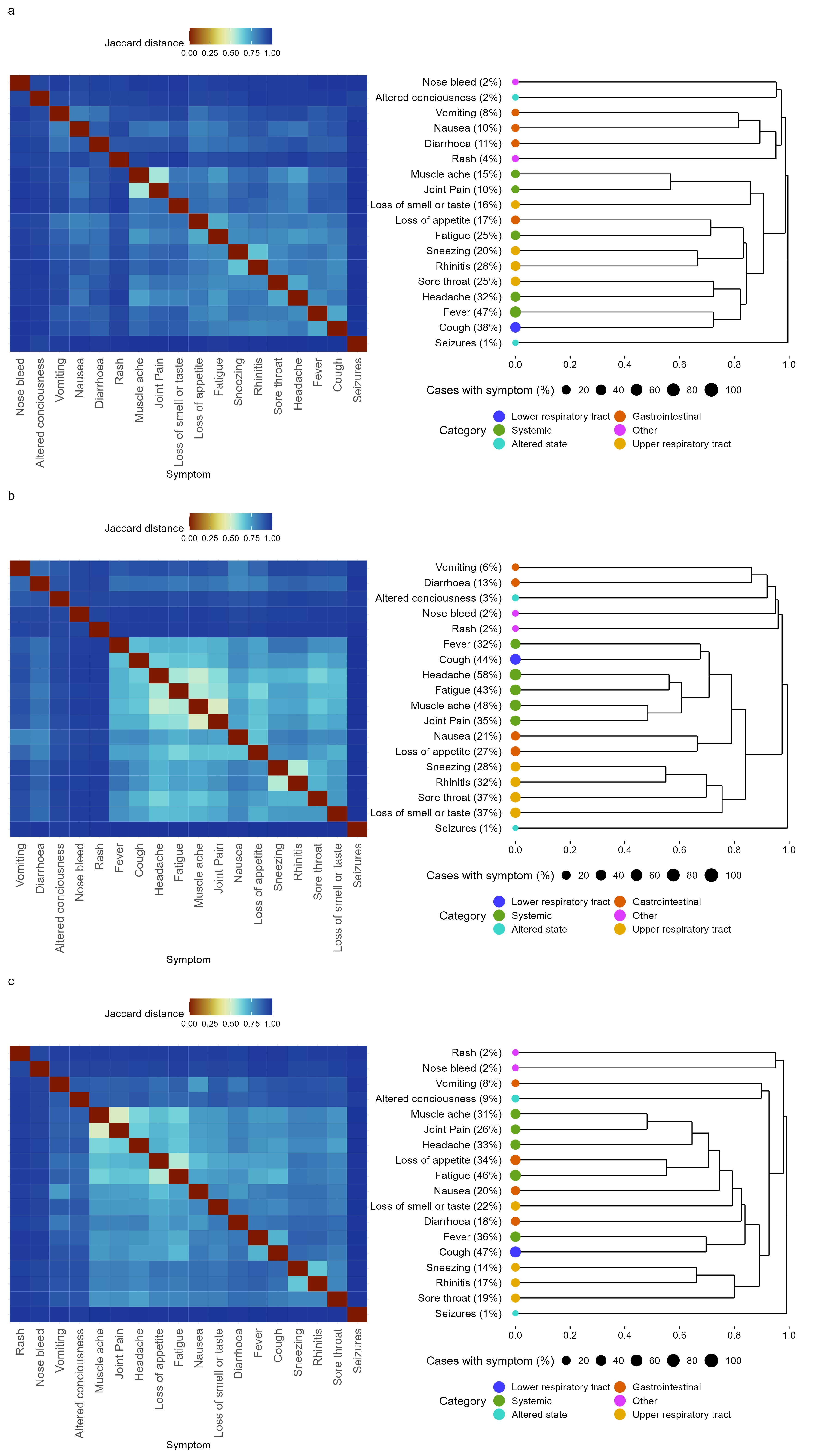}
    \caption{Hierarchical clustering of the SGSS dataset with age stratification. Jaccard distance matrices between symptoms adjacent
to associated dendrograms were obtained through hierarchical clustering under
complete linkage. The symptom category is denoted using coloured points at the
roots of the dendrogram. The central columns give the name of the symptom with
the percentage of symptomatic cases who exhibit symptoms in the dataset. \textbf{a}. Children, \textbf{b}. Adults, \textbf{c}. Elders.}
    \label{fig:SGSS age stratified heirarchical clustering}
\end{figure}

\begin{figure}[H]
    \centering
    \includegraphics[width=0.7\linewidth]{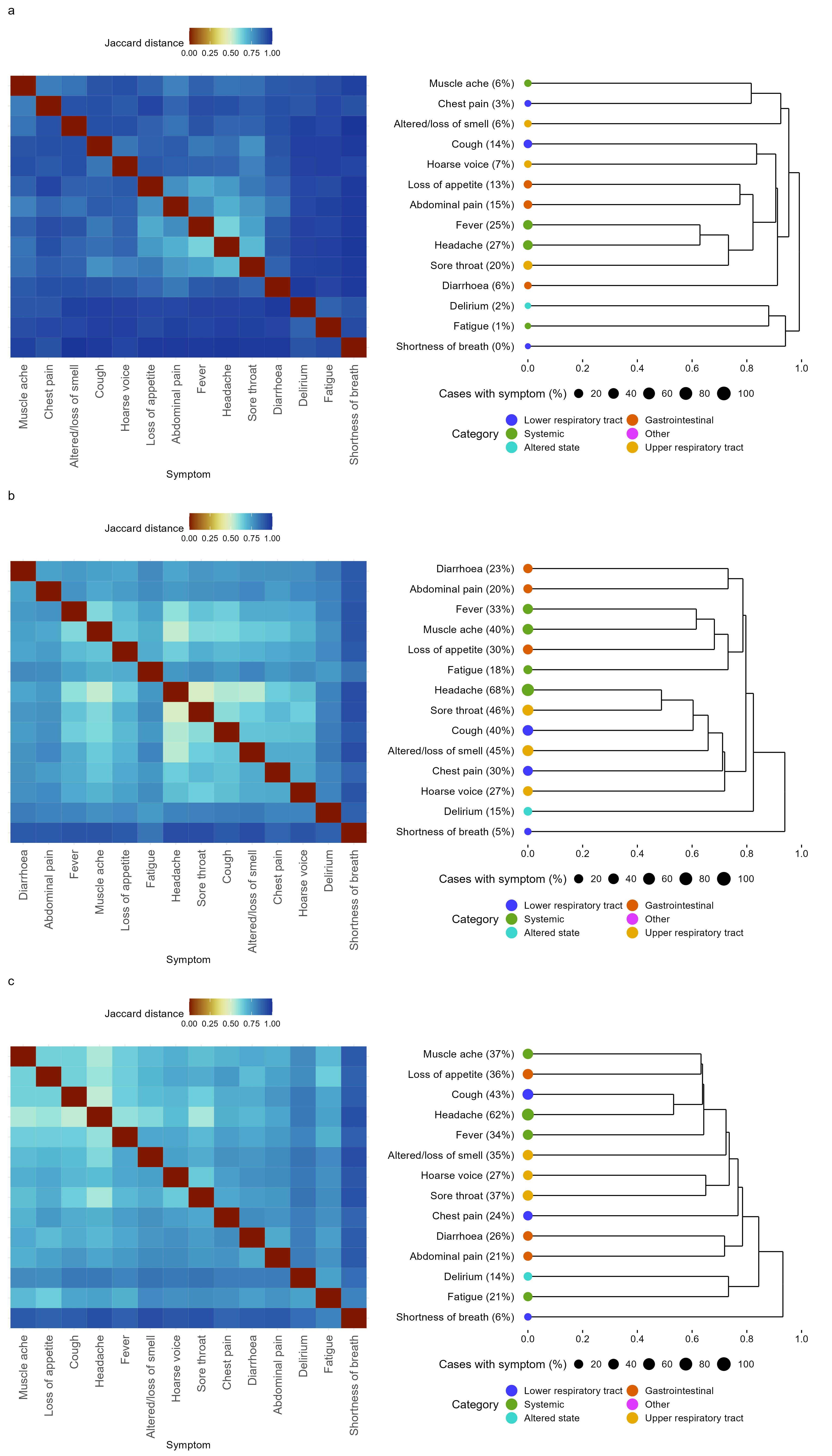}
    \caption{Hierarchical clustering of the COVID Symptom Study dataset with age stratification. Jaccard distance matrices between symptoms adjacent
to associated dendrograms were obtained through hierarchical clustering under
complete linkage. The symptom category is denoted using coloured points at the
roots of the dendrogram. The central columns give the name of the symptom with
the percentage of symptomatic cases who exhibit symptoms in the dataset. \textbf{a}. Children, \textbf{b}. Adults, \textbf{c}. Elders.}
    \label{fig:Zoe age stratified heirarchical clustering}
\end{figure}

\begin{figure}[H]
    \centering
    \includegraphics[width=0.7\linewidth]{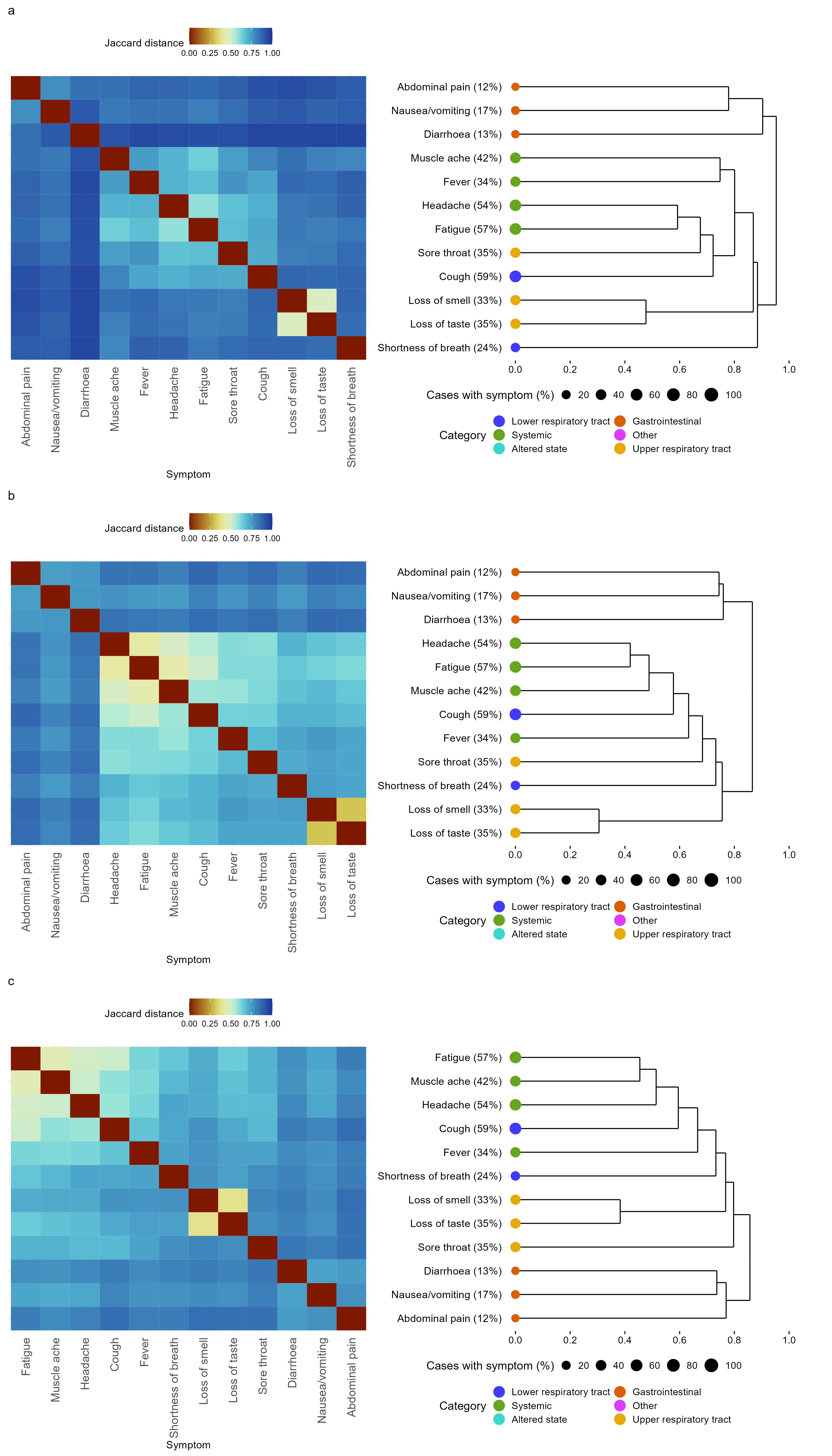}
    \caption{Hierarchical clustering of the COVID-19 Infection Survey dataset with age stratification. Jaccard distance matrices between symptoms adjacent
to associated dendrograms were obtained through hierarchical clustering under
complete linkage. The symptom category is denoted using coloured points at the
roots of the dendrogram. The central columns give the name of the symptom with
the percentage of symptomatic cases who exhibit symptoms in the dataset. \textbf{a}. Children, \textbf{b}. Adults, \textbf{c}. Elders.}
    \label{fig:ONS age stratified heirarchical clustering}
\end{figure}

\begin{figure}[H]
    \centering
    \includegraphics[width=0.4\linewidth]{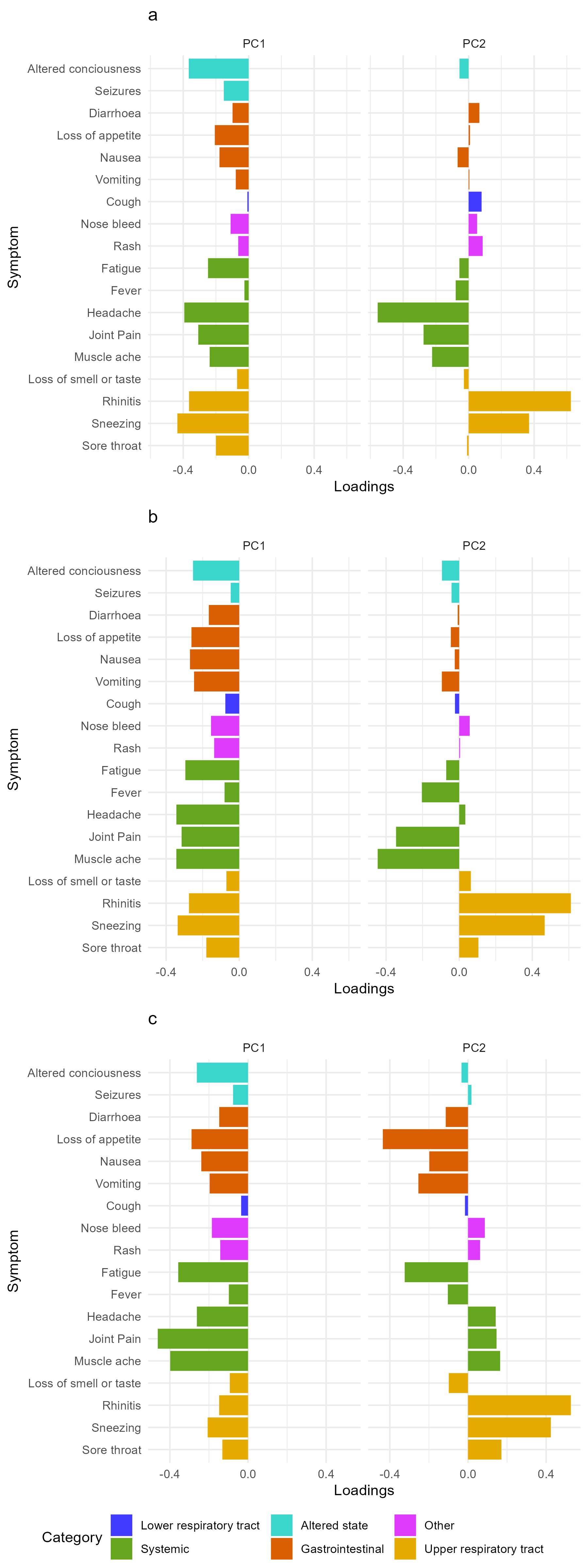}
    \caption{Logistic Principal Components Analysis (LPCA)
results performed on the Pillar 2 dataset with age stratification. For each stratum, elements of the principal components are visualised
as vertical bar plots. Each vector is insensitive to overall multiplication by
$-1$. Symptom categories are labelled by colours.  \textbf{a}. Children, \textbf{b}. Adults, \textbf{c}. Elders.}
    \label{fig:pillar 2 age stratified LPCA}
\end{figure}

\begin{figure}[H]
    \centering
    \includegraphics[width=0.4\linewidth]{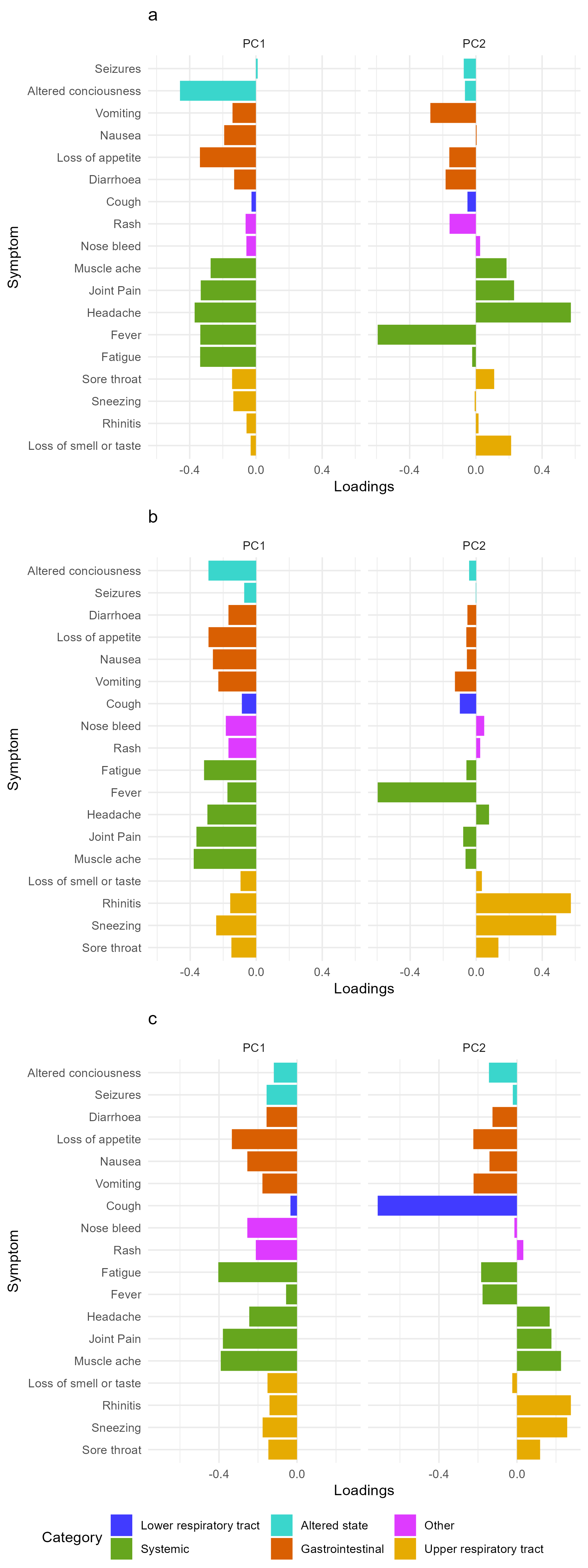}
    \caption{Logistic Principal Components Analysis (LPCA)
    results performed on the SGSS dataset with age stratification. For each stratum, elements of the principal components are visualised
    as vertical bar plots. Each vector is insensitive to overall multiplication by
    $-1$. Symptom categories are labelled by colours.  \textbf{a}. Children, \textbf{b}. Adults, \textbf{c}. Elders.}
    \label{fig:SGSS age stratified LPCA}
\end{figure}

\begin{figure}[H]
    \centering
    \includegraphics[width=0.4\linewidth]{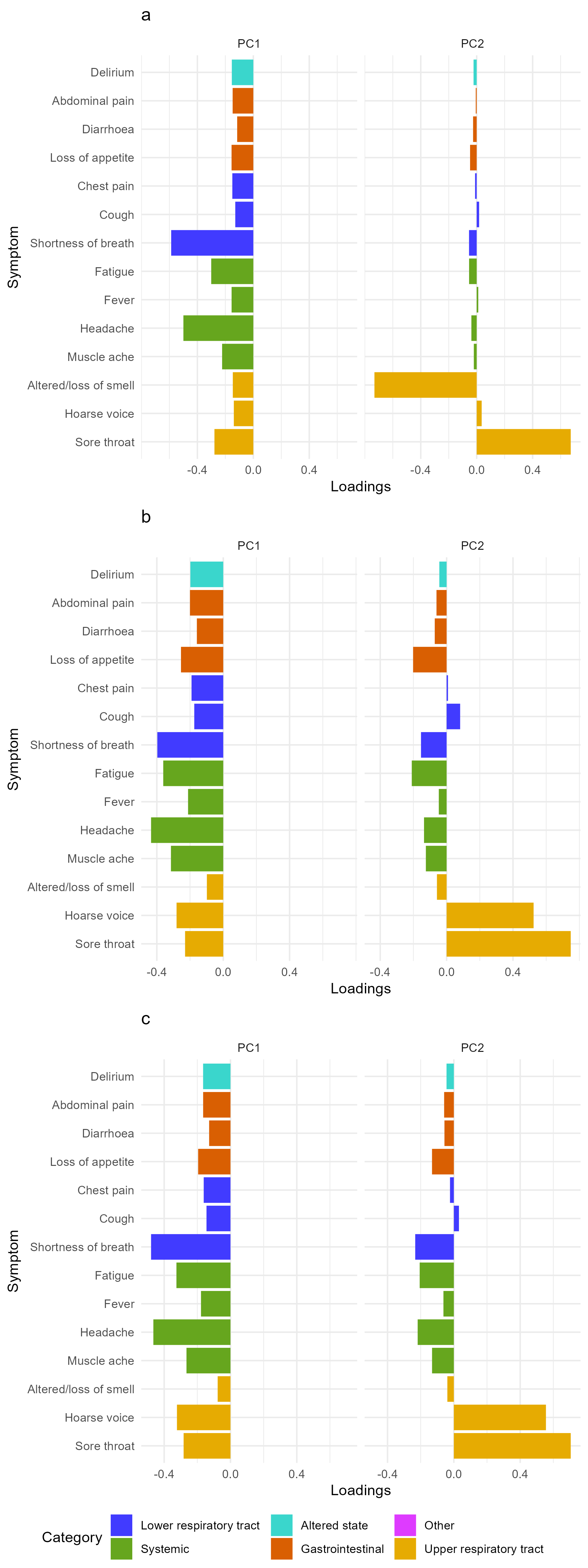}
    \caption{Logistic Principal Components Analysis (LPCA)
results performed on the COVID Symptom Study dataset with age stratification. For each stratum, elements of the principal components are visualised
as vertical bar plots. Each vector is insensitive to overall multiplication by
$-1$. Symptom categories are labelled by colours.  \textbf{a}. Children, \textbf{b}. Adults, \textbf{c}. Elders.}
    \label{fig:Zoe age stratified LPCA}
\end{figure}

\begin{figure}[H]
    \centering
    \includegraphics[width=0.4\linewidth]{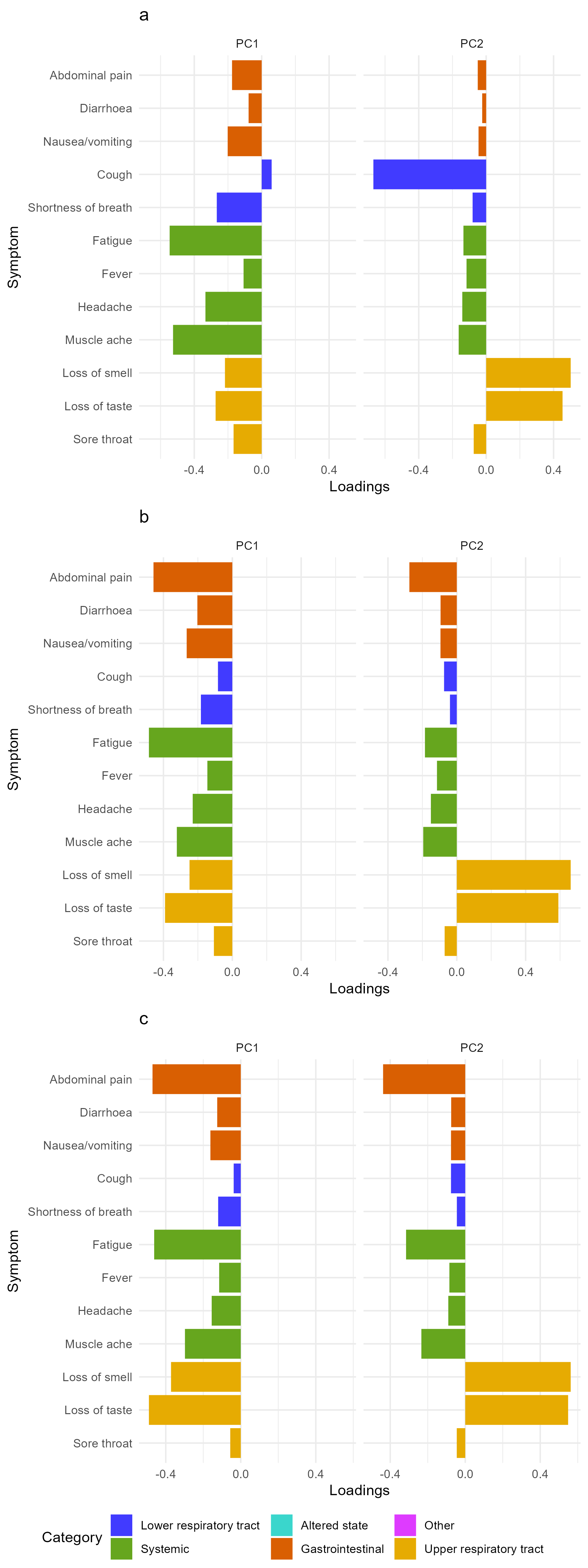}
    \caption{Logistic Principal Components Analysis (LPCA)
results performed on the COVID-19 Infection Survey dataset with age stratification. For each stratum, elements of the principal components are visualised
as vertical bar plots. Each vector is insensitive to overall multiplication by
$-1$. Symptom categories are labelled by colours.  \textbf{a}. Children, \textbf{b}. Adults, \textbf{c}. Elders.}
    \label{fig:ONS age stratified LPCA}
\end{figure}

\begin{figure}[H]
    \centering
    \includegraphics[width = \textwidth]{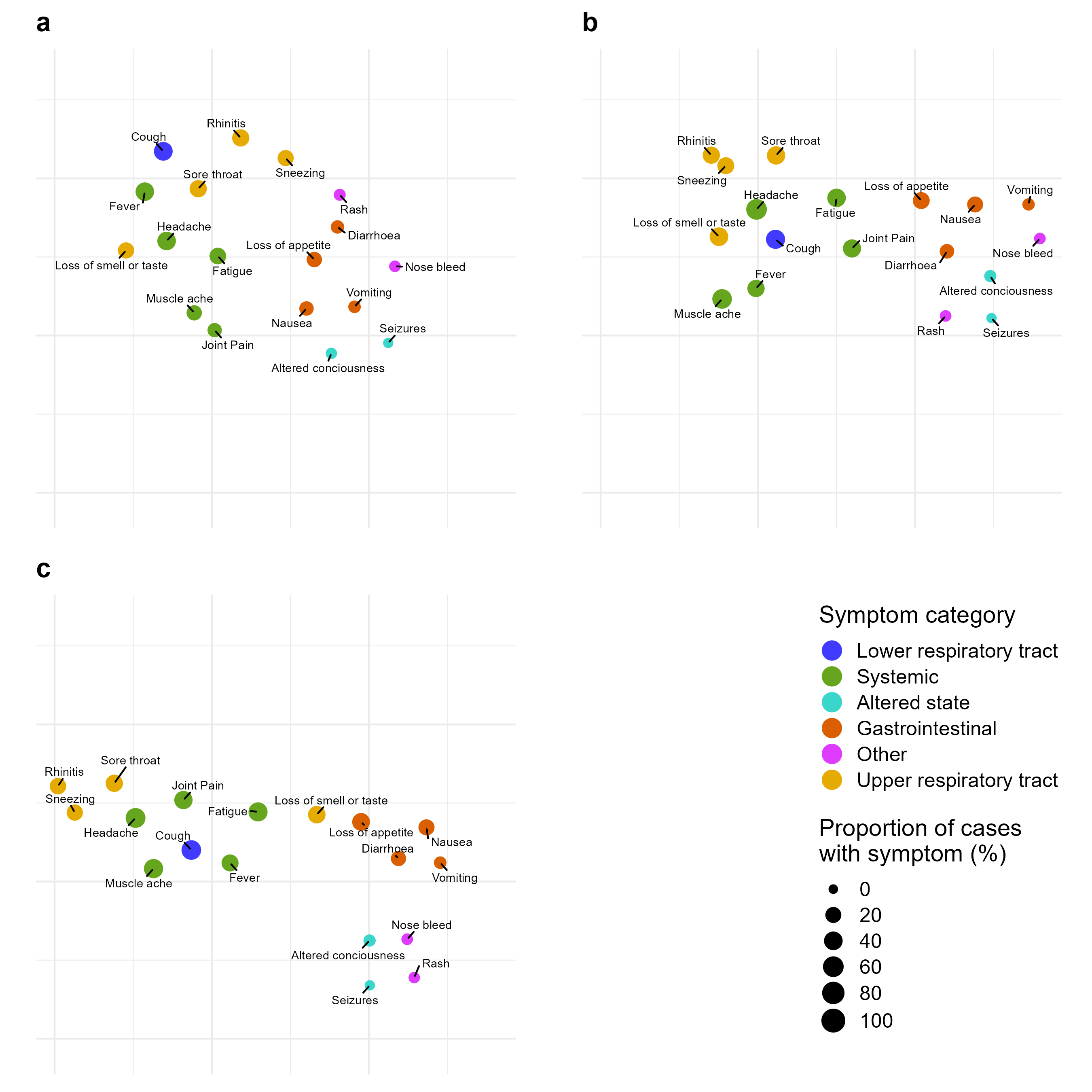}
    \caption{UMAP embeddings of SARS-CoV-2 symptoms performed on Pillar 2 dataset with age stratification. The algorithm
attempts to place combinations of symptoms that commonly co-occur close to each
other. Point size is proportional to the proportion of cases that develop a
given symptom. For this embedding, the parameters were chosen to capture more of
the global structure of symptoms and produces less well-defined clusters. \textbf{a}. Children, \textbf{b}. Adults, \textbf{c}. Elders.}
    \label{fig:Pillar 2 Age stratified UMAP}
\end{figure}

\begin{figure}[H]
    \centering
    \includegraphics[width = \textwidth]{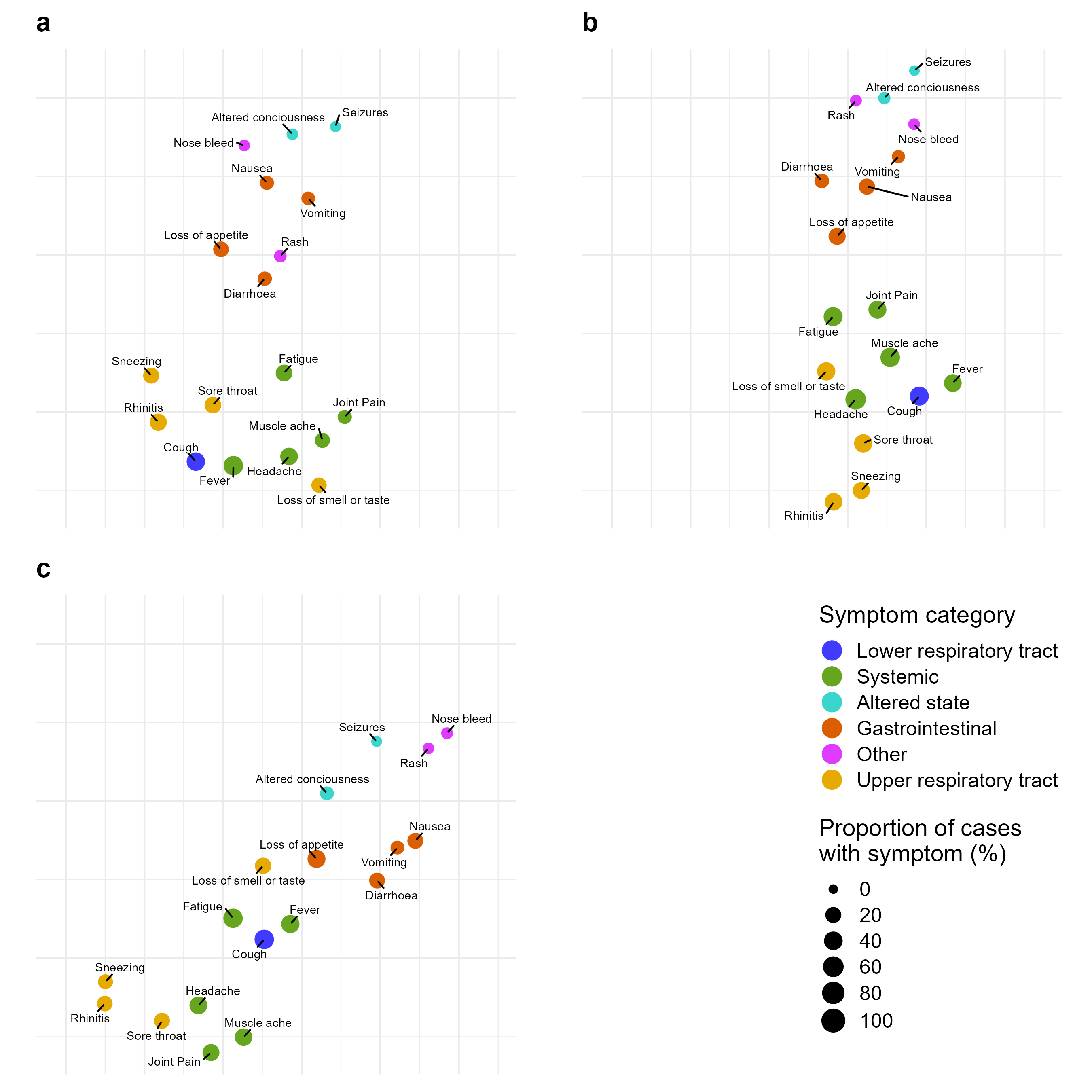}
    \caption{UMAP embeddings of SARS-CoV-2 symptoms performed on SGSS dataset with age stratification. The algorithm
attempts to place combinations of symptoms that commonly co-occur close to each
other. Point size is proportional to the proportion of cases that develop a
given symptom. For this embedding, the parameters were chosen to capture more of
the global structure of symptoms and produces less well-defined clusters. \textbf{a}. Children, \textbf{b}. Adults, \textbf{c}. Elders.}
    \label{fig:SGSS age stratified UMAP (loose)}
\end{figure}

\begin{figure}[H]
    \centering
    \includegraphics[width = \textwidth]{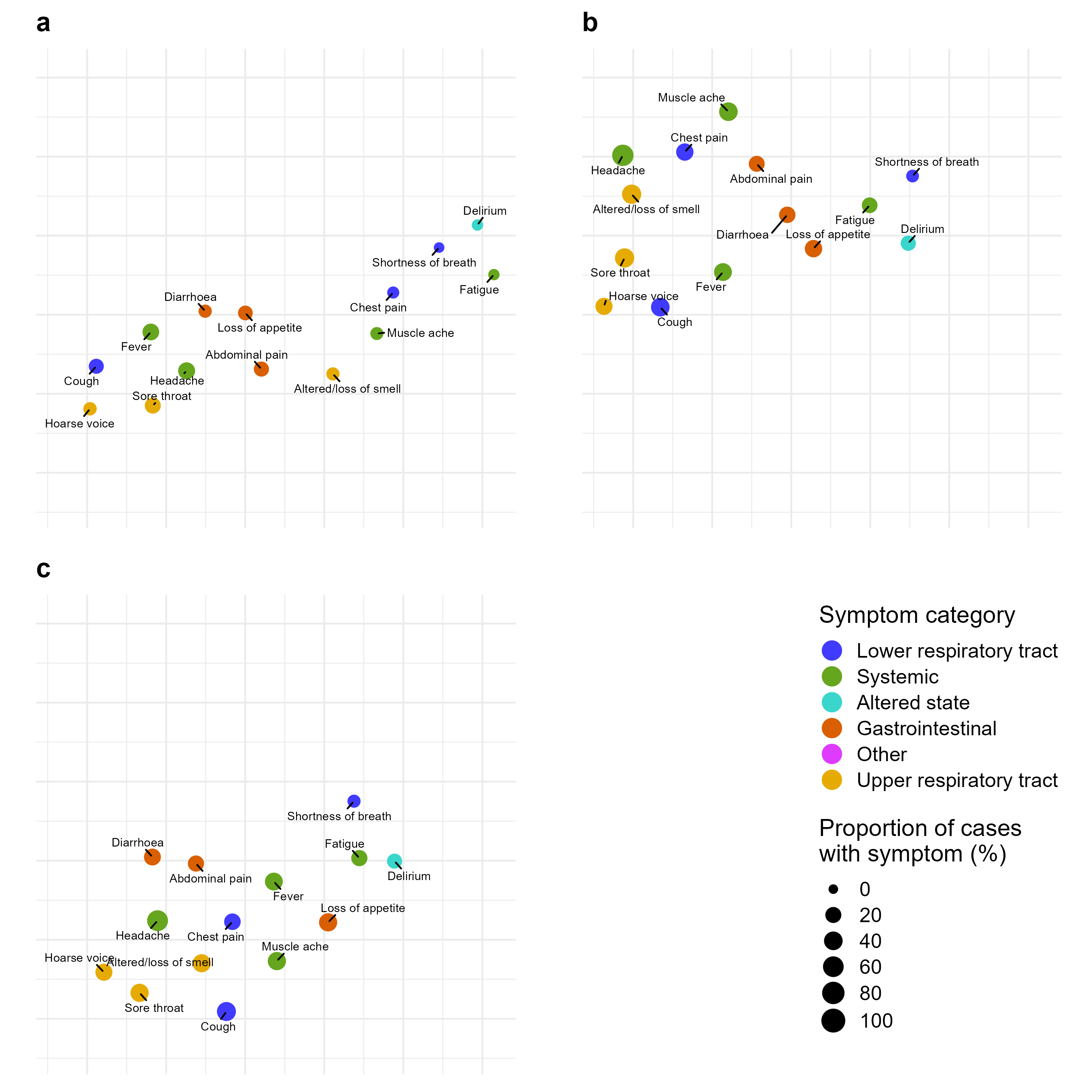}
    \caption{UMAP embeddings of SARS-CoV-2 symptoms performed on COVID Symptom Study dataset with age stratification. The algorithm
attempts to place combinations of symptoms that commonly co-occur close to each
other. Point size is proportional to the proportion of cases that develop a
given symptom. For this embedding, the parameters were chosen to capture more of
the global structure of symptoms and produces less well-defined clusters. \textbf{a}. Children, \textbf{b}. Adults, \textbf{c}. Elders.}
    \label{fig:Zoe age stratified UMAP (loose)}
\end{figure}

\begin{figure}[H]
    \centering
    \includegraphics[width = \textwidth]{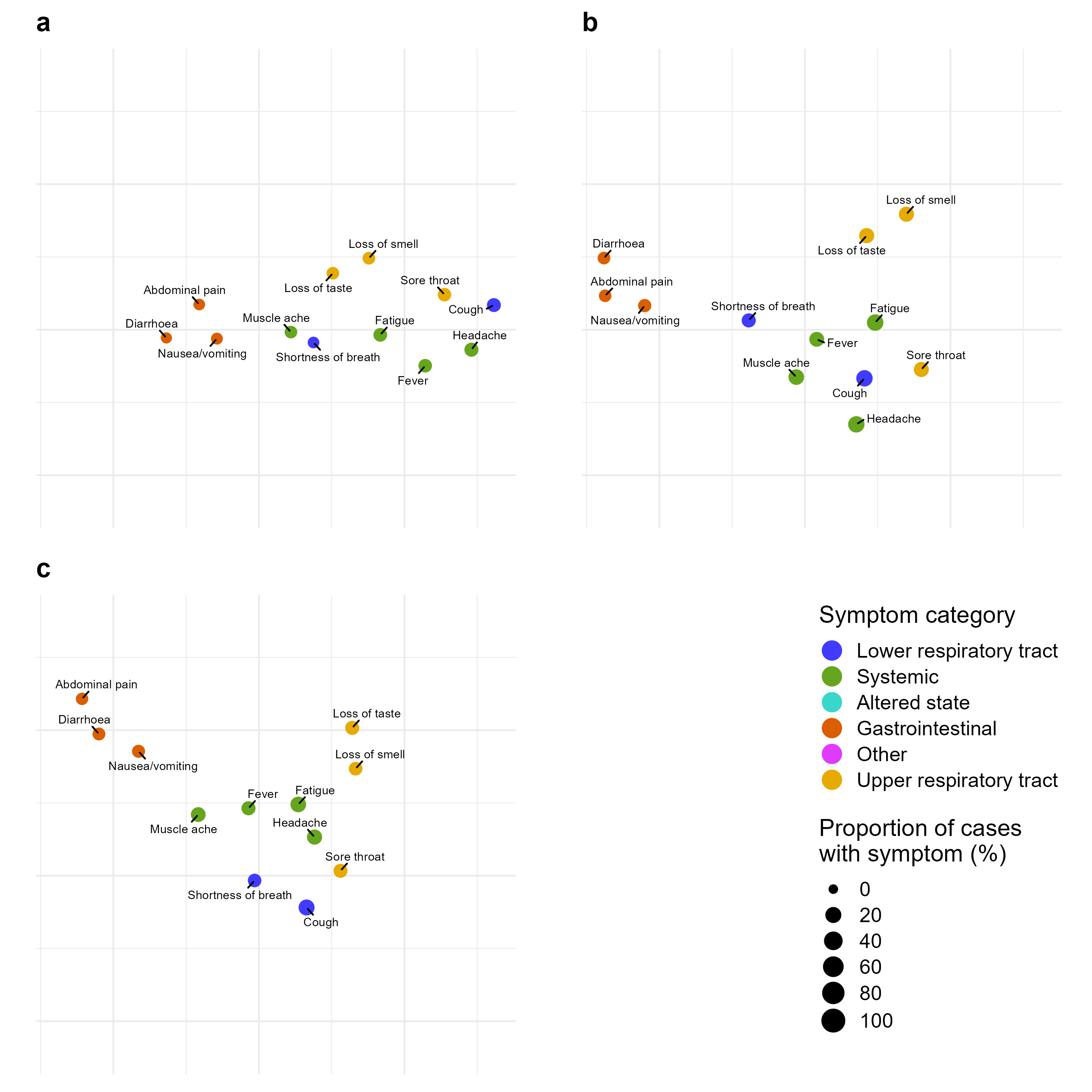}
    \caption{UMAP embeddings of SARS-CoV-2 symptoms performed on COVID-19 Infection Survey dataset with age stratification. The algorithm
attempts to place combinations of symptoms that commonly co-occur close to each
other. Point size is proportional to the proportion of cases that develop a
given symptom. For this embedding, the parameters were chosen to capture more of
the global structure of symptoms and produces less well-defined clusters. \textbf{a}. Children, \textbf{b}. Adults, \textbf{c}. Elders.}
    \label{fig:ONS Age stratified UMAP}
\end{figure}

\begin{figure}[H]
    \centering
    \includegraphics[width = \textwidth]{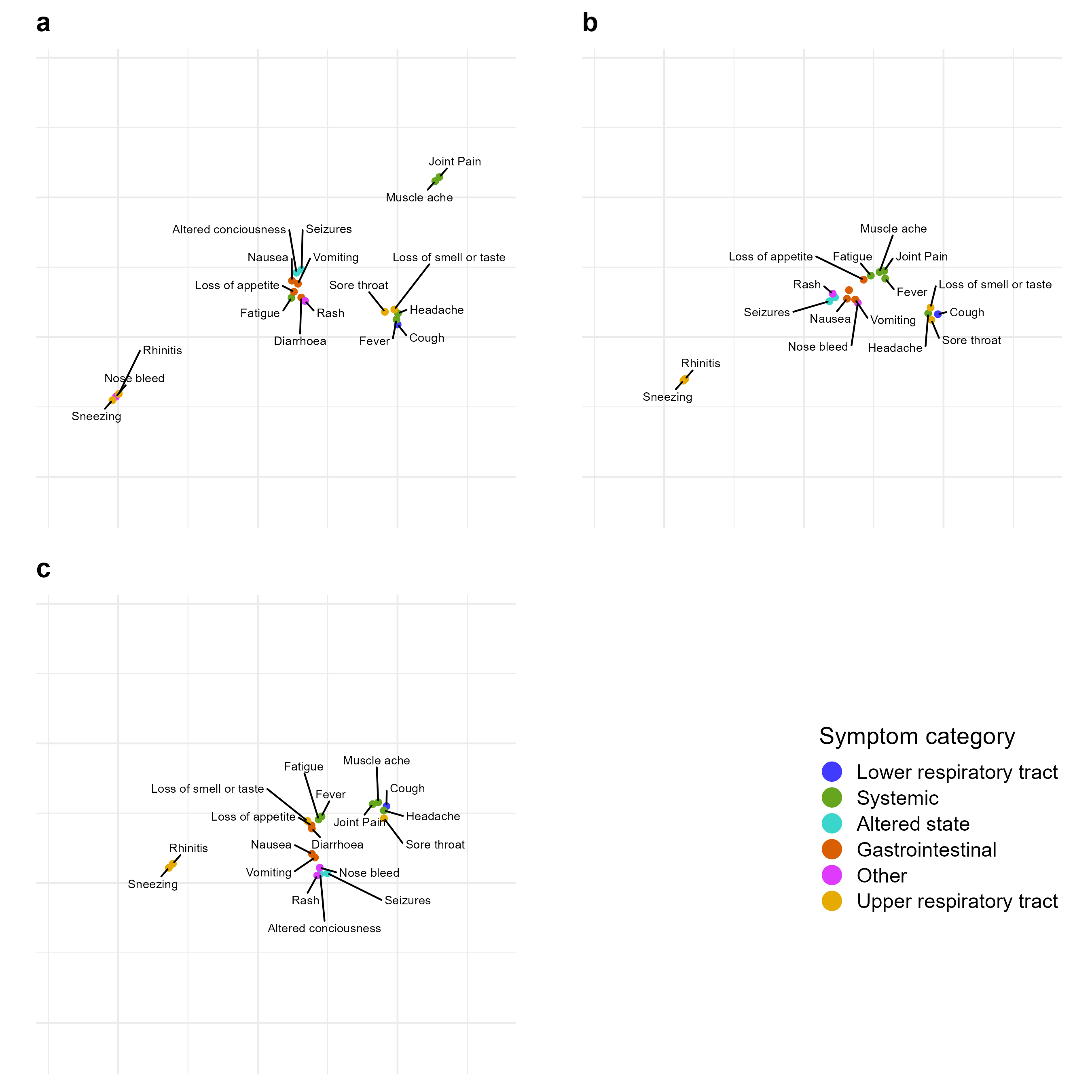}
    \caption{UMAP embeddings of SARS-CoV-2 symptoms performed on Pillar 2 dataset with age stratification. The algorithm
attempts to place combinations of symptoms that commonly co-occur close to each
other. For this embedding, the parameters were chosen to produce well-separated
symptom clusters. \textbf{a}. Children, \textbf{b}. Adults, \textbf{c}. Elders.}
    \label{fig:Pillar 2 Age stratified UMAP (tight)}
\end{figure}

\begin{figure}[H]
    \centering
    \includegraphics[width = \textwidth]{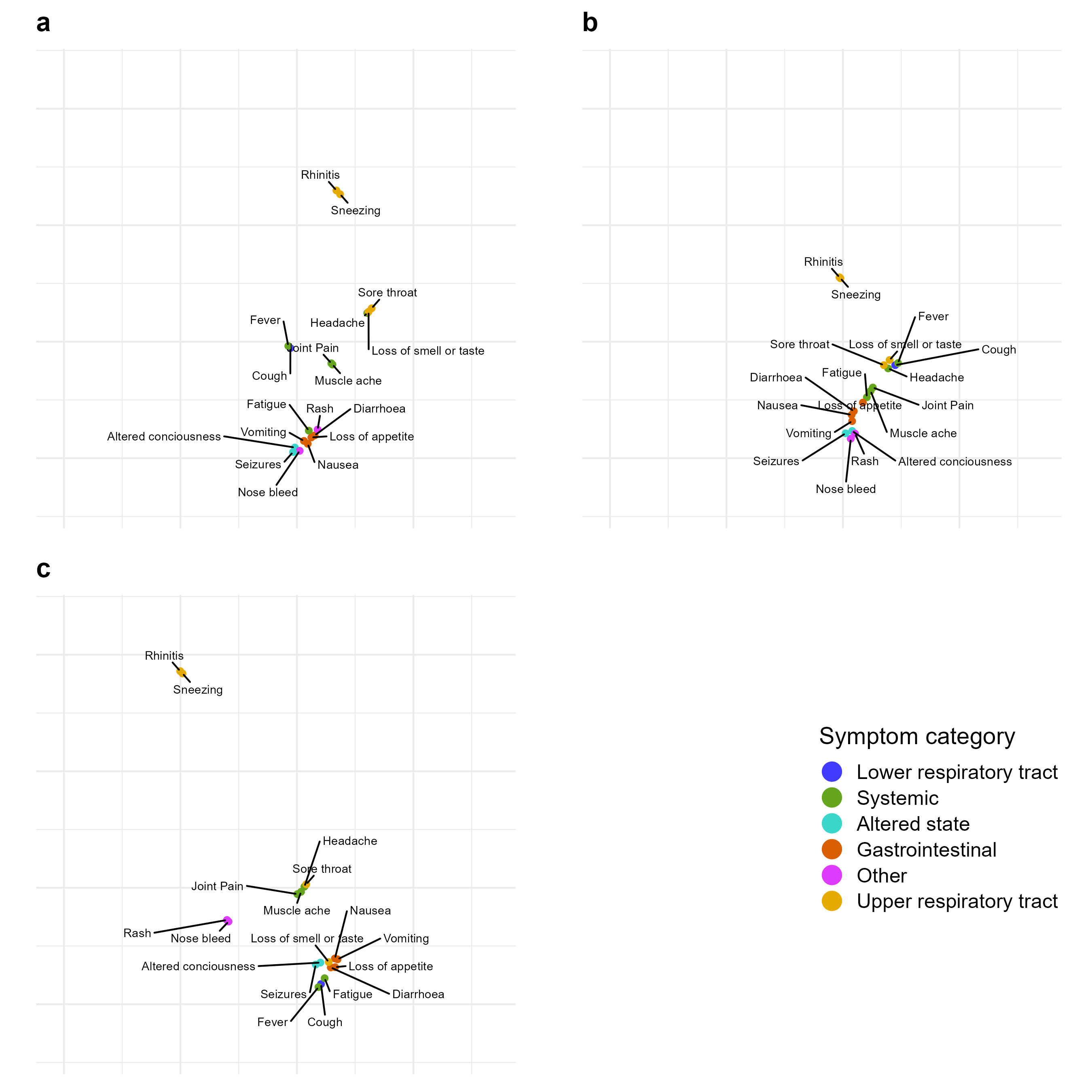}
    \caption{UMAP embeddings of SARS-CoV-2 symptoms performed on SGSS dataset with age stratification. The algorithm
attempts to place combinations of symptoms that commonly co-occur close to each
other. For this embedding, the parameters were chosen to produce well-separated
symptom clusters. \textbf{a}. Children, \textbf{b}. Adults, \textbf{c}. Elders.}
    \label{fig:SGSS age stratified UMAP (tight)}
\end{figure}

\begin{figure}[H]
    \centering
    \includegraphics[width = \textwidth]{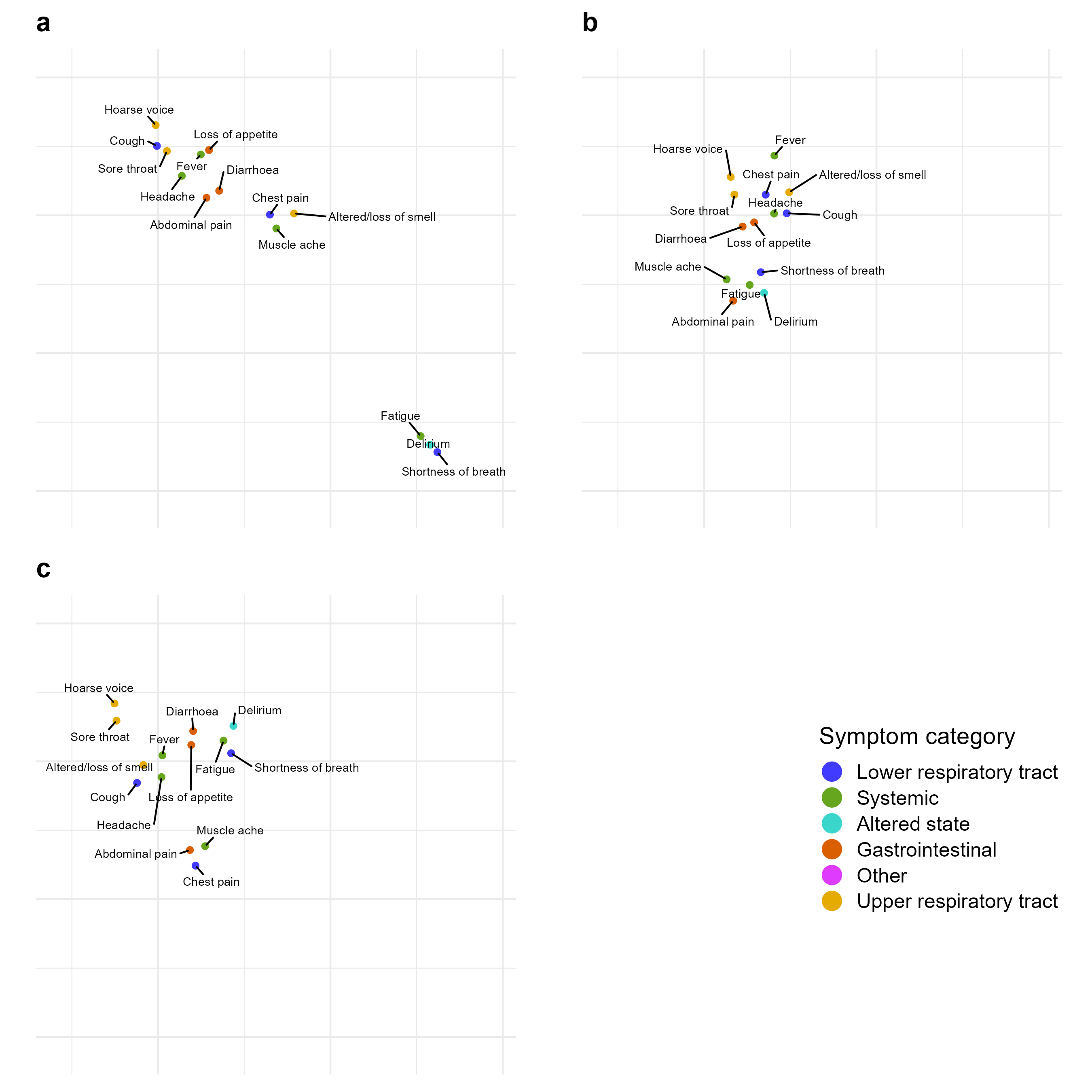}
    \caption{UMAP embeddings of SARS-CoV-2 symptoms performed on COVID Symptom Study dataset with age stratification. The algorithm
attempts to place combinations of symptoms that commonly co-occur close to each
other. For this embedding, the parameters were chosen to produce well-separated
symptom clusters. \textbf{a}. Children, \textbf{b}. Adults, \textbf{c}. Elders.}
    \label{fig:Zoe age stratified UMAP (tight)}
\end{figure}

\begin{figure}[H]
    \centering
    \includegraphics[width = \textwidth]{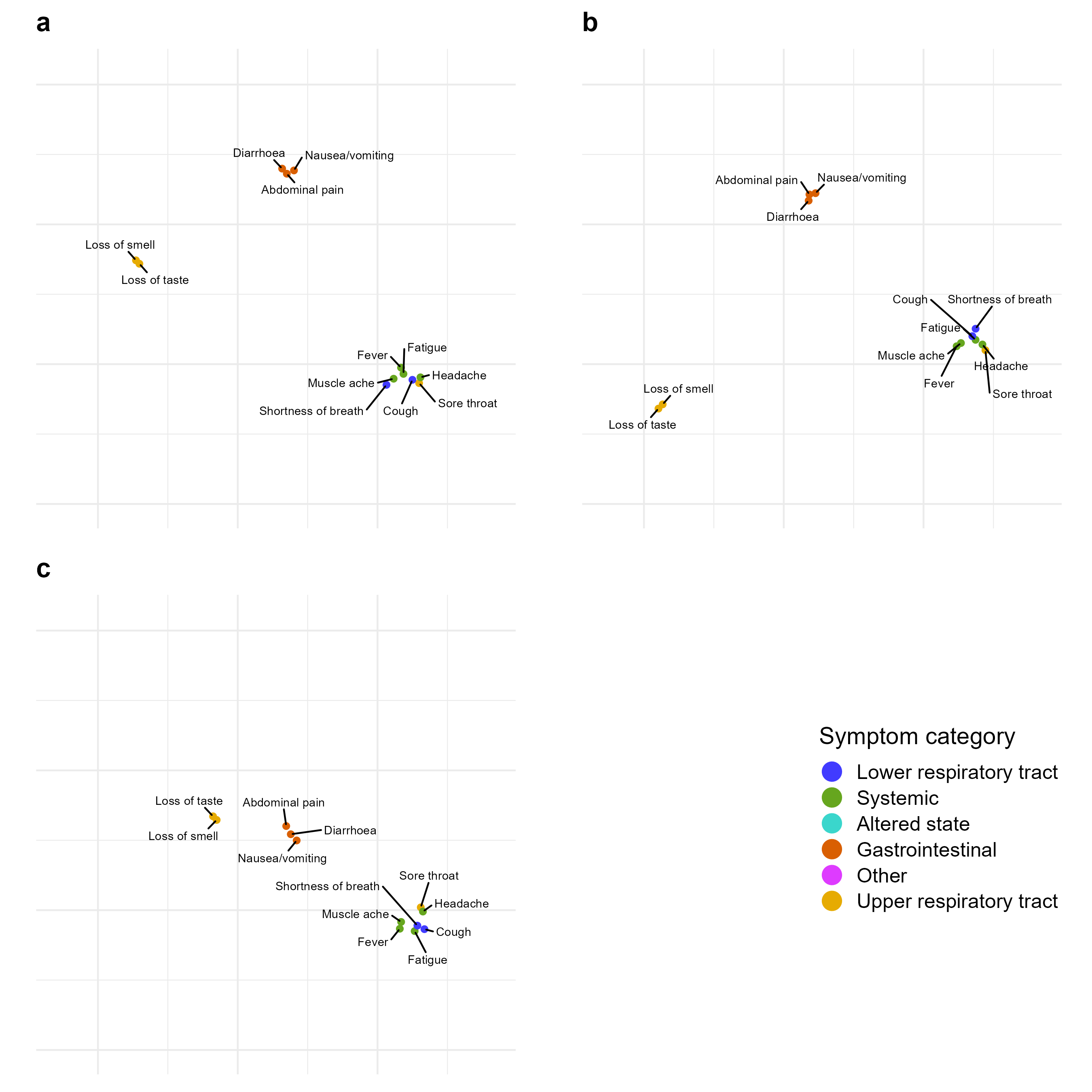}
    \caption{UMAP embeddings of SARS-CoV-2 symptoms performed on COVID-19 Infection Survey dataset with age stratification. The algorithm
attempts to place combinations of symptoms that commonly co-occur close to each
other. For this embedding, the parameters were chosen to produce well-separated
symptom clusters. \textbf{a}. Children, \textbf{b}. Adults, \textbf{c}. Elders.}
    \label{fig:ONS Age stratified UMAP (tight)}
\end{figure}

\begin{figure}[H]
    \centering
    \includegraphics[width = \textwidth]{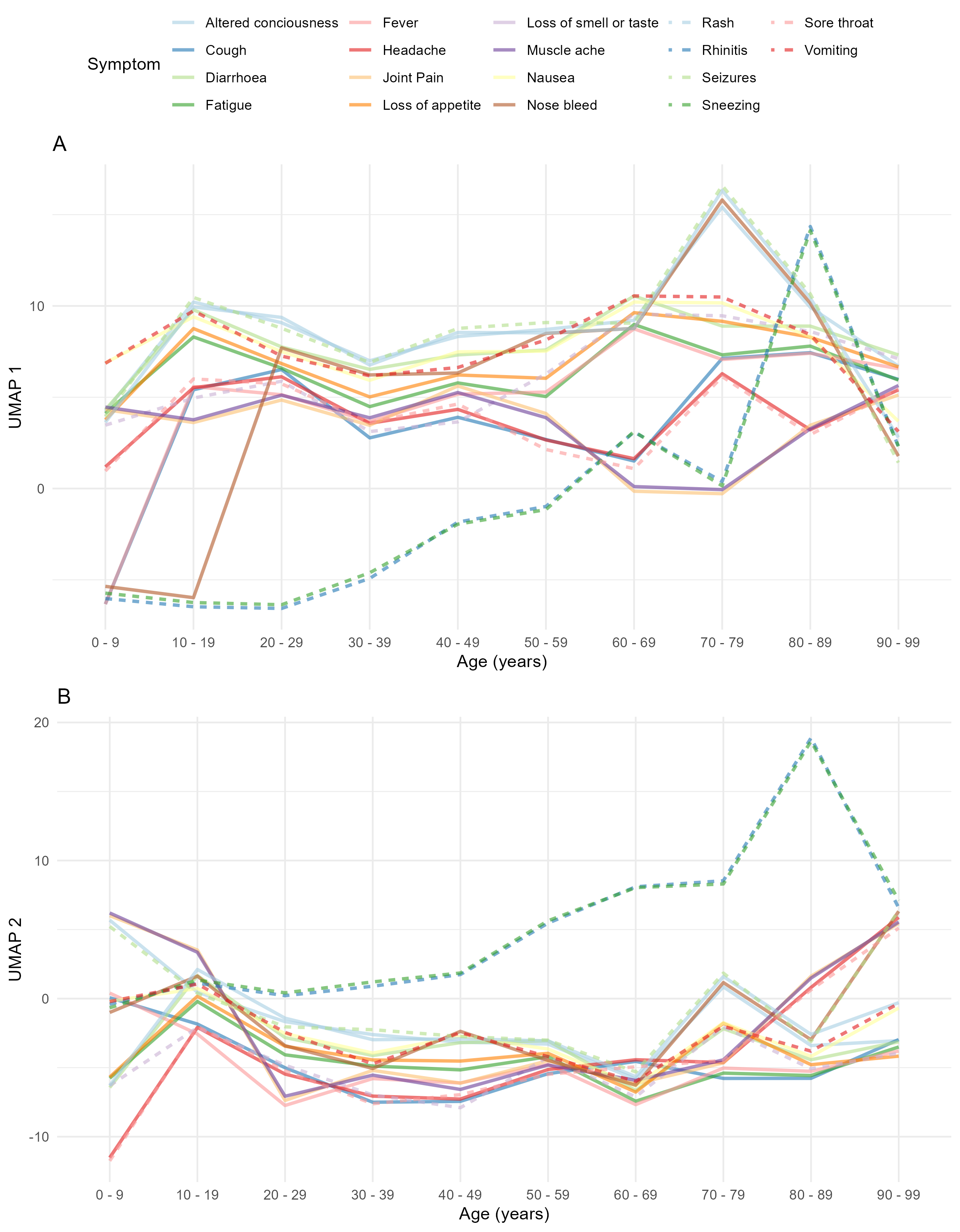}
    \caption{Marginal plots of AlignedUMAP embeddings of SARS-CoV-2, for Pillar 2 data age-stratified into strata of length 10 years. For each strata, an optimal two-
dimensional embedding into Euclidean space, denoted via UMAP 1 and UMAP 2, of the symptoms is found, subject to the loose constraint that each symptom is placed in a similar location in adjacent embeddings.}
    \label{fig:marginal_umap_p2}
\end{figure}

\begin{figure}[H]
    \centering
    \includegraphics[width = \textwidth]{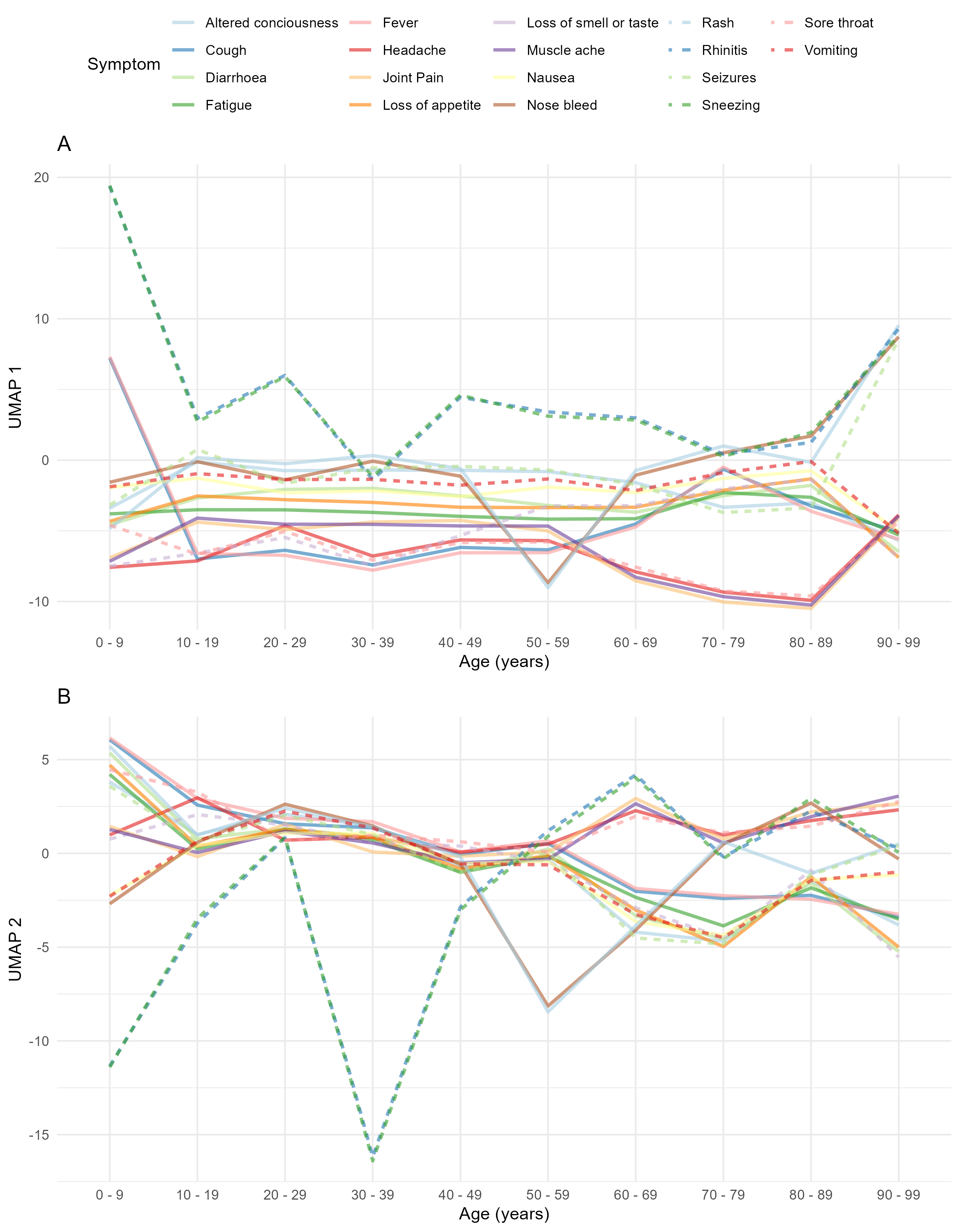}
    \caption{Marginal plots of AlignedUMAP embeddings of SARS-CoV-2, for SGSS data age-stratified into strata of length 10 years. For each strata, an optimal two-
dimensional embedding into Euclidean space, denoted via UMAP 1 and UMAP 2, of the symptoms is found, subject to the loose constraint that each symptom is placed in a similar location in adjacent embeddings.}
    \label{fig:marginal_umap_sgss}
\end{figure}

\begin{figure}[H]
    \centering
    \includegraphics[width = \textwidth]{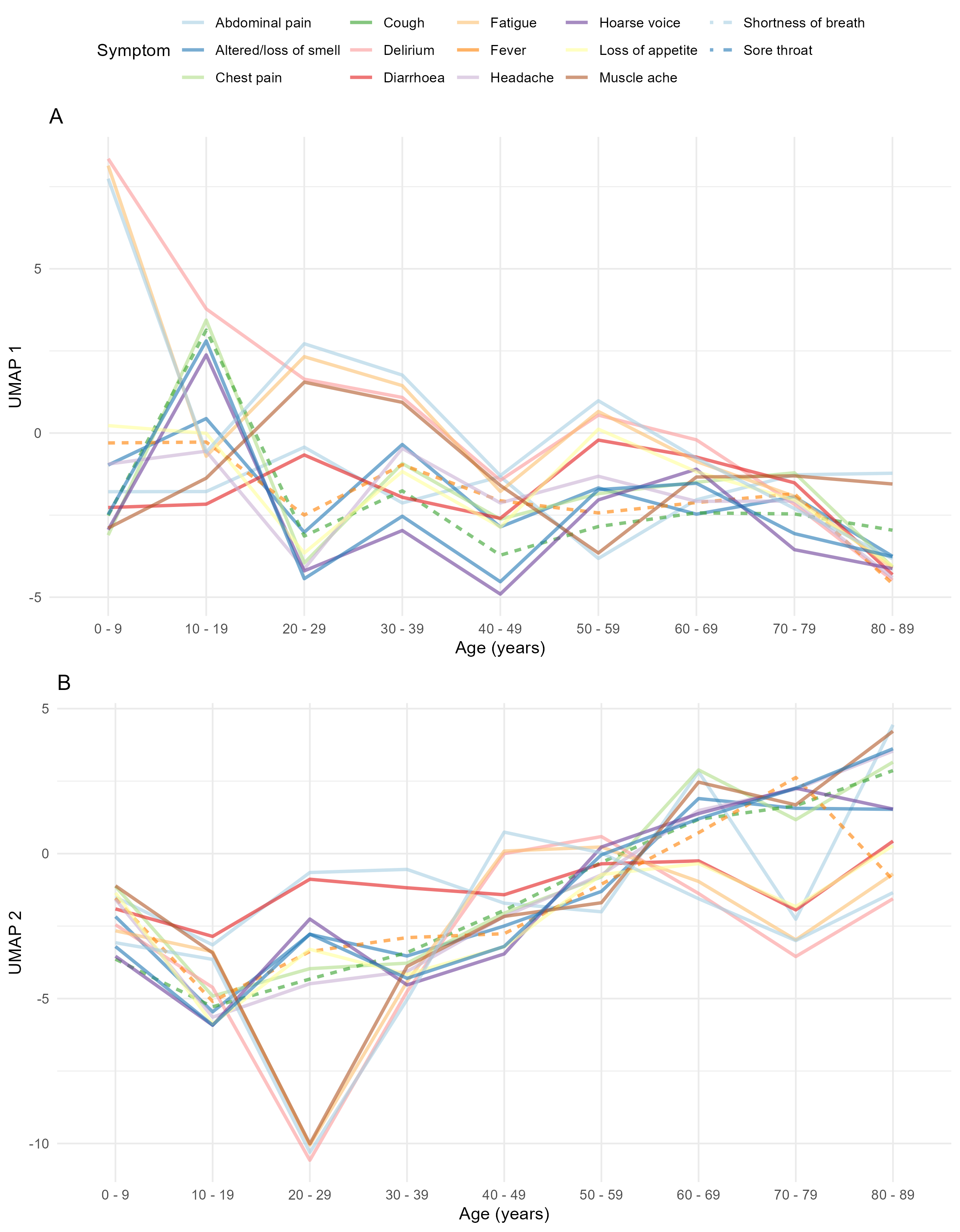}
    \caption{Marginal plots of AlignedUMAP embeddings of SARS-CoV-2, for COVID Symptom Study data age-stratified into strata of length 10 years. For each strata, an optimal two-
dimensional embedding into Euclidean space, denoted via UMAP 1 and UMAP 2, of the symptoms is found, subject to the loose constraint that each symptom is placed in a similar location in adjacent embeddings.}
    \label{fig:marginal_umap_css}
\end{figure}

\begin{figure}[H]
    \centering
    \includegraphics[width = \textwidth]{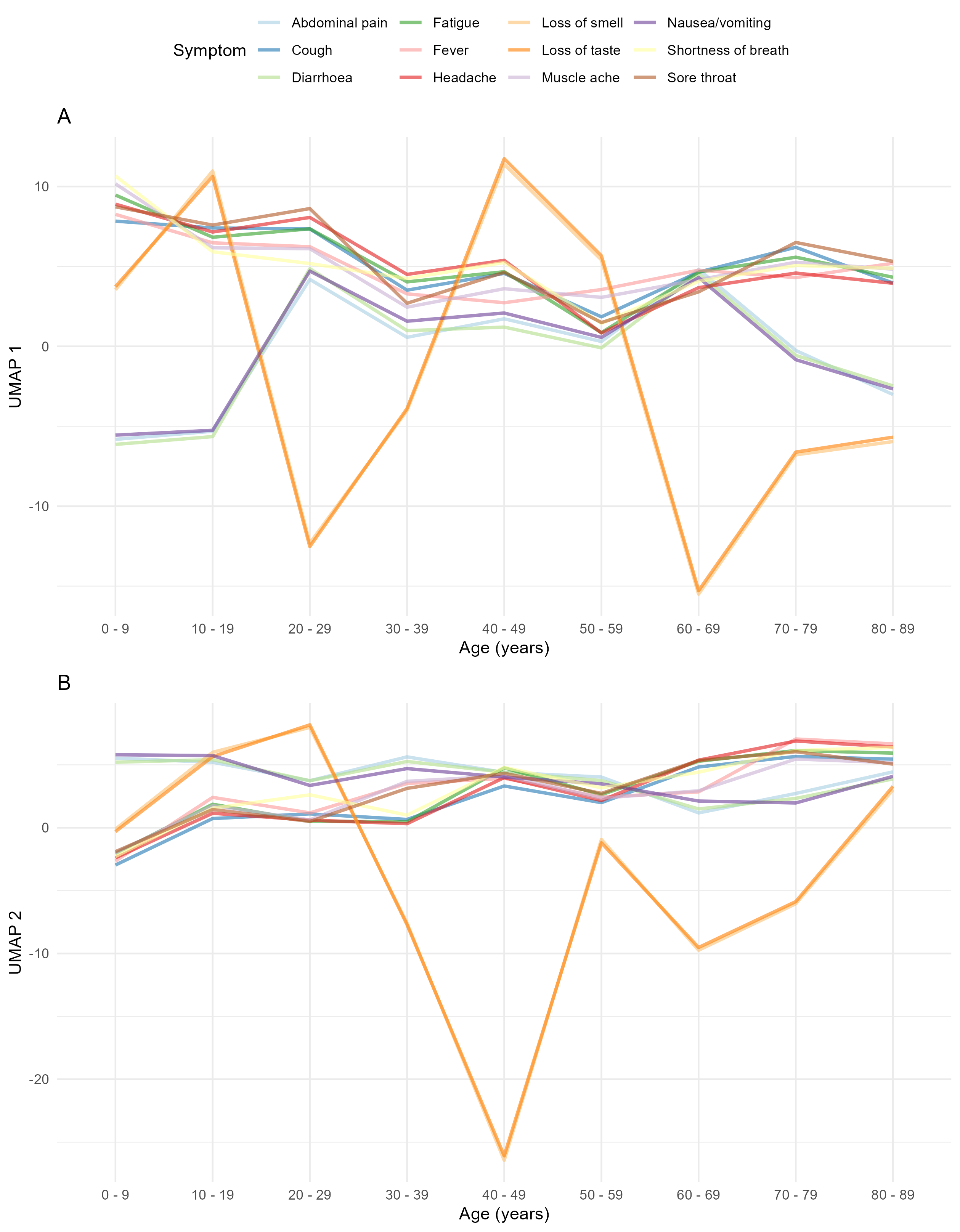}
    \caption{Marginal plots of AlignedUMAP embeddings of SARS-CoV-2, for COVID-19 Infection Survey data age-stratified into strata of length 10 years. For each strata, an optimal two-
dimensional embedding into Euclidean space, denoted via UMAP 1 and UMAP 2, of the symptoms is found, subject to the loose constraint that each symptom is placed in a similar location in adjacent embeddings.}
    \label{fig:marginal_umap_cis}
\end{figure}


\begin{thebibliography}{31}
\providecommand{\natexlab}[1]{#1}
\providecommand{\url}[1]{\texttt{#1}}
\expandafter\ifx\csname urlstyle\endcsname\relax
  \providecommand{\doi}[1]{doi: #1}\else
  \providecommand{\doi}{doi: \begingroup \urlstyle{rm}\Url}\fi

\bibitem[{World Health Organization}(2022)]{WHO_Sitrep}
{World Health Organization}.
\newblock {Coronavirus disease ({COVID}-19) pandemic}, 2022.
\newblock URL
  \url{https://www.who.int/emergencies/diseases/novel-coronavirus-2019}.
\newblock 25 May 2022.

\bibitem[Hale et~al.(2021)Hale, Angrist, Goldszmidt, Kira, Petherick, Phillips,
  Webster, Cameron-Blake, Hallas, Majumdar, and Tatlow]{hale_global_2021}
Thomas Hale, Noam Angrist, Rafael Goldszmidt, Beatriz Kira, Anna Petherick,
  Toby Phillips, Samuel Webster, Emily Cameron-Blake, Laura Hallas, Saptarshi
  Majumdar, and Helen Tatlow.
\newblock A global panel database of pandemic policies ({Oxford} {COVID}-19
  {Government} {Response} {Tracker}).
\newblock \emph{Nature Human Behaviour}, 5\penalty0 (4):\penalty0 529--538,
  2021.

\bibitem[{Google LLC}(2021)]{GoogleMobility}
{Google LLC}.
\newblock {Community Mobility Reports}, 2021.
\newblock URL \url{https://www.google.com/covid19/mobility/}.

\bibitem[Fyles et~al.(2021)Fyles, Fearon, Overton, {University of Manchester
  COVID-19 Modelling Group}, Wingfield, Medley, Hall, Pellis, and
  House]{Fyles:2021}
Martyn Fyles, Elizabeth Fearon, Christopher Overton, {University of Manchester
  COVID-19 Modelling Group}, Tom Wingfield, Graham~F. Medley, Ian Hall, Lorenzo
  Pellis, and Thomas House.
\newblock Using a household-structured branching process to analyse contact
  tracing in the {SARS-CoV-2} pandemic.
\newblock \emph{Philosophical Transactions of the Royal Society {B}: Biological
  Sciences}, 376\penalty0 (1829):\penalty0 20200267, 2021.

\bibitem[Crozier et~al.(2021)Crozier, Dunning, Rajan, Semple, and
  Buchan]{Croziern1625}
Alex Crozier, Jake Dunning, Selina Rajan, Malcolm~G Semple, and Iain~E Buchan.
\newblock Could expanding the {COVID-19} case definition improve the {UK's}
  pandemic response?
\newblock \emph{BMJ}, 374, 2021.
\newblock \doi{10.1136/bmj.n1625}.
\newblock URL \url{https://www.bmj.com/content/374/bmj.n1625}.

\bibitem[Struyf et~al.()Struyf, Deeks, Dinnes, Takwoingi, Davenport, Leeflang,
  Spijker, Hooft, Emperador, Dittrich, Domen, Horn, Bruel, and
  Group]{struyf_signs_2020}
Thomas Struyf, Jonathan~J. Deeks, Jacqueline Dinnes, Yemisi Takwoingi, Clare
  Davenport, Mariska~{MG} Leeflang, René Spijker, Lotty Hooft, Devy Emperador,
  Sabine Dittrich, Julie Domen, Sebastiaan R.~A. Horn, Ann Van~den Bruel, and
  Cochrane {COVID}-19 Diagnostic Test~Accuracy Group.
\newblock Signs and symptoms to determine if a patient presenting in primary
  care or hospital outpatient settings has {COVID}‐19 disease.
\newblock \penalty0 (7).
\newblock ISSN 1465-1858.
\newblock \doi{10.1002/14651858.CD013665}.
\newblock URL
  \url{https://www.cochranelibrary.com/cdsr/doi/10.1002/14651858.CD013665/full}.

\bibitem[Williamson et~al.(2020)Williamson, Walker, Bhaskaran, Bacon, Bates,
  Morton, Curtis, Mehrkar, Evans, Inglesby, Cockburn, McDonald, MacKenna,
  Tomlinson, Douglas, Rentsch, Mathur, Wong, Grieve, Harrison, Forbes,
  Schultze, Croker, Parry, Hester, Harper, Perera, Evans, Smeeth, and
  Goldacre]{williamson_opensafely_2020}
E.~J. Williamson, A.~J. Walker, K.~Bhaskaran, S.~Bacon, C.~Bates, C.~E. Morton,
  H.~J. Curtis, A.~Mehrkar, D.~Evans, P.~Inglesby, J.~Cockburn, H.~I. McDonald,
  B.~MacKenna, L.~Tomlinson, I.~J. Douglas, C.~T. Rentsch, R.~Mathur, A.~Y.~S.
  Wong, R.~Grieve, D.~Harrison, H.~Forbes, A.~Schultze, R.~Croker, J.~Parry,
  F.~Hester, S.~Harper, R.~Perera, S.~J.~W. Evans, L.~Smeeth, and B.~Goldacre.
\newblock {{F}actors associated with {C}{O}{V}{I}{D}-19-related death using
  {O}pen{S}{A}{F}{E}{L}{Y}}.
\newblock \emph{Nature}, 584\penalty0 (7821):\penalty0 430--436, 08 2020.

\bibitem[Clift et~al.(2020)Clift, Coupland, Keogh, Diaz-Ordaz, Williamson,
  Harrison, Hayward, Hemingway, Horby, Mehta, Benger, Khunti, Spiegelhalter,
  Sheikh, Valabhji, Lyons, Robson, Semple, Kee, Johnson, Jebb, Williams, and
  Hippisley-Cox]{clift_2020_living_risk}
A.~K. Clift, C.~A.~C. Coupland, R.~H. Keogh, K.~Diaz-Ordaz, E.~Williamson,
  E.~M. Harrison, A.~Hayward, H.~Hemingway, P.~Horby, N.~Mehta, J.~Benger,
  K.~Khunti, D.~Spiegelhalter, A.~Sheikh, J.~Valabhji, R.~A. Lyons, J.~Robson,
  M.~G. Semple, F.~Kee, P.~Johnson, S.~Jebb, T.~Williams, and J.~Hippisley-Cox.
\newblock {{L}iving risk prediction algorithm ({Q}{C}{O}{V}{I}{D}) for risk of
  hospital admission and mortality from Coronavirus 19 in adults: national
  derivation and validation cohort study}.
\newblock \emph{BMJ}, 371:\penalty0 m3731, 10 2020.

\bibitem[Buitrago-Garcia et~al.()Buitrago-Garcia, Egli-Gany, Counotte,
  Hossmann, Imeri, Ipekci, Salanti, and Low]{buitrago-garcia_occurrence_2020}
Diana Buitrago-Garcia, Dianne Egli-Gany, Michel~J. Counotte, Stefanie Hossmann,
  Hira Imeri, Aziz~Mert Ipekci, Georgia Salanti, and Nicola Low.
\newblock Occurrence and transmission potential of asymptomatic and
  presymptomatic {SARS}-{CoV}-2 infections: A living systematic review and
  meta-analysis.
\newblock 17\penalty0 (9):\penalty0 e1003346.
\newblock ISSN 1549-1676.
\newblock \doi{10.1371/journal.pmed.1003346}.
\newblock URL
  \url{https://journals.plos.org/plosmedicine/article?id=10.1371/journal.pmed.1003346}.

\bibitem[Millar et~al.(2022)Millar, Neyton, Seth, Dunning, Merson, Murthy,
  Russell, Keating, Swets, Sudre, Spector, Ourselin, Steves, Wolf, Docherty,
  Harrison, Openshaw, Semple, Baillie, and ISARIC‑4C]{millar_robust_2020}
Jonathan~E Millar, Lucile Neyton, Sohan Seth, Jake Dunning, Laura Merson,
  Srinivas Murthy, Clark~D Russell, Sean Keating, Maaike Swets, Carole~H Sudre,
  Timothy~D Spector, Sebastien Ourselin, Claire~J Steves, Jonathan Wolf,
  Annemarie~B Docherty, Ewen~M Harrison, Peter~{JM} Openshaw, Malcolm~G Semple,
  J~Kenneth Baillie, and ISARIC‑4C.
\newblock Distinct clinical symptom patterns in patients hospitalised with
  {COVID-19} in an analysis of 59,011 patients in the {ISARIC-4C} study.
\newblock \emph{Scientific Reports}, 12:\penalty0 6843, 2022.
\newblock \doi{10.1038/s41598-022-08032-3}.

\bibitem[Sudre et~al.(2021)Sudre, Lee, Lochlainn, Varsavsky, Murray, Graham,
  Menni, Modat, Bowyer, Nguyen, Drew, Joshi, Ma, Guo, Lo, Ganesh, Buwe, Pujol,
  du~Cadet, Visconti, Freidin, El-Sayed~Moustafa, Falchi, Davies, Gomez, Fall,
  Cardoso, Wolf, Franks, Chan, Spector, Steves, and
  Ourselin]{sudre_symptom_2020}
Carole~H. Sudre, Karla~A. Lee, Mary~Ni Lochlainn, Thomas Varsavsky, Benjamin
  Murray, Mark~S. Graham, Cristina Menni, Marc Modat, Ruth C.~E. Bowyer,
  Long~H. Nguyen, David~A. Drew, Amit~D. Joshi, Wenjie Ma, Chuan-Guo Guo,
  Chun-Han Lo, Sajaysurya Ganesh, Abubakar Buwe, Joan~Capdevila Pujol,
  Julien~Lavigne du~Cadet, Alessia Visconti, Maxim~B. Freidin, Julia~S.
  El-Sayed~Moustafa, Mario Falchi, Richard Davies, Maria~F. Gomez, Tove Fall,
  M.~Jorge Cardoso, Jonathan Wolf, Paul~W. Franks, Andrew~T. Chan, Tim~D.
  Spector, Claire~J. Steves, and S{\'e}bastien Ourselin.
\newblock Symptom clusters in {COVID}-19: A potential clinical prediction tool
  from the {COVID} {S}ymptom {S}tudy app.
\newblock \emph{Science Advances}, 7\penalty0 (12), 2021.
\newblock \doi{10.1126/sciadv.abd4177}.
\newblock URL \url{https://advances.sciencemag.org/content/7/12/eabd4177}.

\bibitem[Elliott et~al.(2021)Elliott, Whitaker, Bodinier, Eales, Riley, Ward,
  Cooke, Darzi, {Chadeau-Hyam}, and Elliott]{Elliott2021.02.10}
J~Elliott, M~Whitaker, B~Bodinier, O~Eales, S~Riley, H~Ward, G~Cooke, A~Darzi,
  M~{Chadeau-Hyam}, and P~Elliott.
\newblock Predictive symptoms for {COVID-19} in the community: {REACT-1} study
  of over 1 million people.
\newblock \emph{PLoS Medicine}, 28:\penalty0 e1003777, 2021.
\newblock \doi{10.1371/journal.pmed.1003777}.

\bibitem[Fragaszy et~al.(2021)Fragaszy, Shrotri, Geismar, Aryee, Beale,
  Braithwaite, Byrne, Fong, Gibbs, Hardelid, Kovar, Lampos, Nastouli,
  Navaratnam, Nguyen, Patel, Aldridge, Hayward, and on~behalf~of Virus
  Watch~Collaborative]{Fragaszy2021.05.14}
Ellen Fragaszy, Madhumita Shrotri, Cyril Geismar, Anna Aryee, Sarah Beale,
  Isobel Braithwaite, Thomas Byrne, Wing Lam~Erica Fong, Jo~Gibbs, Pia
  Hardelid, Jana Kovar, Vasileios Lampos, Eleni Nastouli, Annalan M~D
  Navaratnam, Vincent Nguyen, Parth Patel, Robert~W Aldridge, Andrew Hayward,
  and on~behalf~of Virus Watch~Collaborative.
\newblock Symptom profiles and accuracy of clinical definitions for covid-19 in
  the community. results of the virus watch community cohort.
\newblock \emph{medRxiv}, 2021.
\newblock \doi{10.1101/2021.05.14.21257229}.
\newblock URL
  \url{https://www.medrxiv.org/content/early/2021/06/11/2021.05.14.21257229}.

\bibitem[Hofmann and Zeuzem(2011)]{Hofmann:2011}
Wolf~Peter Hofmann and Stefan Zeuzem.
\newblock A new standard of care for the treatment of chronic {HCV} infection.
\newblock \emph{Nature Reviews Gastroenterology \& Hepatology}, 8\penalty0
  (5):\penalty0 257--264, 2011.

\bibitem[Deliu et~al.(2017)Deliu, Belgrave, Sperrin, Buchan, and
  Custovic]{Deliu:2017}
Matea Deliu, Danielle Belgrave, Matthew Sperrin, Iain Buchan, and Adnan
  Custovic.
\newblock Asthma phenotypes in childhood.
\newblock \emph{Expert Review of Clinical Immunology}, 13\penalty0
  (7):\penalty0 705--713, 2017.

\bibitem[Geifman et~al.(2018)Geifman, Kennedy, Schneider, Buchan, and
  Brinton]{Geifman:2018}
Nophar Geifman, Richard~E. Kennedy, Lon~S. Schneider, Iain Buchan, and
  Roberta~Diaz Brinton.
\newblock Data-driven identification of endophenotypes of {Alzheimer}'s disease
  progression: implications for clinical trials and therapeutic interventions.
\newblock \emph{Alzheimer's Research \& Therapy}, 10:\penalty0 4, 2018.

\bibitem[{NHS}(2021)]{NHSTT:2021}
{NHS}.
\newblock {Get tested for coronavirus (COVID-19)}, 2021.
\newblock URL
  \url{https://www.nhs.uk/conditions/coronavirus-covid-19/testing/get-tested-for-coronavirus/}.

\bibitem[Swann et~al.(2020)Swann, Holden, Turtle, Pollock, Fairfield, Drake,
  Seth, Egan, Hardwick, Halpin, Girvan, Donohue, Pritchard, Patel, Ladhani,
  Sigfrid, Sinha, Olliaro, Nguyen-Van-Tam, Horby, Merson, Carson, Dunning,
  Openshaw, Baillie, Harrison, Docherty, and Semple]{Swannm3249}
Olivia~V Swann, Karl~A Holden, Lance Turtle, Louisa Pollock, Cameron~J
  Fairfield, Thomas~M Drake, Sohan Seth, Conor Egan, Hayley~E Hardwick, Sophie
  Halpin, Michelle Girvan, Chloe Donohue, Mark Pritchard, Latifa~B Patel,
  Shamez Ladhani, Louise Sigfrid, Ian~P Sinha, Piero~L Olliaro, Jonathan~S
  Nguyen-Van-Tam, Peter~W Horby, Laura Merson, Gail Carson, Jake Dunning, Peter
  J~M Openshaw, J~Kenneth Baillie, Ewen~M Harrison, Annemarie~B Docherty, and
  Malcolm~G Semple.
\newblock Clinical characteristics of children and young people admitted to
  hospital with {COVID}-19 in {United Kingdom}: prospective multicentre
  observational cohort study.
\newblock \emph{BMJ}, 370:\penalty0 m3249, 2020.

\bibitem[Drew et~al.(2020)Drew, Nguyen, Steves, Menni, Freydin, Varsavsky,
  Sudre, Cardoso, Ourselin, Wolf, Spector, and Chan]{Drew1362}
David~A. Drew, Long~H. Nguyen, Claire~J. Steves, Cristina Menni, Maxim Freydin,
  Thomas Varsavsky, Carole~H. Sudre, M.~Jorge Cardoso, Sebastien Ourselin,
  Jonathan Wolf, Tim~D. Spector, and Andrew~T. Chan.
\newblock Rapid implementation of mobile technology for real-time epidemiology
  of {COVID}-19.
\newblock \emph{Science}, 368\penalty0 (6497):\penalty0 1362--1367, 2020.
\newblock ISSN 0036-8075.
\newblock \doi{10.1126/science.abc0473}.
\newblock URL \url{https://science.sciencemag.org/content/368/6497/1362}.

\bibitem[Pouwels et~al.(2021)Pouwels, House, Pritchard, Robotham, Birrell,
  Gelman, Vihta, Bowers, Boreham, Thomas, Lewis, Bell, Bell, Newton, Farrar,
  Diamond, Benton, Walker, and {COVID-19 Infection Survey Team}]{ONSKoen}
K~B Pouwels, T~House, E~Pritchard, J~V Robotham, P~J Birrell, A~Gelman, K~D
  Vihta, N~Bowers, I~Boreham, H~Thomas, J~Lewis, I~Bell, J~I Bell, J~N Newton,
  J~Farrar, I~Diamond, P~Benton, A~S Walker, and {COVID-19 Infection Survey
  Team}.
\newblock Community prevalence of {SARS}-{CoV}-2 in {England} from {April} to
  {November}, 2020: results from the {ONS} {Coronavirus} {Infection} {Survey}.
\newblock \emph{Lancet Public Health}, 6(1):e30-e38, Jan 2021.
\newblock \doi{doi: 10.1016/S2468-2667(20)30282-6}.

\bibitem[Landgraf and Lee(2020)]{landgraf2020dimensionality}
Andrew~J Landgraf and Yoonkyung Lee.
\newblock Dimensionality reduction for binary data through the projection of
  natural parameters.
\newblock \emph{Journal of Multivariate Analysis}, 180:\penalty0 104668, 2020.

\bibitem[{McInnes, L and Healy, J,}(2018)]{UMAP}
{McInnes, L and Healy, J,}.
\newblock {UMAP: Uniform Manifold Approximation and Projection for Dimension
  Reduction}, 2018.
\newblock [arXiv:1802.03426].

\bibitem[{McInnes} et~al.(){McInnes}, Healy, and Melville]{mcinnes_umap_2021}
Leland {McInnes}, John Healy, and James Melville.
\newblock {UMAP: Uniform Manifold Approximation and Projection} for dimension
  reduction — umap 0.5 documentation.
\newblock URL \url{https://umap-learn.readthedocs.io/en/latest/index.html}.

\bibitem[Coenen et~al.()Coenen, Pearce, and PAIR]{understanding_umap}
Andy Coenen, Adam Pearce, and Google PAIR.
\newblock Understanding {UMAP}.
\newblock URL \url{https://pair-code.github.io/understanding-umap/}.
\newblock Accessed 2023-02-02.

\bibitem[Lyu et~al.()Lyu, Wu, Wang, Huang, Wu, Cao, Zhao, Cao, Hu, Chen, Wang,
  Su, Zhang, Peng, Li, Cao, Hong, and Fang]{31035214}
Dongbin Lyu, Zhiguo Wu, Yun Wang, Qinte Huang, Zhenling Wu, Tongdan Cao, Jie
  Zhao, Yonghua Cao, Yingyan Hu, Jun Chen, Yong Wang, Yousong Su, Chen Zhang,
  Daihui Peng, Zezhi Li, Lan Cao, Wu~Hong, and Yiru Fang.
\newblock Disagreement and factors between symptom on self-report and clinician
  rating of major depressive disorder: {A} report of a national survey in
  {China}.
\newblock 253:\penalty0 141--146.
\newblock ISSN 0165-0327.
\newblock \doi{10.1016/j.jad.2019.04.073}.
\newblock URL
  \url{https://www.sciencedirect.com/science/article/pii/S0165032718331823}.

\bibitem[Silverstein et~al.()Silverstein, Faraone, Alperin, Biederman, Spencer,
  and Adler]{29172673}
Michael~J. Silverstein, Stephen~V. Faraone, Samuel Alperin, Joseph Biederman,
  Thomas~J. Spencer, and Lenard~A. Adler.
\newblock How informative are self-reported adult
  attention-deficit/hyperactivity disorder symptoms? {A}n examination of the
  agreement between the adult attention-deficit/hyperactivity disorder
  self-report scale v1.1 and adult attention-deficit/hyperactivity disorder
  investigator symptom rating scale.
\newblock 28\penalty0 (5):\penalty0 339--349.
\newblock ISSN 1044-5463.
\newblock \doi{10.1089/cap.2017.0082}.
\newblock URL \url{http://www.liebertpub.com/doi/10.1089/cap.2017.0082}.
\newblock Publisher: Mary Ann Liebert, Inc., publishers.

\bibitem[Chan et~al.()Chan, Sun, Aitchison, and Sivapalan]{33416510}
Eric~C. Chan, Yuting Sun, Katherine~J. Aitchison, and Sudhakar Sivapalan.
\newblock Mobile app–based self-report questionnaires for the assessment and
  monitoring of bipolar disorder: {S}ystematic review.
\newblock 5\penalty0 (1):\penalty0 e13770.
\newblock \doi{10.2196/13770}.
\newblock URL \url{https://formative.jmir.org/2021/1/e13770}.
\newblock Company: {JMIR} Formative Research Distributor: {JMIR} Formative
  Research Institution: {JMIR} Formative Research Label: {JMIR} Formative
  Research Publisher: {JMIR} Publications Inc., Toronto, Canada.

\bibitem[Wilson et~al.()Wilson, Bopp, Papalia, and Bopp]{31293064}
Oliver W.~A. Wilson, Christopher~M. Bopp, Zack Papalia, and Melissa Bopp.
\newblock Objective vs self-report assessment of height, weight and body mass
  index: {R}elationships with adiposity, aerobic fitness and physical activity.
\newblock 9\penalty0 (5):\penalty0 e12331.
\newblock ISSN 1758-8111.
\newblock \doi{10.1111/cob.12331}.
\newblock URL \url{https://onlinelibrary.wiley.com/doi/abs/10.1111/cob.12331}.
\newblock \_eprint: https://onlinelibrary.wiley.com/doi/pdf/10.1111/cob.12331.

\bibitem[Tomlinson et~al.()Tomlinson, Plenert, Dadzie, Loves, Cook, Schechter,
  Furtado, Dupuis, and Sung]{32567173}
Deborah Tomlinson, Erin Plenert, Grace Dadzie, Robyn Loves, Sadie Cook, Tal
  Schechter, Jennifer Furtado, L.~Lee Dupuis, and Lillian Sung.
\newblock Discordance between pediatric self-report and parent proxy-report
  symptom scores and creation of a dyad symptom screening tool (co-{SSPedi}).
\newblock 9\penalty0 (15):\penalty0 5526--5534.
\newblock ISSN 2045-7634.
\newblock \doi{10.1002/cam4.3235}.
\newblock URL \url{https://onlinelibrary.wiley.com/doi/abs/10.1002/cam4.3235}.
\newblock \_eprint: https://onlinelibrary.wiley.com/doi/pdf/10.1002/cam4.3235.

\bibitem[Hastie et~al.(2009)Hastie, Tibshirani, and Friedman.]{Hastie:2009}
T~Hastie, R~Tibshirani, and J~Friedman.
\newblock \emph{The Elements of Statistical Learning: Data Mining, Inference,
  and Prediction.}
\newblock Springer-Verlag, New York, 2nd edition, 2009.

\bibitem[Antonelli et~al.(2021)Antonelli, Penfold, Merino, Sudre, Molteni,
  Berry, Canas, Graham, Klaser, Modat, Murray, Kerfoot, Chen, Deng, Österdahl,
  Cheetham, Drew, Nguyen, Pujol, Hu, Selvachandran, Polidori, May, Wolf, Chan,
  Hammers, Duncan, Spector, Ourselin, and Steves]{antonelli_risk_2021}
Michela Antonelli, Rose~S. Penfold, Jordi Merino, Carole~H. Sudre, Erika
  Molteni, Sarah Berry, Liane~S. Canas, Mark~S. Graham, Kerstin Klaser, Marc
  Modat, Benjamin Murray, Eric Kerfoot, Liyuan Chen, Jie Deng, Marc~F.
  Österdahl, Nathan~J. Cheetham, David~A. Drew, Long~H. Nguyen, Joan~Capdevila
  Pujol, Christina Hu, Somesh Selvachandran, Lorenzo Polidori, Anna May,
  Jonathan Wolf, Andrew~T. Chan, Alexander Hammers, Emma~L. Duncan, Tim~D.
  Spector, Sebastien Ourselin, and Claire~J. Steves.
\newblock Risk factors and disease profile of post-vaccination {SARS}-{CoV}-2
  infection in {UK} users of the {COVID} {S}ymptom {S}tudy app: a prospective,
  community-based, nested, case-control study.
\newblock \emph{The Lancet Infectious Diseases}, 2021.

\end{thebibliography}
\end{document}